\documentclass[amsmath,amssymb,aps,prd,twocolumn,superscriptaddress,showpacs,preprintnumbers]{revtex4}

\usepackage[dvipdfmx]{graphicx}
\usepackage{multirow}
\usepackage{times}
\usepackage[dvipdfmx]{color}
\allowdisplaybreaks[1]
\usepackage[normalem]{ulem}  
\renewcommand\sout{\bgroup \color{red} \ULdepth=-.5ex \ULset}

\newcommand{\va}{\boldsymbol a}

\begin{document}
\preprint{KUNS-2565}

\title{
Relativistic Causal Hydrodynamics
Derived  from Boltzmann Equation: a novel reduction theoretical approach
}

\author{Kyosuke Tsumura}
\email[]{kyosuke.tsumura@fujifilm.com}
\affiliation{Analysis Technology Center,
  Research \& Development Management Headquarters,
  Fujifilm Corporation,
  Kanagawa 250-0193, Japan.}
\author{Yuta Kikuchi}
\email[]{kikuchi@ruby.scphys.kyoto-u.ac.jp}
\affiliation{Department of Physics, Faculty of Science, Kyoto University,
Kyoto 606-8502, Japan.}
\author{Teiji Kunihiro}
\email[]{kunihiro@ruby.scphys.kyoto-u.ac.jp}
\affiliation{Department of Physics, Faculty of Science, Kyoto University,
Kyoto 606-8502, Japan.}

\date{\today}

\begin{abstract}
We derive the second-order hydrodynamic equation
and the microscopic formulae of the relaxation times
as well as the transport coefficients systematically from the relativistic Boltzmann equation.
Our derivation is based on a novel development of the renormalization-group method,
a powerful reduction theory of dynamical systems,
which has been applied successfully to derive 
the non-relativistic second-order hydrodynamic equation.
Our theory nicely gives a compact expression
of the deviation of the distribution function 
in terms of the linearized collision operator,
which is different from those used as
an ansatz in the conventional fourteen-moment method.
It is confirmed that
the resultant microscopic expressions of the transport coefficients 
coincide with those derived
in the Chapman-Enskog expansion method.
Furthermore, we show that 
the microscopic expressions of the relaxation times
have natural and physically plausible forms.
We prove that
the propagating velocities of the fluctuations of the hydrodynamical variables
do not exceed the light velocity,
and hence our second-order equation ensures the desired causality.
It is also confirmed that
the equilibrium state is stable for any perturbation described by our equation.
\end{abstract}

\maketitle

\setcounter{equation}{0}
\section{
  Introduction
}
\label{sec:sec1}
The experiments of relativistic heavy ion collision at the Relativistic Heavy Ion Collider (RHIC)
at Brookhaven National Laboratory and the Large Hadron Collider (LHC) at CERN
seem to have created a hot matter that is most likely to be composed of quarks and gluons,
i.e., the quark-gluon plasma (QGP) \cite{Shuryak:2003xe,Gyulassy:2004zy}.
One of the most surprising findings is that
the created matter is well described by the hydrodynamics with tiny dissipation
\cite{Bass:2000ib,Teaney:2000cw,Teaney:2003kp,Hirano:2005wx,Hirano:2005xf,Nonaka:2006yn,Baier:2006um,
Baier:2006sr,Baier:2006gy,Romatschke:2007jx,Romatschke:2007mq,Bozek:2011gq,Hirano:2012kj}. 
Such a finding prompted
an interest in the origin of the viscosity in the gauge theories
and also the dissipative hydrodynamic equation.
The relativistic dissipative hydrodynamic equation
is also utilized in analyzing  various high-energy astrophysical phenomena \cite{balsara2001total}
including the accelerated expansion of the universe
by bulk viscosity of dark matter and/or dark energy \cite{Fabris:2005ts,Colistete:2007xi}.
  
One must say, however, that
the theory of relativistic hydrodynamics for a viscous fluid has not been established
on a firm ground yet,
although there have been many important studies
since Eckart's pioneering work \cite{Eckart:1940te}:
A naive relativistic extension  of the Navier-Stokes equation
has fundamental problems such as ambiguity of flow velocity \cite{Eckart:1940te,landau1959lifshitz,Van:2011yn,Osada:2011gx}, 
existence of unphysical instabilities \cite{Hiscock:1985zz}, 
and lack of causality \cite{Israel:1976tn,Israel:1979wp},
the last of which motivated people to introduce
the second-order hydrodynamic equation.
The way of formulation of the second-order hydrodynamics is, however, controversial
and an established equation has not been obtained
although some suggestive and promising approaches have been proposed. 

It is worth emphasizing here that
the second-order hydrodynamic equation that is free from the causality problem
even in the non-relativistic regime is not yet established, either.
The causality problem inherent in the first-order equation,
which we call generically the Navier-Stokes equation,
appears as the instantaneous propagation of information,
which is attributed to parabolicity of the equation \cite{stewart1971non,cattaneo1958form,Muller:1967zza,muller1993extended}.
In the seminal paper by Grad \cite{grad1949kinetic},
he showed that
the causality problem could be circumvented
by the moment method, which now bears his name:
It is found that
the thirteen-moment approximation to the functional forms
of the distribution function
leads to a hyperbolic non-relativistic equation
satisfying the causality, with finite propagation speeds of physical quantities.
Here,
we make a sideremark that
the description by the Grad equation may be called \textit{mesoscopic} \cite{jou1996extended,dedeurwaerdere1996foundations}
since it occupies an intermediate level between
the descriptions by the Navier-Stokes equation and the Boltzmann equation;
see also \cite{balescu1988transport}.
It should be noted, however, that
the dynamics described by the Grad equation has been recently shown inconsistent with
the underlying Boltzmann equation
in the mesoscopic scales of space and time
\cite{torrilhon2009special}.
Indeed,
Grad's moment method lacks a principle for determining
the functional form of the distribution function
that is consistent with the underlying Boltzmann equation,
and then
it is inevitable for the moment method to adopt an ad-hoc but seemingly plausible ansatz for it.
Although there are subsequent attempts
to construct the equation that respects both of the causality and the consistency
with the Boltzmann equation in the mesoscopic regime 
\cite{struchtrup2003regularization,levermore1996moment,torrilhon2010hyperbolic,ottinger2010thermodynamically},
the consistency between the resultant equations
and the mesoscopic dynamics of the Boltzmann equation remains unclear.

Nevertheless, 
many attempts were made to extend Grad's moment method
to establish a mesoscopic description of a {\em relativistic} system
\cite{Israel:1976tn,Israel:1979wp,Huovinen:2008te,Molnar:2009pq,El:2009vj,Bouras:2010hm,Denicol:2010br,
Denicol:2010xn,Denicol:2012cn,Pu:2009fj},
but with only a partial success, as anticipated.
For instance,
the celebrated Israel-Stewart equation \cite{Israel:1979wp},
which is a typical second-order relativistic hydrodynamic equation
derived from the Boltzmann equation based on the moment method with fourteen moments employed
is found to be incomplete if not incorrect
because the solutions behave differently
from those of the relativistic Boltzmann equation quantitatively \cite{Huovinen:2008te,Molnar:2009pq,El:2009vj}.
The incompleteness or incorrectness can be traced back
to ambiguous heuristic assumptions inherent in the moment method.
Quite recently, however,
some heuristic but promising methods have been proposed \cite{Denicol:2010xn,Denicol:2012cn,Jaiswal:2013npa}
to get rid of such drawbacks,
and it seems that
the resultant solutions indeed become closer to that of the Boltzmann equation.
Although their results are encouraging,
one must say that
their derivation are still based on plausible but ambiguous assumptions that require a microscopic foundation.
In fact, the constitutive equations contain the second-order spatial derivatives
of the hydrodynamical variables,
which necessarily leads to the parabolicity that should have been avoided.

Recently,
the mesoscopic dynamics or the second-order dynamics that respects the causality
has been extracted from the Boltzmann equation for the non-relativistic case
in the classical regime
without recourse to any ansatz for the functional forms of the distribution function
by two of the present authors \cite{Tsumura:2013uma}:
There ``the renormalization-group (RG) method''
\cite{Chen:1994zza,Chen:1995ena,Kunihiro:1995zt,Kunihiro:1996rs,kunihiro1998dynamical,Kunihiro:1997uy,Kunihiro:1998jp,Boyanovsky:1998aa,goto1999lie,Ei:1999pk,Boyanovsky:1999cy,ziane2000certain,nozaki2001renormalization,Hatta:2001ui,Boyanovsky:2003ui,Kunihiro:2005dd,
deville2008analysis,chiba2008c,chiba2009extension}
was adopted 
as a powerful method of the reduction theory of dynamical systems 
to reduce the Boltzmann equation to the 
mesoscopic dynamics.
The basic observation in these works is that
the first-order hydrodynamics is the slow dynamics achieved asymptotically
in the kinetic equation \cite{Hatta:2001ui,Kunihiro:2005dd}: 
The asymptotic dynamics is described  by the  zero modes of the linearized collision operator, 
which happen to be temperature, density, and fluid velocity,
i.e., the hydrodynamic variables.
In terms of the reduction theory of dynamical systems
\cite{kuramoto1989reduction},
it means that the hydrodynamic variables constitute
the natural coordinates of the invariant/attractive manifold
of the space of the distribution function
in which the asymptotic dynamics in the hydrodynamical regime is described.
In the RG method,
the hydrodynamic variables, which is now the would-be zero modes,
acquire the time dependence by the RG equation.
The resultant evolution equation
is nothing but the hydrodynamic equation, the Navier-Stokes equation.
Then,
the extension to the extraction of the mesoscopic dynamics
consists of developing the way to include some excited (fast) modes properly 
as additional components of the invariant/attractive manifold;
note that
the mesoscopic dynamics is faster than that described by the Navier-Stokes equation.
In \cite{Tsumura:2013uma},
the following natural conditions are found to give the adequate excited modes
to be incorporated in the hydrodynamical variables in the classical and non-relativistic case:
(A)~the resultant dynamics should be consistent with the reduced dynamics obtained 
by employing only the zero modes in the asymptotic regime;
(B)~the resultant dynamics should be as simple as possible
because we are interested to reduce the dynamics to a simpler one;
the term ``simple" means that
the resultant dynamics is described with a fewer number of dynamical variables
and is given by an equation composed of a fewer number of terms.
Here, we note that
the latter principle (B) is one of the fundamental principle
of the reduction theory of the dynamics as emphasized by Kuramoto \cite{kuramoto1989reduction}.
It was shown  that
these conditions lead to a concise scheme called the \textit{doublet scheme},
and that the resultant equation with thirteen dynamical variables
satisfies the causality in an apparent way
and has the same form as that of the Grad equation 
but with different  microscopic formulae of the transport coefficients and relaxation times;
the expressions of the transport coefficients
coincide with those  by the Chapman-Enskog method \cite{chapman1970mathematical},
the novel formulae of the relaxation times
allow a natural physical interpretation as the relaxation times.
This is an encouraging result!

A comment is in order here
on the relation between this work and \cite{Tsumura:2009vm} 
in which the two of the present authors (K.T. and T.K.) derived
a second-order hydrodynamic equation
from the relativistic Boltzmann equation 
on the basis of the RG method.
The derivation presented in \cite{Tsumura:2009vm},
however, contained an inconclusive part which is, in retrospect, incorrect, unfortunately.
Indeed the functional form of the excited modes was
not determined so as to solve the Boltzmann equation
but that adopted in the Israel-Stewart fourteen-moment method  
was mistakenly used as a possible solution:
It is known that the Israel-Stewart ansatz does not solve the relativistic Boltzmann equation.
In this work, 
we first find a proper solution to the relativistic Boltzmann equation
in the relevant kinetic regime
on the basis of an elaborated doublet scheme in the RG method,
and thus \textit{derive} the functional form of the excited modes 
that is consistent with the underlying Boltzmann equation. 
Then simply applying the RG equation,
we obtain
the second-order relativistic hydrodynamic equation,
which accordingly gives the correct asymptotic dynamics 
of the Boltzmann equation in the mesoscopic regime.

The present paper is an extension of the previous work \cite{Tsumura:2013uma}
to a relativistic case with the full quantum statistics as well as classical one.
We here remark that preliminary results in the classical case 
were announced in \cite{Tsumura:2012gq}.
Needless to say,
the quantum statistics is essential
in investigating the behavior of a quantum fluid composed of bosons and/or fermions.  
In the present paper,
we shall give a detailed and complete account of
the derivations of the causal hydrodynamic equations within the quantum and classical
statistics together with those of the microscopic expressions
of the transport coefficients and relaxation times.
We shall also show that
a concise and natural derivation is possible
for the excited modes that is given by the doublet scheme \cite{Tsumura:2013uma}
on the basis of the very principle of the reduction 
theory of the dynamics.

Moreover, we prove that
the propagating velocities of the fluctuations of the hydrodynamical variables
do not exceed the light velocity,
and hence our seconder-order equation ensures the causality as desired. 
It is also shown
that the equilibrium state is stable for any perturbation described by our equation.
We give a compact expression
of the deviation of the distribution function to 
be used in the fourteen-moment method.

This paper is organized as follows:
In Sec. \ref{sec:sec2},
we briefly summarize the basics of the relativistic Boltzmann equation.
In Sec. \ref{sec:sec3},
we derive the causal hydrodynamic equation
by applying the doublet scheme in the RG method,
and give the microscopic representations of the transport coefficients and relaxation times.
Then the basic properties of the resultant equation including the causality are
shown together with a comparison of the microscopic expressions
with those given by other methods. 
We devote Sec. \ref{sec:sec4} to a summary and concluding remarks.
In Appendix \ref{sec:app3},
we derive the functional forms of the excited modes
which are given by the faithful solution of the Boltzmann equation.
In Appendix \ref{sec:app1},
the explicit solution is given
for a linear equation with a time-dependent inhomogeneous term appearing in the text.
In Appendix \ref{sec:app2},
we present a detailed and lengthy derivation of the relaxation equations,
which shows how the microscopic expressions of the relaxation times and lengths are obtained.
In Appendix \ref{sec:app4},
we give a proof that our second-order equation is really causal
and that the static solution is stable against any fluctuations.

In this paper,
we use the natural unit, i.e., $\hbar = c = k_{\mathrm{B}} = 1$,
and the Minkowski metric $g^{\mu\nu} = \mathrm{diag}(+1,-1,-1,-1)$.

\section{
  Preliminary
}
\label{sec:sec2}
In this section,
we summarize the basic facts about
the relativistic Boltzmann equation \cite{de1980relatlvlatlc}.

\subsection{
  Basics of relativistic Boltzmann equation
}
\label{sec:ChapA-2-1}
The relativistic Boltzmann equation reads \cite{de1980relatlvlatlc,cercignani2002relativistic}
\begin{align}
  \label{eq:ChapA-2-1-001}
  p^\mu  \partial_\mu f_p(x) = C[f]_p(x),
\end{align}
where $f_p(x)$ denotes the one-particle distribution function
with $p^{\mu}$ being the four-momentum of the on-shell particle, i.e.,
$p^\mu p_\mu = p^2 = m^2$ and $p^0 > 0$.
The right-hand-side term $C[f]_p(x)$ denotes the collision integral
\begin{align}
  \label{eq:ChapA-2-1-002}
  C[f]_p(x) &\equiv \frac{1}{2!}\int\mathrm{d}p_1\mathrm{d}p_2\mathrm{d}p_3
  \omega(p, p_1|p_2,p_3)
  \nonumber\\
  &\times ( (1+af_p(x))(1+af_{p_1}(x))f_{p_2}(x) f_{p_3}(x)
  \nonumber\\
  &- f_p(x) f_{p_1}(x)(1+af_{p_2}(x))(1+af_{p_3}(x))),
\end{align}
where $\omega(p , p_1|p_2 , p_3)$ is
the transition probability due to the microscopic two-particle interaction
with the symmetry property
\begin{align}
  \label{eq:ChapA-2-1-003}
  &\omega(p, p_1|p_2,p_3) = \omega(p_2, p_3|p, p_1)
  \nonumber\\
  &= \omega(p_1,p|p_3, p_2) = \omega(p_3 , p_2|p_1 ,p),
\end{align}
and the energy-momentum conservation
\begin{align}
  \label{eq:ChapA-2-1-004}
  \omega(p , p_1|p_2 , p_3) \propto \delta^4(p + p_1 - p_2 - p_3),
\end{align}
and $a$ represents the quantum statistical effect, i.e., $a=+1$ for boson, 
$a=-1$ for fermion, and $a=0$ for the Boltzmann gas.
In the following,
we suppress the arguments $x$,
and abbreviate an integration measure as 
\begin{align}
  \mathrm{d}p \equiv \mathrm{d}^3\boldsymbol{p}/[(2\pi)^3 p^0],
\end{align}
with $\boldsymbol{p}$ being the spatial components of the four momentum $p^\mu$
when no misunderstanding is expected.

For an arbitrary vector $\varphi_p$
\cite{footnote:vector},
the collision integral satisfies
the following identity thanks to the above-mentioned symmetry properties,
\begin{align}
  \label{eq:ChapA-2-1-006}
  \int\mathrm{d}p \varphi_p C[f]_p
  &= \frac{1}{2!}\frac{1}{4}
  \int\mathrm{d}p\mathrm{d}p_1\mathrm{d}p_2\mathrm{d}p_3 \,
  \omega(p \,,\, p_1|p_2 \,,\, p_3)
  \nonumber\\
  &\times  (\varphi_p+\varphi_{p_1}- \varphi_{p_2} -\varphi_{p_3})
  \nonumber\\
  &\times  ( (1+af_p)(1+af_{p_1})f_{p_2}  f_{p_3}
  \nonumber\\ 
  &- f_p f_{p_1}(1+af_{p_2})(1+af_{p_3})).
\end{align}
Substituting $(1,p^\mu)$ into $\varphi_p$ in Eq. (\ref{eq:ChapA-2-1-006}),
we find that $(1,p^\mu)$ are collision invariants satisfying
\begin{align}
  \label{eq:ChapA-2-1-007}
  \int\mathrm{d}p  \frac{1}{p^0}  C[f]_p &= 0,
  \\
  \label{eq:ChapA-2-1-008}
  \int\mathrm{d}p \frac{1}{p^0}  p^\mu  C[f]_p &= 0,
\end{align}
due to the particle-number and energy-momentum conservation in the collision process,
respectively.
We note that
the function
$\varphi_{0p} \equiv \alpha(x) + p^\mu \beta_\mu(x)$ 
is also a collision invariant
where $\alpha (x)$ and $\beta^\mu (x)$ are arbitrary functions of $x$.

Owing to the particle-number and energy-momentum conservation in the collision process leading
to Eqs. (\ref{eq:ChapA-2-1-007}) and (\ref{eq:ChapA-2-1-008}),
we have the balance equations
\begin{align}
  \label{eq:ChapA-2-1-009}
  \partial_\mu N^\mu
  &= 0,
  \\
  \label{eq:ChapA-2-1-010}
  \partial_\nu T^{\mu\nu}
  &= 0,
\end{align}
where
the particle current $N^\mu$ and the energy-momentum tensor $T^{\mu\nu}$ 
are defined by
\begin{align}
  \label{eq:particle_current}
  N^\mu &\equiv
  \int\mathrm{d}p\, p^\mu  f_p,
  \\
  \label{eq:energy-momentum_tensor}
  T^{\mu\nu} &\equiv
  \int\mathrm{d}p\, p^\mu p^\nu  f_p,
\end{align}
respectively.
It should be noted that
any dynamical properties are not contained in these equations
unless the evolution of $f_p$
has been obtained as a solution to Eq. (\ref{eq:ChapA-2-1-001}).

In the Boltzmann theory,
the entropy current may be defined \cite{de1980relatlvlatlc} by
\begin{align}
  \label{eq:ChapA-2-1-011}
  S^\mu \equiv - \int\mathrm{d}p \, p^\mu 
  \Bigg[ f_p\ln  f_p - \frac{(1+af_p)\ln(1+af_p)}{a} \Bigg].
\end{align}
The entropy current $S^\mu$ satisfies the divergence equation
\begin{align}
  \label{eq:ChapA-2-1-012}
  \partial_\mu S^\mu =  - \int\mathrm{d}p \, C[f]_p  
  \ln \Bigg[ \frac{f_p}{1+af_p} \Bigg],
\end{align}
because of Eq. (\ref{eq:ChapA-2-1-001}).
One sees that
$S^\mu$ is conserved only if $\ln (f_p/(1+af_p))$ is a collision invariant,
i.e., 
$\ln (f_p/(1+af_p)) = \varphi_{0p} = \alpha(x) + p^\mu \beta_\mu(x)$.
One thus finds \cite{de1980relatlvlatlc,cercignani2002relativistic} that
the entropy-conserving distribution function can be parametrized as
\begin{align}
  \label{eq:ChapA-2-1-013}
  f_p
  = \frac{1}{\mathrm{e}^{(p^\mu  u_\mu-\mu)/T}-a}
  \equiv f^{\mathrm{eq}}_p,
\end{align}
where
$T$, $\mu$, and $u^\mu$ 
may depend on the space and time, and are interpreted as
the local temperature,
chemical potential,
and flow velocity, respectively,
with the normalization 
\begin{align}
  \label{eq:normalization_of_flow}
  u^\mu  u_\mu = 1.
\end{align}
Thus the function (\ref{eq:ChapA-2-1-013})
is identified with the local equilibrium distribution function.
We see that the collision integral identically vanishes 
for the local equilibrium distribution $f^{\mathrm{eq}}_p$ as
\begin{align}
  \label{eq:ChapA-2-1-015}
  C[f^{\mathrm{eq}}]_p = 0,
\end{align}
owning to the detailed balance
\begin{align}
  &\omega (p , p_1|p_2 , p_3) 
  \big[ (1+af^{\mathrm{eq}}_{p})(1+af^{\mathrm{eq}}_{p_1})
  f^{\mathrm{eq}}_{p_2} f^{\mathrm{eq}}_{p_3} 
  \nonumber\\
  &- f^{\mathrm{eq}}_p  f^{\mathrm{eq}}_{p_1}
  (1+af^{\mathrm{eq}}_{p_2}) (1+af^{\mathrm{eq}}_{p_3}) \big] = 0,
\end{align}
guaranteed by the energy-momentum conservation (\ref{eq:ChapA-2-1-004}).

Substituting $f_p = f^{\mathrm{eq}}_{p}$
into Eqs. (\ref{eq:particle_current}) and (\ref{eq:energy-momentum_tensor}),
we have
\begin{align}
  \label{eq:Chap0-001}
  N^{\mu} &= n \, u^\mu \equiv N^\mu_0,
  \\
  \label{eq:Chap0-002}
  T^{\mu\nu} &= e u^\mu u^\nu - P \Delta^{\mu\nu} \equiv T^{\mu\nu}_0,
\end{align}
with
\begin{align}
  \label{eq:ChapA-2-2-004}
  \Delta^{\mu\nu} \equiv g^{\mu\nu} - u^\mu u^\nu.
\end{align}
Here, 
$n$, $e$, and $P$ denote the particle-number density, internal energy, and pressure, respectively,
whose microscopic representations
are given by
\begin{align}
  \label{eq:Chap0-004}
  n &\equiv \int\mathrm{d}p 
  f^{\mathrm{eq}}_p (p \cdot u)
  \nonumber\\
  &= (2\pi)^{-3}  4\pi  m^3  \sum_{k=1}^{\infty}a^{k-1}\mathrm{e}^{k\mu/T}
   (km/T)^{-1}  K_2(km/T),
 \\
  \label{eq:Chap0-005}
  e &\equiv \int\mathrm{d}p f^{\mathrm{eq}}_p (p\cdot u)^2 
  \nonumber\\
  &= m n \Bigg[
    \frac{\sum_{k=1}^{\infty} a^{k-1}\mathrm{e}^{k\mu/T} (km/T)^{-1} K_3(km/T)}
    {\sum_{l=1}^{\infty} a^{l-1}\mathrm{e}^{l\mu/T} (lm/T)^{-1}  K_2(lm/T)} 
  \nonumber\\
  &- \frac{\sum_{k=1}^{\infty} a^{k-1}\mathrm{e}^{k\mu/T} (km/T)^{-2}  K_2(km/T)}
  {\sum_{l=1}^{\infty} a^{l-1}\mathrm{e}^{l\mu/T} (lm/T)^{-1}  K_2(lm/T)} \Bigg],
 \\
  \label{eq:Chap0-006}
  P &\equiv \int\mathrm{d}p  f^{\mathrm{eq}}_p  (-p^\mu p^\nu  \Delta_{\mu\nu}/3) 
  \nonumber\\
  &= m n
  \frac{\sum_{k=1}^{\infty} a^{k-1}\mathrm{e}^{k\mu/T} (km/T)^{-2} K_2(km/T)}
  {\sum_{l=1}^{\infty} a^{l-1}\mathrm{e}^{l\mu/T} (lm/T)^{-1} K_2(lm/T)},
\end{align}
with $K_2(z)$ and $K_3(z)$ being
the second- and third-order modified Bessel functions.
Setting $a=0$ in the above expressions,
we can check that the classical expressions
for $n$, $e$, and $P$ \cite{de1980relatlvlatlc} are reproduced.
We note that
$N^{\mu}_0$ and $T^{\mu\nu}_0$ in Eqs. (\ref{eq:Chap0-001}) and (\ref{eq:Chap0-002})
are identical to those in the relativistic Euler equation,
which describes the fluid dynamics
without dissipative effects,
and $n$, $e$, and $P$
defined by Eqs. (\ref{eq:Chap0-004})-(\ref{eq:Chap0-006})
are the equations of state of the dilute gas.
Since the entropy-conserving distribution function $f^{\mathrm{eq}}_{p}$
reproduces the relativistic Euler equation,
we find that the dissipative effects are attributable to the deviation of $f_{p}$
from $f^{\mathrm{eq}}_{p}$.


\section{
  Relativistic causal hydrodynamics
  by doublet scheme in RG method
}
\label{sec:sec3}
In this section,
we derive the causal relativistic hydrodynamic equation as
the mesoscopic dynamics
from the relativistic Boltzmann equation (\ref{eq:ChapA-3-1-015}):
The derivation is based on
the doublet scheme in the RG method developed 
for the non-relativistic case
in \cite{Tsumura:2013uma};
the present formulation is an extension
to the relativistic case 
and given in a simplified and more transparent manner.
We examine some properties of the resultant equation,
concerning
the frame, the stability of the equilibrium state, and the causality
as well as the microscopic representations
of the transport coefficients and the relaxation times.
It will be noted that
our formalism solving the Boltzmann equation gives the compact expression
of the perturbed distribution function,
which may be used in the moment method 
as the proper ansatz of the distribution function
that can lead to the hydrodynamic equation consistent with the Boltzmann equation.

\subsection{
  Reduced dynamics by RG method
}
\label{sec:ChapBr-4-2}

\subsubsection{
  Macroscopic-frame vector
}
Since we are interested in the hydrodynamic regime
to be realized asymptotically
where the time and space dependence of the physical quantities are small,
we try to solve Eq. (\ref{eq:ChapA-2-1-001})
in the situation where the space-time variation of $f_p(x)$ is small
and the space-time scales are coarse-grained from those in the kinetic regime.
To make a coarse graining with the Lorentz covariance being retained,
we introduce a  time-like Lorentz vector denoted by
 $ \va^\mu$
with $\va^2 > 0$ and $\va^0 > 0$ \cite{Tsumura:2007ji,Tsumura:2011cj},
which may depend on $x^\mu$;
$\va^\mu = \va^\mu(x)$.
Thus, $\va^\mu$
specifies the covariant but macroscopic coordinate system where 
the local rest frame of the flow velocity and/or the flow velocity itself are defined:
Since such a coordinate system is called \textit{frame},
we call $\va^\mu$ the \textit{macroscopic frame vector}.
In fact, with the use of $\va^\mu$,
we define the covariant and macroscopic coordinate system $(\tau,\sigma^\mu)$
from the space-time coordinate $x^\mu$ as
$\mathrm{d}\tau \equiv \va^\mu \mathrm{d}x_\mu$ and 
$\mathrm{d}\sigma^\mu \equiv ( g^{\mu\nu} - \va^\mu\va^\nu/\va^2)\mathrm{d}x_\nu$,
which lead to
derivatives given by
$\partial/\partial\tau = (\va^\mu/\va^2) \partial_\mu$ and
$\partial/\partial\sigma_\mu = ( g^{\mu\nu} - \va^\mu \va^\nu/\va^2 )\partial_\nu$.

Then, the relativistic Boltzmann equation (\ref{eq:ChapA-2-1-001}) 
in the new coordinate system $(\tau,\sigma^\mu)$
is written as
\begin{align}
  \label{eq:ChapA-3-1-011}
  p \cdot \va(\tau,\sigma)\frac{\partial}{\partial \tau} f_p(\tau,\sigma)
  + p^\mu \frac{\partial}{\partial\sigma^\mu} f_p(\tau,\sigma)
  = C[f]_p(\tau,\sigma),
\end{align}
where $\va^\mu(\tau,\sigma) \equiv \va^\mu(x)$
and $f_p(\tau,\sigma) \equiv f_p(x)$.
We remark the prefactor of the time derivative
is a Lorentz scalar and positive definite;
$p \cdot \va(\tau,\sigma) > 0$,
which is easily verified by taking the rest frame of $p^0$.

Since we are interested in a hydrodynamic solution
to Eq. (\ref{eq:ChapA-3-1-011}) as mentioned above,
we suppose that
the time variation of $\va^\mu(\tau,\sigma)$ is much smaller than that of the microscopic processes
and hence $\va^\mu(\tau,\sigma)$ has no $\tau$ dependence, i.e.,
\begin{align}
  \label{eq:ChapA-3-1-014}
  \va^\mu(\tau,\sigma)= \va^\mu(\sigma).
\end{align}
Then, with the use of Eq. (\ref{eq:ChapA-3-1-014}),
we shall convert Eq. (\ref{eq:ChapA-3-1-011}) into
\begin{align}
  \label{eq:ChapA-3-1-015}
  \frac{\partial}{\partial \tau} f_p(\tau,\sigma)
  &=\frac{1}{p \cdot \va(\sigma)} C[f]_p(\tau,\sigma)
  \nonumber\\
  &- \epsilon \frac{1}{p \cdot \va(\sigma)}
  p^\mu\frac{\partial}{\partial\sigma^\mu} f_p(\tau,\sigma).
\end{align}
Here,
the parameter $\epsilon$ is introduced
for characterizing the 
smallness of the inhomogeneity of the distribution function,
which may be identified with  
the ratio of the mean free path over the characteristic macroscopic length,
i.e., the Knudsen number.
Since $\epsilon$ appears
in front of the second term of the right-hand side of Eq. (\ref{eq:ChapA-3-1-015}),
the relativistic Boltzmann equation
has a form to which the perturbative expansion is applicable.

In the present analysis based on the RG method, 
the perturbative expansion of the distribution function  
with respect to $\epsilon$ is first performed
with the zeroth-order solution being the local equilibrium one,
which has no dissipative effects.
The dissipative effects
are taken into account in the higher orders;
the spatial inhomogeneity as the perturbation
gives rise to a deformation
of the distribution function which is responsible for the dissipative effects. 
Note that the deformation also can trigger a relaxation toward
the local equilibrium state. 
Thus,
the above rewrite of the equation with $\epsilon$ reflects 
the physical assumption
that only the spatial inhomogeneity
plays dual roles as the origin of the dissipation 
and the cause of a relaxation to the local equilibrium state.
It is noteworthy that
our RG method applied to the non-relativistic Boltzmann equation
with the corresponding assumption successfully leads 
to
the non-relativistic causal hydrodynamic equation
\cite{Tsumura:2013uma},
which means that
the present approach 
is simply a relativistic generalization of the non-relativistic case.

\subsubsection{
  Construction of approximate solution around arbitrary initial time
}

In accordance with the general formulation
of the RG method \cite{Kunihiro:1995zt,Kunihiro:1997uy,Ei:1999pk},
let $f_p(\tau ,\sigma)$ be 
an exact solution yet to be obtained with an initial condition set up, say at $\tau=-\infty$.
Then we pick up an arbitrary time $\tau=\tau_0$ 
in the (asymptotic) hydrodynamic regime,
and try to obtain the perturbative solution $\tilde{f}_p$ to Eq. (\ref{eq:ChapA-3-1-015})
around the time $\tau = \tau_0$ with the initial condition
\begin{align}
  \label{eq:ChapA-4-1-001}
  \tilde{f}_p(\tau = \tau_0,\sigma; \tau_0) = f_p(\tau_0,\sigma),
\end{align}
where we have made explicit that the solution has the $\tau_0$ dependence.
The initial value or the exact solution as well as the perturbative solution are expanded
with respect to $\epsilon$ as follows;
\begin{align}
  \label{eq:ChapA-4-1-002}
  \tilde{f}_p(\tau, \sigma; \tau_0)
  &= \tilde{f}_p^{(0)}(\tau, \sigma;\tau_0)
  + \epsilon \tilde{f}_p^{(1)}(\tau, \sigma;\tau_0)
  \nonumber\\
  &+ \epsilon^2 \tilde{f}_p^{(2)}(\tau,\sigma; \tau_0)
  + \cdots,
  \\
  \label{eq:ChapA-4-1-003}
  f_p(\tau_0, \sigma) &= f_p^{(0)}(\tau_0, \sigma)
  + \epsilon f_p^{(1)}(\tau_0, \sigma)
  \nonumber\\
  &+ \epsilon^2 f_p^{(2)}(\tau_0 , \sigma) + \cdots.
\end{align}
The respective initial conditions at $\tau = \tau_0$ are set up as
\begin{align}
  \label{eq:ChapA-4-1-004}
  \tilde{f}_p^{(l)}(\tau_0 ,\sigma ;\tau_0)
  = f_p^{(l)}(\tau_0,\sigma),
  \,\,\,\,\,\, l = 0, 1, 2,\cdots.
\end{align}
In the expansion, the zeroth-order value
$\tilde{f}_p^{(0)}(\tau_0 ,\sigma; \tau_0) = f_p^{(0)}(\tau_0 ,\sigma)$
is supposed to be as close as possible to an exact solution $f_p(\tau, \sigma)$.
In the RG method,
the globally valid solution is constructed by patching the
local solutions $\tilde{f}_p^{(0)}(\tau ,\sigma; \tau_0)$
which are only valid around $\tau=\tau_0$,
which is tantamount to
making an envelope curve of the perturbative solutions with $\tau_0$ being the parameter
characterizing the perturbative trajectories \cite{Kunihiro:1995zt,Kunihiro:1997uy}.

Substituting the above expansions into Eq. (\ref{eq:ChapA-3-1-015})
we obtain the series of the perturbative equations with respect to $\epsilon$, 
where the macroscopic frame vector is now replaced by 
a $\tau$-independent but $\tau_0$-dependent one \cite{Tsumura:2007ji,Tsumura:2011cj}
\begin{align}
  \label{eq:ChapA-4-1-005}
  \va^\mu(\sigma)
  =
  \va^\mu(\sigma ;\tau_0).
\end{align}

We have now a hierarchy of equations in order by order of $\epsilon$. 
As is mentioned before,
our strategy to obtain the mesoscopic dynamics
is constructing it as a minimal extension of the hydrodynamic one
that is to be realized {\em asymptotically after a long time}
within the Boltzmann equation.
Notice that the hydrodynamics is a closed slow dynamics described solely by 
the would-be zero modes of the linearized collision operator 
corresponding to the conservation laws.
The {\em slowest} dynamics will be given as a {\em stationary} solution, 
which actually exists for the zeroth order equation;
the stationary solution is nothing but the local equilibrium one \cite{Tsumura:2007ji,Tsumura:2011cj}.
In our way of the solution of the Boltzmann equation 
on the perturbation theory with the single 
expansion parameter $\epsilon$, 
the deviation of the distribution function from the local equilibrium one is
caused by the spatial inhomogeneity 
as given by the perturbative term in Eq. (\ref{eq:ChapA-3-1-015}) and 
hence is proportional to $\epsilon$.
We shall show that this setting of the analysis successfully solves the 
Boltzmann equation in a consistent way and leads to the mesoscopic dynamics.

With the above order counting in mind, let us construct the perturbative solution in 
the asymptotic regime  order by order.
The zeroth-order equation  reads
\begin{align}
  \label{eq:ChapA-4-1-006}
  \frac{\partial}{\partial \tau} \tilde{f}^{(0)}_p(\tau ,\sigma ;\tau_0)
  = \frac{1}{p \cdot \va(\sigma ;\tau_0)}
  C[\tilde{f}^{(0)}]_p(\tau ,\sigma ;\tau_0).
\end{align}
Since we are interested in the slow motion
which would be realized asymptotically as $\tau \rightarrow \infty$,
we should take the following stationary solution,
\begin{align}
  \label{eq:ChapA-4-1-007}
  \frac{\partial}{\partial \tau}\tilde{f}_p^{(0)}(\tau ,\sigma ;\tau_0) = 0,
\end{align}
which is realized when
$\tilde{f}_p^{(0)}(\tau ,\sigma ;\tau_0)$ is the fixed point,
\begin{align}
  \label{eq:ChapA-4-1-008}
  \frac{1}{p \cdot \va(\sigma ;\tau_0)} 
  C[\tilde{f}^{(0)}]_p(\tau ,\sigma ;\tau_0) = 0,
\end{align}
for arbitrary $\sigma$.
We see that
Eq. (\ref{eq:ChapA-4-1-008}) is identical to Eq. (\ref{eq:ChapA-2-1-015}),
and hence
$\tilde{f}_p^{(0)}(\tau ,\sigma ; \tau_0)$
is found to be the local equilibrium distribution function (\ref{eq:ChapA-2-1-013}):
\begin{align}
  \label{eq:ChapA-4-1-009}
  \tilde{f}_p^{(0)}(\tau ,\sigma ;\tau_0)
  &= f^{\mathrm{eq}}_p(\sigma ;\tau_0)
  \nonumber\\
  &=\frac{1}{\mathrm{e}^{[p^\mu u_\mu(\sigma ;\tau_0)-\mu(\sigma ;\tau_0)]
  /T(\sigma ; \tau_0)}-a},
\end{align}
with $u^\mu(\sigma ;\tau_0) u_\mu(\sigma ;\tau_0) = 1$,
which implies that
\begin{align}
  \label{eq:ChapA-4-1-010}
  f_p^{(0)}(\tau_0 ,\sigma)
  = \tilde{f}_p^{(0)}(\tau = \tau_0 ,\sigma ;\tau_0)
  = f^{\mathrm{eq}}_p(\sigma ;\tau_0).
\end{align}
The five would-be integral constants
$T(\sigma ;\tau_0)$, $\mu(\sigma ;\tau_0)$, and $u_\mu(\sigma ;\tau_0)$
are independent of $\tau$ but may depend on $\tau_0$ as well as $\sigma$,
and the local temperature, local chemical potential, and flow velocity can be naturally obtained.
For the sake of the convenience,
we define the following quantity:
\begin{align}
  \bar{f}^{\mathrm{eq}}_p(\sigma ;\tau_0)
  &\equiv
  1 + a f^{\mathrm{eq}}_p(\sigma ;\tau_0)\nonumber\\
  &=\frac{\mathrm{e}^{[p^\mu u_\mu(\sigma ;\tau_0)-\mu(\sigma ;\tau_0)]
  /T(\sigma ; \tau_0)}}{\mathrm{e}^{[p^\mu u_\mu(\sigma ;\tau_0)-\mu(\sigma ;\tau_0)]
  /T(\sigma ; \tau_0)}-a}.
\end{align}
We remark about an explicit form of $\va^\mu(\sigma ;\tau_0)$
that should be a Lorentz four vector described by the hydrodynamic variables
$T(\sigma ;\tau_0)$, $\mu(\sigma ;\tau_0)$, and $u_\mu(\sigma ;\tau_0)$ and their derivatives.
In the case of the first-order hydrodynamic equation,
it was shown \cite{Tsumura:2012ss} that
as long as such a $\va^\mu(\sigma ;\tau_0)$ is independent of the momentum $p^\mu$,
the leading terms of the resultant equation perfectly agree with those obtained with the choice
\begin{align}
  \label{eq:a_equal_u}
  \va^\mu(\sigma ;\tau_0) = u^\mu(\sigma ;\tau_0).
\end{align}
In the present work,
we will present the analysis
that is based on this choice,
and derive the second-order hydrodynamic equation as
a natural extension of the first-order one obtained in \cite{Tsumura:2012ss}.
In the following,
we suppress
the coordinate arguments $(\sigma ;\tau_0)$
when no misunderstanding is expected.

The choice $\va^\mu = u^\mu$ leads to the following identities
\begin{align}
  \frac{\partial}{\partial\tau} &= u^\mu \partial_\mu,
  \\
  \frac{\partial}{\partial\sigma_\mu} &= \Delta^{\mu\nu} \partial_\mu \equiv \nabla^\mu.
\end{align}
Note that $\partial/\partial\tau$ and $\nabla^\mu$
are the Lorentz-covariant temporal and spacial derivatives,
respectively.

Now that the preliminary set up is over,
let us move to the analysis of the first-order equation.
Inserting the expansion (\ref{eq:ChapA-4-1-002}) into
Eq. (\ref{eq:ChapA-3-1-015}) with
the setting (\ref{eq:a_equal_u}),
we have
the first-order equation as
\begin{align}
  \label{eq:first-order-eq}
  \frac{\partial}{\partial\tau} \tilde{f}^{(1)}_p(\tau)
  &= \int\mathrm{d}q f^{\mathrm{eq}}_{p}\bar{f}^{\mathrm{eq}}_{p} \hat{L}_{pq} 
  (f^{\mathrm{eq}}_{q} \bar{f}^{\mathrm{eq}}_{q})^{-1} 
  \tilde{f}^{(1)}_{q}(\tau)\nonumber\\
  &+ f^{\mathrm{eq}}_{p}\bar{f}^{\mathrm{eq}}_{p} F_{0p},
\end{align}
where $\hat{L}_{pq}$ is the linearized collision operator
\begin{align}
  \label{eq:ChapBr-2-2-012}
  \hat{L}_{pq}
  &\equiv (f^{\mathrm{eq}}_{p} \bar{f}^{\mathrm{eq}}_{p})^{-1} 
  \frac{1}{p\cdot u}
  \frac{\delta}{\delta f_{q}}C[f]_{p}\Bigg|_{f=f^{\mathrm{eq}}}
  f^{\mathrm{eq}}_{q} \bar{f}^{\mathrm{eq}}_{q}
  \nonumber\\
  &= - \frac{1}{2!} \frac{1}{p\cdot u} \int\mathrm{d}p_1\mathrm{d}p_2\mathrm{d}p_3 
  \omega(p,p_1|p_2,p_3)
  \nonumber\\
  &\times\frac{\bar{f}^{\mathrm{eq}}_{p_1} f^{\mathrm{eq}}_{p_2}f^{\mathrm{eq}}_{p_3}}
  {f^{\mathrm{eq}}_{p}} 
  (\delta_{pq} + \delta_{p_1q} - \delta_{p_2q} - \delta_{p_3q}),
\end{align}
and
$F_{0p}$ is an inhomogeneous term
\begin{align}
  \label{eq:F0}
  F_{0p} \equiv - (f^{\mathrm{eq}}_p \bar{f}^{\mathrm{eq}}_{p})^{-1} 
  \frac{1}{p\cdot u} p \cdot \nabla f^{\mathrm{eq}}_p.
\end{align}
For the sake of simplicity,
we rewrite Eq.(\ref{eq:first-order-eq})
in a vector form
\begin{align}
  \frac{\partial}{\partial\tau} \tilde{f}^{(1)}(\tau)
  = f^{\mathrm{eq}}\bar{f}^{\mathrm{eq}} \hat{L} 
  (f^{\mathrm{eq}} \bar{f}^{\mathrm{eq}})^{-1} 
  \tilde{f}^{(1)}(\tau) 
  + f^{\mathrm{eq}}\bar{f}^{\mathrm{eq}} F_{0},
\end{align}
where
we have treated $f^{\mathrm{eq}}_p$ and $\bar{f}^{\mathrm{eq}}_p$ as a diagonal matrix.

The linearized collision operator has some remarkable properties
that play important roles in the following analysis.
To see this, let us define 
an inner product for two arbitrary functions $\psi_p$ and $\chi_p$ by
\begin{align}
  \langle \psi , \chi \rangle
  \equiv \int\mathrm{d}p \, (p\cdot u) 
  f^{\mathrm{eq}}_p \bar{f}^{\mathrm{eq}}_p \psi_p \chi_p.
\end{align}
This inner product is a generalization of the one introduced in \cite{Tsumura:2007ji,Tsumura:2011cj}
for the classical statistics to the quantum one.
This inner product respects
the positive definiteness as
\begin{align}
  \label{eq:inner_positive}
  \langle \psi , \psi {\rangle} > 0,
  \,\,\,\quad {\rm for}\,\,\psi_p \ne 0,
\end{align}
because $(p\cdot u)$ in the inner product is positive-definite.
Then we find that
$\hat{L}$ is self-adjoint with respect to this inner product
\begin{align}
  \label{eq:self-adjoint}
  \langle \psi ,\hat{L} \chi \rangle
  &=
  - \frac{1}{4} \frac{1}{2!} \int\mathrm{d}p\mathrm{d}p_1\mathrm{d}p_2\mathrm{d}p_3 
  \omega(p,p_1|p_2,p_3)
  \nonumber\\
  &\times f^{\mathrm{eq}}_{p} f^{\mathrm{eq}}_{p_1}
  \bar{f}^{\mathrm{eq}}_{p_2} \bar{f}^{\mathrm{eq}}_{p_3} 
  (\psi_p + \psi_{p_1} - \psi_{p_2} - \psi_{p_3})
  \nonumber\\
  &\times(\chi_p + \chi_{p_1} - \chi_{p_2} - \chi_{p_3})
  \nonumber\\
  &= \langle \hat{L} \psi , \chi {\rangle},
\end{align}
and non-positive definite
\begin{align}
  \label{eq:non-positive}
  \langle \psi , \hat{L} \psi\rangle
  &=
  - \frac{1}{4} \frac{1}{2!} \int\mathrm{d}p\mathrm{d}p_1\mathrm{d}p_2\mathrm{d}p_3 
  \omega(p,p_1|p_2,p_3)
  \nonumber\\
  &\times f^{\mathrm{eq}}_{p} f^{\mathrm{eq}}_{p_1} 
  \bar{f}^{\mathrm{eq}}_{p_2} \bar{f}^{\mathrm{eq}}_{p_3} 
  (\psi_p + \psi_{p_1} - \psi_{p_2} - \psi_{p_3})^2
  \nonumber\\
  &\le 0,
\end{align}
with $\psi_p$ and $\chi_p$ being arbitrary vectors. 
The operator $\hat{L}$ has the five eigenvectors belonging to the zero eigenvalue;
\begin{align}
\label{eq:ChapA-4-2-013}
  \big[ \hat{L} \varphi_{0}^\alpha \big]_p = 0,
\end{align}
with
\begin{align}
  \label{eq:ChapA-4-2-012}
  \varphi_{0p}^\alpha \equiv \left\{
  \begin{array}{ll}
    \displaystyle{p^\mu,} & \displaystyle{\alpha = \mu,} \\[1mm]
    \displaystyle{1,}     & \displaystyle{\alpha = 4.}
  \end{array}
  \right.
\end{align}
We note that
$\varphi_{0p}^\alpha$ with $\alpha=0,\,\cdots,\,4$ are the collision invariants,
and span the kernel of $\hat{L}$.
We call $\varphi_{0p}^\alpha$ the zero modes.

To represent the solution to the first-order equation (\ref{eq:first-order-eq}) in a comprehensive way,
we define the projection operator $P_0$ onto the kernel of $\hat{L}$
which is called the P${}_0$ space and the projection operator $Q_0$
onto the Q${}_0$ space complement to the P${}_0$ space:
\begin{align}
  \label{eq:ChapA-4-2-014}
  \big[ P_0  \psi \big]_p 
  &\equiv
  \varphi_{0p}^\alpha  \eta^{-1}_{0\alpha\beta}
  \langle  \varphi_0^\beta ,\psi \rangle,
  \\
  \label{eq:ChapA-4-2-015}
  Q_0 &\equiv 1 - P_0,
\end{align}
where
$\eta^{-1}_{0\alpha\beta}$ is the inverse matrix of
the the P-space metric matrix $\eta_0^{\alpha\beta}$
defined by
\begin{align}
  \label{eq:ChapA-4-2-016}
  \eta_0^{\alpha\beta} \equiv \langle  \varphi_0^\alpha ,\varphi_0^\beta \rangle.
\end{align}

Now the solution to (\ref{eq:first-order-eq})
is given in terms of $P_0$ and $Q_0$ as
\begin{align}
  \label{eq:first-order_solution}
  \tilde{f}^{(1)}(\tau,\sigma;\tau_0)
  &=f^{\mathrm{eq}} \bar{f}^{\mathrm{eq}} \Big[
  \mathrm{e}^{\hat{L}(\tau-\tau_0)}  \Psi
  + (\tau-\tau_0)P_0 F_0
  \nonumber\\
  &+ (\mathrm{e}^{\hat{L}(\tau-\tau_0)} - 1)  \hat{L}^{-1} Q_0  F_0
  \Big],
\end{align}
with
\begin{align}
  \label{eq:first-order_initialvalue}
  f^{(1)}(\sigma;\tau_0) = \tilde{f}^{(1)}(\tau=\tau_0,\sigma;\tau_0)
  = f^{\mathrm{eq}} \bar{f}^{\mathrm{eq}} \Psi,
\end{align}
where $\Psi$ is the integral constant.
Here,
the second and third terms in Eq. (\ref{eq:first-order_solution}) 
describe the motion caused by the perturbation term $F_0$,
i.e., the spatial inhomogeneity,
while 
the first term can be identified with
the deviation from the stationary solution $f^{\mathrm{eq}}$,
which should be constructed in the perturbative expansion
with respect to the ratio
of the deviation from $f^{\mathrm{eq}}$ to $f^{\mathrm{eq}}$.
In fact,
the sum of $f^{\mathrm{eq}}$ and the first term, i.e.,
$f^{\mathrm{eq}}(1 + \epsilon \bar{f}^{\mathrm{eq}} \mathrm{e}^{\hat{L}(\tau-\tau_0)} \Psi)$, 
is nothing but the time-dependent solution to Eq. (\ref{eq:ChapA-4-1-006})
valid up to $O(\epsilon)$.
It is obvious that
this solution relaxes to $f^{\mathrm{eq}}(1 + \epsilon \bar{f}^{\mathrm{eq}} P_0 \Psi)$ in the asymptotic regime,
because
$f^{\mathrm{eq}} \bar{f}^{\mathrm{eq}} \mathrm{e}^{\hat{L}(\tau-\tau_0)} Q_0 \Psi$ vanishes
as $\tau \rightarrow \infty$.
In order to obtain the time-dependent solution
that describes the relaxation process to the stationary solution $f^{\mathrm{eq}}$,
we must suppose that $P_0 \Psi = 0$,
i.e.,
$\Psi$ contains no zero modes.
This is a kind of the matching condition.
Indeed,
if $\Psi$ were to contain zero modes,
such zero modes could be eliminated
by the redefinition of the zeroth-order initial value
specified by the local temperature $T(\sigma; \tau_0)$, chemical potential 
$\mu (\sigma; \tau_0)$, and 
flow velocity $u_\mu(\sigma; \tau_0)$.
In fact, $\delta f^{\mathrm{eq}}_p\equiv -f^{\mathrm{eq}}_p \bar{f}^{\mathrm{eq}}_p (p^\mu \alpha_\mu + \beta)$
can be written as a sum of the derivatives
 of $f^{\mathrm{eq}}_p$
with respect to
$T$, 
$\mu$, and 
$u_\mu$
with the identification,
$\alpha_{\mu} = \delta(u_\mu/T)=\delta u_{\mu}/T+u_{\mu}\delta(1/T)$ 
and $\beta = -\delta(\mu/T)=-\delta{\mu}/T-\mu\delta(1/T)$, which leads to
$\delta f^{\mathrm{eq}}_p =-f^{\mathrm{eq}}_p\bar{f}^{\mathrm{eq}}_p
(p^\mu \delta (u_\mu/T) - \delta (\mu/T))$. 
We note that the transverse component of $\alpha_{\mu}$ is proportional to 
$\delta u_{\mu}$.
Thus we see that
the possible existence of the zero modes in $\Psi$
would be renormalized into the local temperature, chemical potential, and flow velocity,
and absorbed into the redefinition of the initial distribution function at local equilibrium.

We note the appearance of the secular term proportional
to $\tau-\tau_0$ in Eq. (\ref{eq:first-order_solution}),
which apparently invalidate the perturbative solution when $|\tau-\tau_0|$ becomes large.

For later convenience,
let us expand $\mathrm{e}^{(\tau-\tau_0)\hat{L}}$ with respect to $\tau-\tau_0$
and retain the terms up to the first order as,
\begin{align}
 \label{eq:first-order_solution_expanded}
 \tilde{f}^{(1)}(\tau,\sigma;\tau_0)
 &\simeq f^{\mathrm{eq}} \bar{f}^{\mathrm{eq}}
 \Big[\Psi+(\tau-\tau_0)\hat{L}\Psi
 +(\tau-\tau_0)P_0F_0
 \nonumber \\
 &+(\tau-\tau_0)Q_0F_0\Big].
\end{align}
Here the neglected terms of $O((\tau-\tau_0)^2)$ are irrelevant
when we impose the RG equation,
which can be identified with the envelope equation \cite{Kunihiro:1995zt}
and thus the global solution is constructed
by patching the tangent line of the perturbative solution
at the arbitrary initial time $\tau=\tau_0$, as mentioned before. 

\begin{figure}
 \centering
 \includegraphics[width=5cm]{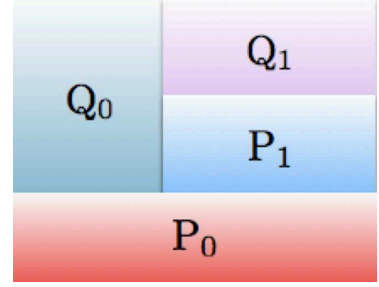}
 \caption{
 Decomposition of the solution space of the Boltzmann equation.
 The P${}_0$ space is the kernel of the linearized collision operator,
 while the Q${}_0$ space is spanned by excited mode,
 which is decomposed into the P${}_1$ and Q${}_1$ spaces.
 }
 \label{fig:1}
\end{figure}

Now the problem is how to extend the vector space beyond that spanned
by the zero modes to accommodate the excited modes
that are responsible for the mesoscopic dynamics
and should consist of the basic variables together with the zero modes
to describes the second-order hydrodynamics.
The vector space to which the excited modes belong are called the P${}_1$ space.
Here one should note that 
the P${}_1$ space is a subspace of the Q${}_0$ space,
as shown in Fig. \ref{fig:1}.
To this end,
let us see what the first-order solution (\ref{eq:first-order_solution_expanded})
tells us how to extend the vector space.
In fact,
to do that we only have to make the following requirement:
The tangent spaces of the perturbative solution at $\tau=\tau_0$
become as small as possible to simplify the obtained equation.
Instead of the two requirements (A) and (B) introduced in Sec. \ref{sec:sec1},
we utilize here
this one requirement
to determine
explicit forms of the vector $\Psi$ and the P${}_1$ space.
We note that
although the resultant forms of them are the same as those obtained with (A) and (B),
the derivation of them becomes more natural and straightforward.
Simplicity of the obtained equation is
one of the basic principles in the reduction theory of dynamical systems.
Here,
we note that
such tangent spaces are spanned by the terms proportional to $\tau-\tau_0$ in Eq. (\ref{eq:first-order_solution_expanded}),
while the P${}_1$ space is spanned by all the terms except for the zero modes in Eq. (\ref{eq:first-order_solution_expanded}).

Thus,
we can reduce this requirement to the following two conditions;
\begin{itemize}
 \item $\hat{L} \Psi$ and $Q_0F_0$ should belong to a common vector space.
 \item The P${}_1$ space is spanned by independent components of $\hat{L} \Psi$ and $\Psi$.
\end{itemize}
The first condition is restated as
that $\Psi$ and $\hat{L}^{-1}Q_0F_0$ should belong to a common vector space.
Therefore let us calculate $\hat{L}^{-1} Q_0 F_0$ and
examine the structure of the vector space to which it belongs.
This explicit calculation of the deformation of the distribution function 
 constitutes one of the central parts of the 
present work,  contrasting to the moment method in which some seemingly plausible ansatz is
adopted without any explicit solution.
A straightforward but somewhat tedious calculation of it
is worked out in Appendix \ref{sec:app3}:
 The result is given as
\begin{align}
  \label{eq:varphi1}
  \big[ \hat{L}^{-1} Q_0 F_0 \big]_p
  &= \frac{1}{T} \Bigg[ \big[ \hat{L}^{-1}\hat{\Pi} \big]_p (- \nabla\cdot u)
  - \big[ \hat{L}^{-1} \hat{J}^\mu\big]_p \frac{T}{h} \nabla_\mu \frac{\mu}{T}
  \nonumber\\
  &+ \big[ \hat{L}^{-1}\hat{\pi}^{\mu\nu}\big]_p \Delta_{\mu\nu\rho\sigma}\nabla^\rho u^\sigma
  \Bigg].
\end{align}
Here,
$\hat{\Pi}_p$, $\hat{J}^\mu_p$, and $\hat{\pi}^{\mu\nu}_p$
are microscopic representations of dissipative currents
whose definitions are given by
\begin{align}
  \label{eq:nine_vector_field}
  (\hat{\Pi}_p,\hat{J}^\mu_p,\hat{\pi}^{\mu\nu}_p) 
  = \frac{1}{p\cdot u}(\Pi_p,J^\mu_p,\pi^{\mu\nu}_p),
\end{align}
with
\begin{align}
  \Pi_p &\equiv (p\cdot u)^2\Bigg[
  \frac{1}{3} - \left.\frac{\partial P}{\partial e}\right|_{n}
  \Bigg]
  +(p\cdot u)
  \left.\frac{\partial P}{\partial n}\right|_{e}
  -\frac{1}{3}m^2,\\
  J^\mu_p &\equiv - \Delta^{\mu\nu} \, p_\nu ((p\cdot u) - h),\\
  \pi^{\mu\nu}_p &\equiv \Delta^{\mu\nu\rho\sigma}\,p_\rho\,p_\sigma.
\end{align}
In the above equations,
we have introduced the enthalpy per particle $h$ and the projection matrix $\Delta^{\mu\nu\rho\sigma}$
given by
\begin{align}
  \label{eq:hath}
  h &\equiv (e + P)/n,
  \\
   \Delta^{\mu\nu\rho\sigma}
  &\equiv 1/2(\Delta^{\mu\rho}\Delta^{\nu\sigma} + \Delta^{\mu\sigma}\Delta^{\nu\rho}
  - 2/3 \Delta^{\mu\nu}\Delta^{\rho\sigma}),
\end{align}
respectively.
It is notable that
the coefficients of the nine vectors $\big[ \hat{L}^{-1}\,\hat{\Pi} \big]_p$,
$\big[ \hat{L}^{-1}\hat{J}^\mu \big]_p$,
and $\big[ \hat{L}^{-1}\hat{\pi}^{\mu\nu} \big]_p$
are linearly independent, i.e., the following statement is true;
\begin{align}
  &\alpha (- T^{-1} \nabla\cdot u)
  + \beta^\mu (- h^{-1} \nabla_\mu (\mu/T))
  \nonumber\\
  &+ \gamma^{\mu\nu} (T^{-1} \Delta_{\mu\nu\rho\sigma}\nabla^\rho u^\sigma)
  = 0,\,\,\,\forall\,\,\,T,\,\mu,\,u^\mu
  \nonumber\\
  &\rightarrow
  \alpha = \beta^\mu = \gamma^{\mu\nu} = 0.
\end{align}
Thus,
we can take the following nine vectors
\begin{align}
  \label{eq:independentvectors_in_Psi}
  \big[ \hat{L}^{-1}\hat{\Pi} \big]_p,\,\,\,
  \big[ \hat{L}^{-1}\hat{J}^\mu \big]_p,\,\,\,
  \big[ \hat{L}^{-1}\hat{\pi}^{\mu\nu} \big]_p,
\end{align}
as a set of the bases of the vector space 
that $\big[ \hat{L}^{-1} Q_0 F_0 \big]_p$ and hence $\Psi$ belong to. 
Here we note that the above Lorentz vector and the tensor are transverse; 
\begin{align}
  \label{eq:transversality-of-bases1}
  \big[ \hat{L}^{-1}\hat{J}^\mu \big]_p
  &= \Delta^{\mu\nu} \big[ \hat{L}^{-1}\hat{J}_\nu \big]_p,
  \\
  \label{eq:transversality-of-bases2}
  \big[ \hat{L}^{-1}\hat{\pi}^{\mu\nu} \big]_p
  &= \Delta^{\mu\nu\rho\sigma} \big[ \hat{L}^{-1}\hat{\pi}_{\rho\sigma} \big]_p.
\end{align}

Thus we now see that $\Psi$ can be written as a linear combination of these bases as
\begin{align}
  \label{eq:Psi_in_RG}
  \Psi_{p} &=
  \Bigg[
    \frac{\big[ \hat{L}^{-1}\hat{\Pi} \big]_p}{\langle \hat{\Pi},\hat{L}^{-1}\hat{\Pi} {\rangle}}
  \Bigg] \Pi
  +\Bigg[
    \frac{h\big[ \hat{L}^{-1} \hat{J}^\mu \big]_p}{\frac{1}{3}\langle \hat{J}^\nu, \hat{L}^{-1}\hat{J}_\nu {\rangle}}
  \Bigg]  J_\mu
  \nonumber\\
  &+\Bigg[
    \frac{\big[ \hat{L}^{-1}\hat{\pi}^{\mu\nu} \big]_p}{\frac{1}{5}\langle \hat{\pi}^{\rho\sigma}, \hat{L}^{-1}\hat{\pi}_{\rho\sigma}{\rangle}}
  \Bigg] \pi_{\mu\nu}.
\end{align}
Here
we have introduced the following
nine
vectors as mere coefficients of the basis vectors: 
\begin{align}
  \Pi(\sigma ;\tau_0),\,\,\,J^\mu(\sigma ;\tau_0),\,\,\,\pi^{\mu\nu}(\sigma ;\tau_0).
\end{align}
We stress that the form of
$\Psi$ given in Eq. (\ref{eq:Psi_in_RG}) is the most generic expression 
that makes $\hat{L}\Psi$ and $Q_0F_0$ belong to the common space.

As is clear now,
we see that the P${}_1$ space is identified
with the vector space spanned by
$\hat{\Pi}_p$,
$\hat{J}^\mu_p$,
$\hat{\pi}^{\mu\nu}_p$,
$\big[\hat{L}^{-1}\hat{\Pi}\big]_p$,
$\big[\hat{L}^{-1}\hat{J}^\mu\big]_p$,
and $\big[\hat{L}^{-1}\hat{\pi}^{\mu\nu}\big]_p$.
The sets of $\hat{\Pi}$ and $\hat{L}^{-1}\hat{\Pi}$,
$\hat{J}^\mu$ and $\hat{L}^{-1}\hat{J}^\mu$, and $\hat{\pi}^{\mu\nu}$ and $\hat{L}^{-1}\hat{\pi}^{\mu\nu}$
are called the \textit{doublet modes} \cite{Tsumura:2013uma}.
The Q${}_0$ space is now decomposed
into the P${}_1$ space spanned by the doublet modes and the Q${}_1$ space
which is the complement to the P${}_0$ and P${}_1$ spaces.
The corresponding projection operators are denoted as $P_1$ and $Q_1$,
respectively.

Now we find that
the coefficients $J^\mu$ and $\pi^{\mu\nu}$ in Eq. (\ref{eq:Psi_in_RG}) 
are taken to be transverse without loss of generality; i.e.,
\begin{align}
  \label{eq:constraint01}
  J^\mu &= \Delta^{\mu\nu}J_\nu,
  \\
  \label{eq:constraint02}
  \pi^{\mu\nu} &= \Delta^{\mu\nu\rho\sigma}\pi_{\rho\sigma},
\end{align}
because of Eqs. (\ref{eq:transversality-of-bases1}) and (\ref{eq:transversality-of-bases2}).
The properties (\ref{eq:constraint01}) and (\ref{eq:constraint02})
lead to the following identities:
\begin{align}
  \label{eq:constraint1}
  u_\mu J^\mu
  &= u_\mu \pi^{\mu\nu} = \Delta_{\mu\nu} \pi^{\mu\nu} = 0,
  \\
  \label{eq:constraint2}
  \pi^{\mu\nu} &= \pi^{\nu\mu}.
\end{align}
It will be found that
$\Pi$, $J^\mu$, and $\pi^{\mu\nu}$ 
can be identified with
the bulk pressure, thermal flux, and stress pressure, respectively.

The second-order equation is written as
\begin{align}
  \label{eq:2nd-ordereq}
  \frac{\partial}{\partial\tau} \tilde{f}^{(2)}(\tau)
  &=f^{\mathrm{eq}} \bar{f}^{\mathrm{eq}} \hat{L}  
  (f^{\mathrm{eq}} \bar{f}^{\mathrm{eq}})^{-1} \tilde{f}^{(2)}(\tau) 
  \nonumber\\
  &+ f^{\mathrm{eq}} \bar{f}^{\mathrm{eq}} K(\tau-\tau_0),
\end{align}
with the time-dependent inhomogeneous term given by
\begin{align}
  \label{eq:K}
  K(\tau-\tau_0) &\equiv
  F_1 \tilde{f}^{(1)}(\tau) 
+ \frac{1}{2} B 
  \Big[ (f^{\mathrm{eq}} \bar{f}^{\mathrm{eq}})^{-1}\tilde{f}^{(1)}(\tau)) \Big]^2\nonumber\\
  &=
  F_1 f^{\mathrm{eq}} \bar{f}^{\mathrm{eq}} \Big[
  \mathrm{e}^{\hat{L}(\tau-\tau_0)}  \Psi
  + (\tau-\tau_0)P_0 F_0\nonumber\\
  &+ (\mathrm{e}^{\hat{L}(\tau-\tau_0)} - 1)  \hat{L}^{-1} Q_0  F_0
  \Big]\nonumber\\
  &+ \frac{1}{2} B 
  \Big[ \mathrm{e}^{\hat{L}(\tau-\tau_0)}  \Psi
  + (\tau-\tau_0)P_0 F_0\nonumber\\
  &+ (\mathrm{e}^{\hat{L}(\tau-\tau_0)} - 1)  \hat{L}^{-1} Q_0  F_0 \Big]^2.
\end{align}
Here,
$F_1$ and $B$ are matrices and their components are given by
\begin{align}
  \label{eq:ChapBr-4-1-040}
  F_{1pq} &\equiv - (f^{\mathrm{eq}}_p \bar{f}^{\mathrm{eq}}_{p})^{-1}
  \frac{1}{p \cdot u} p\cdot\nabla \delta_{pq},
  \\
  \label{eq:ChapBr-4-1-039}
  B_{pqr} &\equiv
   (f^{\mathrm{eq}}_p \bar{f}^{\mathrm{eq}}_p)^{-1}
  \frac{1}{p \cdot u} \frac{\delta^2}{\delta f_q \delta f_r} C[f]_p
  \Bigg|_{f = f^{\mathrm{eq}}} 
  f^{\mathrm{eq}}_q \bar{f}^{\mathrm{eq}}_q 
  f^{\mathrm{eq}}_r \bar{f}^{\mathrm{eq}}_r.
\end{align}
In Eq. (\ref{eq:K}), we have used the notation
\begin{align}
  \big[B\psi\chi\big]_p
  = \int\mathrm{d}q\mathrm{d}r B_{pqr}\psi_q\chi_r.
\end{align}

The solution to Eq. (\ref{eq:2nd-ordereq}) around $\tau \sim \tau_0$ is found to 
take the following form
\begin{align}
  \label{eq:2nd-ordersol}
  \tilde{f}^{(2)}(\tau, \sigma; \tau_0)
  &=
  f^{\mathrm{eq}} \bar{f}^{\mathrm{eq}}
  \Big[ (\tau-\tau_0)P_0\nonumber\\
  &+ (\tau-\tau_0)
  (\hat{L}-\partial/\partial s)
  P_1\mathcal{G}(s) Q_0
  \nonumber\\
  &- (1+(\tau-\tau_0)\partial/\partial s) Q_1 \mathcal{G}(s) Q_0\Big]\,K(s)\Big|_{s=0},
\end{align}
the initial value of which reads
\begin{align}
  \label{eq:2nd-orderinit}
  f^{(2)}(\sigma;\tau_0) &= \tilde{f}^{(2)}(\tau=\tau_0, \sigma; \tau_0)\nonumber\\
  &= -f^{\mathrm{eq}} \bar{f}^{\mathrm{eq}}
  Q_1\mathcal{G}(s)Q_0K(s)\Big|_{s=0}.
\end{align}
The derivation of this solution is presented in Appendix \ref{sec:app1},
where the complete expression of the solution not restricted to $\tau \sim \tau_0$ is given:
In Eq. (\ref{eq:2nd-ordersol}),
we have retained only terms up to the first order of ($\tau-\tau_0$),
and introduced a ``propagator" defined by 
\begin{align}
 \label{eq:propagator}
 \mathcal{G}(s)\equiv (\hat{L}-\partial/\partial s)^{-1}.
\end{align} 
We notice again the appearance of secular terms in Eq. (\ref{eq:2nd-ordersol}).

Summing up
the perturbative solutions up to the second order with respect to $\epsilon$,
we have the full expression of
the approximate solution around $\tau \sim \tau_0$ to the second order:
\begin{align}
  \label{eq:solution}
  &\tilde{f}(\tau, \sigma; \tau_0)\nonumber\\
  &=
  f^{\mathrm{eq}}
  +\epsilon f^{\mathrm{eq}} \bar{f}^{\mathrm{eq}}
  \Big[(1+(\tau-\tau_0)\hat{L})  \Psi + (\tau-\tau_0) F_0 \Big] 
  \nonumber\\
  &+ \epsilon^2f^{\mathrm{eq}} \bar{f}^{\mathrm{eq}}
  \Big[ (\tau-\tau_0)P_0
  + (\tau-\tau_0)
  (\hat{L}-\partial/\partial s)
  P_1\mathcal{G}(s)Q_0
  \nonumber\\
  &- (1+(\tau-\tau_0)\partial/\partial s)Q_1\mathcal{G}(s)Q_0\Big]\,K(s)\Big|_{s=0},
\end{align}
with the initial value
\begin{align}
  \label{eq:initial}
  f(\sigma;\tau_0)
  &= f^{\mathrm{eq}}+\epsilon f^{\mathrm{eq}} \bar{f}^{\mathrm{eq}} \Psi\nonumber\\
  &- \epsilon^2 f^{\mathrm{eq}} \bar{f}^{\mathrm{eq}} 
  Q_1 \mathcal{G}(s) Q_0 K(s)\Big|_{s=0}.
\end{align}
We note that 
the possible appearance of the fast motion caused by the Q${}_1$ space 
in Eq. \eqref{eq:solution}
is avoided by an appropriate choice of the initial value \eqref{eq:initial},
as in the first-order solution;
see Appendix \ref{sec:app1} for the detail.

A couple of remarks are in order here:
\begin{enumerate}
\item 
In the present approach,
we are solving the Boltzmann equation \eqref{eq:ChapA-2-1-001} as faithfully as possible,
in contrast to
the Israel-Stewart fourteen-moment method \cite{Israel:1979wp},
in which an ansatz for the solution is imposed in the form
$f = f^{\mathrm{eq}} +\epsilon f^{\mathrm{eq}} \bar{f}^{\mathrm{eq}}\Psi^{\mathrm{14M}}$
with $\Psi^{\mathrm{14M}} = a + b^\mu p_\mu + c^{\mu\nu} p_\mu p_\nu$.
Here, the coefficients $a$, $b^\mu$, and $c^{\mu\nu}$ are
definite functions of $T$, $\mu$, $u^\mu$, $\Pi$, $J^\mu$, and $\pi^{\mu\nu}$ \cite{Israel:1979wp}.
It is interesting that
our initial value $\Psi$ given in Eq. \eqref{eq:Psi_in_RG}
provides a foundation of the fourteen-moment method but with a novel form of
$\Psi^{\mathrm{14M}}$.
\item
Expanding $\mathcal{G}(s)Q_0$ in terms of $\hat{L}^{-1}{\partial}/{\partial s}$,
the term $\mathcal{G}(s)Q_0 K(s)|_{s=0}$ in Eqs. (\ref{eq:solution}) and (\ref{eq:initial})
is
reduced to
the form of infinite series as
\begin{align}
&\mathcal{G}(s)Q_0 K(s)\Big|_{s=0}\nonumber\\
&= \sum_{n=0}^{\infty} \hat{L}^{-1-n}Q_0\frac{\partial^n}{\partial s^n}K(s)\Big|_{s=0},
\end{align}
because
$\partial^n K(s)/\partial s^n|_{s=0}$ does not vanish for
any
$n$;
see Eq. (\ref{eq:K}).
Admittedly the existence of such
an infinite number of terms 
would be
undesirable
for the construction of the
(closed)
mesoscopic dynamics.
It will be found, however,  that
an averaging procedure for obtaining the mesoscopic dynamics nicely leads
to a cancellation of all the terms but single term in  the resultant equation of motion
thanks
to
the self-adjointness of $\hat{L}$
and
the structure of the P${}_1$ space spanned by the doublet modes;
see Eq. (\ref{eq:important_identity}) below.

%
\end{enumerate}

\subsubsection{
  RG improvement
  of perturbative expansion
}
We note that
the solution (\ref{eq:solution}) contains secular terms
that apparently invalidate the perturbative expansion
for $\tau$ away from the initial time $\tau_0$.
The point of the RG method lies in the fact that
we can utilize the secular terms to obtain
an asymptotic solution valid in a global domain.
Now we see that
$\tilde{f}_p(\tau,\sigma ;\tau_0)$ in Eq. (\ref{eq:solution})
provides a family of curves parameterized with $\tau_0$.
They are all on the exact solution $f_p(\sigma ;\tau)$ given by Eq. (\ref{eq:initial})
at $\tau=\tau_0$ up to $O(\epsilon^2)$,
but only valid locally for $\tau$ near $\tau_0$.
Thus,
it is conceivable that
the \textit{envelope} of the family of curves,
which is in contact with each local solution at $\tau = \tau_0$,
will give a global solution in our asymptotic situation
\cite{Kunihiro:1995zt,Kunihiro:1996rs,Ei:1999pk,Hatta:2001ui,Kunihiro:2005dd}.
According to the classical theory of envelopes,
the envelope that is in contact with any curve in the family at $\tau=\tau_0$ is obtained by
\begin{align}
  \label{eq:ChapA-4-5-001}
  \frac{\partial}{\partial\tau_0}
  \tilde{f}_p(\tau,\sigma ;\tau_0) \Bigg|_{\tau_0 = \tau} = 0,
\end{align}
where the subscript $p$ is restored for later convenience.
Equation (\ref{eq:ChapA-4-5-001})
is called the renormalization group equation \cite{Chen:1994zza},
and has
also
the meaning of the envelope equation \cite{Kunihiro:1995zt}. 
We call Eq.(\ref{eq:ChapA-4-5-001}) the RG/Envelope or RG/E equation following \cite{Kunihiro:1996rs}.
Now Eq.(\ref{eq:ChapA-4-5-001}) is really reduced to
\begin{align}
  \label{eq:RGeq}
  &\frac{\partial}{\partial\tau}\Big( f^{\mathrm{eq}}
  (1 + \epsilon\bar{f}^{\mathrm{eq}}\Psi) \Big)
  - \epsilon f^{\mathrm{eq}}\bar{f}^{\mathrm{eq}}
  \Big[ \hat{L} \Psi + P_0 F_0 + Q_0 F_0  \Big]
  \nonumber\\
  &-\epsilon^2 f^{\mathrm{eq}}\bar{f}^{\mathrm{eq}}
  \Bigg[ P_0 +
(\hat{L}-\partial/\partial s)
P_1\mathcal{G}(s)Q_0\nonumber\\
  &- (\partial/\partial s)\,Q_1\mathcal{G}(s)Q_0
  \Bigg]K(s)\Big|_{s=0} + O(\epsilon^3) = 0.
\end{align}
It is noted that
Eq. (\ref{eq:RGeq}) gives
the equation of motion governing the dynamics of the would-be fourteen integral constants
$T(\sigma;\tau)$, $\mu(\sigma;\tau)$, $u^\mu(\sigma;\tau)$,
$\Pi(\sigma;\tau)$, $J^\mu(\sigma;\tau)$, and $\pi^{\mu\nu}(\sigma;\tau)$.
The envelope function
is given by the initial value (\ref{eq:initial}) 
with the replacement of $\tau_0=\tau$ as
\begin{align}
  \label{eq:envelope}
  f^{\mathrm{G}}_p(\tau, \sigma) 
  &\equiv \tilde{f}_p(\tau, \sigma; \tau_0 = \tau)  \nonumber\\
  &\equiv f_p(\sigma; \tau_0 = \tau)  \nonumber\\
  &= f^{\mathrm{eq}}(1 + \epsilon \bar{f}^{\mathrm{eq}}\Psi)\nonumber\\
  &- \epsilon^2f^{\mathrm{eq}} \bar{f}^{\mathrm{eq}}
  Q_1 \mathcal{G}(s)Q_0K(s)\Big|_{s=0}\Bigg|_{\tau_0 = \tau}
  + O(\epsilon^3),
\end{align}
where the exact solution to
the RG/E equation 
(\ref{eq:RGeq}) is to be inserted.
We note that the envelope function
$f^{\mathrm{G}}_p(\tau, \sigma)$
is
actually the global solution that solves
the Boltzmann equation (\ref{eq:ChapA-3-1-015}) up to
$O(\epsilon^2)$ in a global domain in the asymptotic regime:
Indeed, for arbitrary $\tau(=\tau_0)$ in the global domain in the asymptotic regime,
we have
\begin{align}
\label{eq:proof-global-1}
\frac{\partial}{\partial \tau}f^{\mathrm{G}}_p(\tau, \sigma)
&=\frac{\partial}{\partial \tau}\tilde{f}_p(\tau, \sigma; \tau_0)\Bigg|_{\tau_0 = \tau} 
 + \frac{\partial}{\partial \tau_0}\tilde{f}_p(\tau, \sigma; \tau_0)\Bigg|_{\tau_0 = \tau} \nonumber \\
&= \frac{\partial}{\partial \tau}\tilde{f}_p(\tau, \sigma; \tau_0)\Bigg|_{\tau_0 = \tau},
\end{align}
where the RG/E equation (\ref{eq:ChapA-4-5-001}) has been used.
Furthermore, 
since $\tilde{f}_p(\tau, \sigma; \tau_0)$ solves Eq. (\ref{eq:ChapA-3-1-015})
with $\va^\mu(\sigma) = u^\mu(\sigma;\tau_0)$ up to $O(\epsilon^2)$,
the r.h.s. of Eq. (\ref{eq:proof-global-1}) reads
\begin{align}
\frac{\partial}{\partial \tau}\tilde{f}_p(\tau, \sigma; \tau_0)
&=
\frac{1}{p \cdot u(\sigma;\tau_0)} C[\tilde{f}]_p(\tau,\sigma;\tau_0)\nonumber\\
&- \epsilon \frac{1}{p \cdot u(\sigma;\tau_0)}
p^\mu\frac{\partial}{\partial\sigma^\mu} \tilde{f}_p(\tau,\sigma;\tau_0)
+ O(\epsilon^3).
\end{align}
Then inserting the definition of $f^{\mathrm{G}}_p(\tau,\sigma)$ given in 
the first line of Eq. (\ref{eq:envelope}),
we have
\begin{align}
\frac{\partial}{\partial \tau}f_p^{\mathrm{G}}(\tau, \sigma)
&=\frac{1}{p \cdot u(\sigma;\tau)} C[f^{\mathrm{G}}]_p(\tau,\sigma)
 \nonumber\\
  &- \epsilon \frac{1}{p \cdot u(\sigma;\tau)}
  p^\mu\frac{\partial}{\partial\sigma^\mu} f^{\mathrm{G}}_p(\tau,\sigma) +O(\epsilon^3).
\end{align}
This concludes the proof that the envelope function  
$f^{\mathrm{G}}_p(\tau, \sigma)$ is  the global solution to
the Boltzmann equation (\ref{eq:ChapA-3-1-015}) up to
$O(\epsilon^2)$ in a global domain.

It is noteworthy that
we have derived the mesoscopic dynamics
of the relativistic Boltzmann equation (\ref{eq:ChapA-3-1-015})
in the form of the pair of Eqs. (\ref{eq:RGeq}) and (\ref{eq:envelope}).
It is to be noted that
an infinite number of terms,
produced by $\mathcal{G}(s)$, are
included
both
in
the RG/E equation and the envelope function.

We observe that
the RG/E equation (\ref{eq:RGeq})
includes 
fast modes that should not be identified as the hydrodynamic modes even in the
second order ones. 
While these modes could be incorporated to make a Langevnized  hydrodynamic equation, 
we average out them to have the genuine hydrodynamic equation in the second order.
This averaging can be made
by taking 
the inner product of Eq. (\ref{eq:RGeq}) with the zero modes $\varphi^\alpha_{0p}$
and the excited modes $\big[ \hat{L}^{-1} (\hat{\Pi},\hat{J}^\mu,\hat{\pi}^{\mu\nu}) \big]_p$
used in the definition of $\Psi_p$.
The first averaging 
leads to
\begin{align}
  \label{eq:ChapB-RHD1}
  &\int\mathrm{d}p \varphi^{\alpha}_{0p}
  \Bigg[(p\cdot u)\frac{\partial}{\partial\tau} + \epsilon p\cdot\nabla\Bigg]
  \Bigg[ f^\mathrm{eq}_p 
  ( 1 + \epsilon \bar{f}^{\mathrm{eq}}_p \Psi_p) \Bigg]\nonumber\\
  &= 0 + O(\epsilon^3),
\end{align}
and the second averaging
\begin{align}
  \label{eq:ChapB-RHD2}
  &\int\mathrm{d}p \big[ \hat{L}^{-1}  (\hat{\Pi},\hat{J}^\mu,\hat{\pi}^{\mu\nu}) \big]_p 
  \Bigg[ (p\cdot u)  \frac{\partial}{\partial\tau}
  \nonumber\\
  &+ \epsilon p\cdot\nabla \Bigg]
  \Bigg[ f^\mathrm{eq}_p (1 + \epsilon \bar{f}^{\mathrm{eq}}_p \Psi_p) \Bigg]
  \nonumber\\
  &= \epsilon \langle \hat{L}^{-1} (\hat{\Pi},\hat{J}^\mu,\hat{\pi}^{\mu\nu}),
  \hat{L} \Psi \rangle
  \nonumber\\
  &+ \epsilon^2 \frac{1}{2}\langle \hat{L}^{-1} (\hat{\Pi},\hat{J}^\mu,\hat{\pi}^{\mu\nu}) ,
  B \Psi^2\rangle
  + O(\epsilon^3).
\end{align}
Here we have used the identity given by
\begin{align}
  \label{eq:important_identity}
  &\langle \hat{L}^{-1}(\hat{\Pi},\hat{J}^\mu,\hat{\pi}^{\mu\nu}),
  (\hat{L}-\partial/\partial s)P_1
  \mathcal{G}(s)Q_0 K(s)\Big|_{s=0}\rangle
  \nonumber\\
  &=
  \langle (\hat{L}-\partial/\partial s)\hat{L}^{-1}(\hat{\Pi},\hat{J}^\mu,\hat{\pi}^{\mu\nu}),
  P_1
  \mathcal{G}(s)Q_0 K(s)\Big|_{s=0}\rangle
  \nonumber\\
  &=
  \langle (\hat{L}-\partial/\partial s)\hat{L}^{-1}(\hat{\Pi},\hat{J}^\mu,\hat{\pi}^{\mu\nu}),
  \mathcal{G}(s)Q_0 K(s)\Big|_{s=0}\rangle
  \nonumber\\
  &=
  \langle \hat{L}^{-1}(\hat{\Pi},\hat{J}^\mu,\hat{\pi}^{\mu\nu}),
  (\hat{L}-\partial/\partial s)\mathcal{G}(s)Q_0 K(s)\Big|_{s=0}\rangle
  \nonumber\\
  &=
  \langle \hat{L}^{-1}(\hat{\Pi},\hat{J}^\mu,\hat{\pi}^{\mu\nu}),
  Q_0 K(s)\Big|_{s=0}\rangle
  \nonumber\\
  &=
  \langle \hat{L}^{-1} (\hat{\Pi},\hat{J}^\mu,\hat{\pi}^{\mu\nu}),
  K(0) \rangle
  \nonumber\\
  &=
  \langle \hat{L}^{-1}(\hat{\Pi},\hat{J}^\mu,\hat{\pi}^{\mu\nu}),
  F_1f^{\mathrm{eq}}\bar{f}^{\mathrm{eq}} \Psi
  \rangle
  \nonumber\\
  &+
  \frac{1}{2}\langle \hat{L}^{-1}(\hat{\Pi},\hat{J}^\mu,\hat{\pi}^{\mu\nu}),
  B  \Psi^2\rangle,
\end{align}
where
utilized are
the self-adjointness of $\hat{L}_{pq}$ shown in Eq. (\ref{eq:self-adjoint}),
the structure of the P${}_1$ space spanned by the doublet modes,
i.e.,
the pairs of
$\hat{\Pi}_p$ and $\big[ \hat{L}^{-1}  \hat{\Pi} \big]_p$, 
$\hat{J}^\mu_p$ and $\big[ \hat{L}^{-1}  \hat{J}^\mu \big]_p$, 
and
$\hat{\pi}^{\mu\nu}_p$ and $\big[ \hat{L}^{-1} \hat{\pi}^{\mu\nu} \big]_p$,
and the equality $K(0) = F_1 f^{\mathrm{eq}} \bar{f}^{\mathrm{eq}} \Psi + B\Psi^2/2$
derived from Eq. (\ref{eq:K}).

Thus
the pair of
Eqs. (\ref{eq:ChapB-RHD1}) and (\ref{eq:ChapB-RHD2}) 
constitutes the hydrodynamic equation in the second order, i.e.,
the equation of motion governing $T$, $\mu$, $u^\mu$,
$\Pi$, $J^\mu$, and $\pi^{\mu\nu}$.
It is to be noted that this pair of equations
is
free from an infinite number of terms in contrast to the RG/E equation (\ref{eq:RGeq})
and much simpler than it.
We stress that
this simplification 
through
the averaging by
$\hat{L}^{-1} (\hat{\Pi},\hat{J}^\mu,\hat{\pi}^{\mu\nu})$
is due to
the self-adjointness of $\hat{L}$
and the structure of the P${}_1$ space spanned by the doublet modes
$(\hat{\Pi},\hat{J}^\mu,\hat{\pi}^{\mu\nu})$
and $\hat{L}^{-1} (\hat{\Pi},\hat{J}^\mu,\hat{\pi}^{\mu\nu})$.


\subsection{
  Properties of the reduced dynamics
}

We now put back to $\epsilon=1$.
Noting that  
$(p\cdot u)\frac{\partial}{\partial\tau} + p\cdot\nabla = p^\mu\partial_\mu$,
we find  that Eq. (\ref{eq:ChapB-RHD1}) finally takes the following form
\begin{align}
  \label{eq:balanceeqbyRG}
  \partial_\mu J^{\mu\alpha}_{\mathrm{hydro}} = 0,
\end{align}
with
\begin{align}
  \label{eq:currentsbyRG}
  J^{\mu\alpha}_{\mathrm{hydro}}
  &\equiv \int\mathrm{d}p p^\mu \varphi^\alpha_{0p}
  f^\mathrm{eq}_p  ( 1 + \bar{f}^{\mathrm{eq}}_p\Psi_p)
  \nonumber\\
  &= \left\{
  \begin{array}{ll}
    \displaystyle{e u^\mu u^\nu - (P + \Pi) \Delta^{\mu\nu} + \pi^{\mu\nu},} & \displaystyle{\alpha = \nu,} \\[2mm]
    \displaystyle{n u^\mu + J^\mu,} & \displaystyle{\alpha = 4.}
  \end{array}
  \right.
\end{align}
We remark that
Eq. (\ref{eq:balanceeqbyRG})
is nothing but the balance equations
and $J^{\mu\nu}_{\mathrm{hydro}}$ and 
$J^{\mu 4}_{\mathrm{hydro}}$ can be identified with
the energy-momentum tensor $T^{\mu\nu}$
and particle current $N^\mu$ in the Landau-Lifshitz frame, respectively.
Indeed,
we can derive the same expression as $J^{\mu\alpha}_{\mathrm{hydro}}$
by substituting the distribution function
$f^{\mathrm{G}}(\tau, \sigma)$
in Eq. (\ref{eq:envelope})
into the definitions of $T^{\mu\nu}$ and $N^\mu$
given by Eqs. (\ref{eq:energy-momentum_tensor}) and (\ref{eq:particle_current}).

After a straightforward manipulation
whose details are presented in Appendix \ref{sec:app2},
we can reduce
Eq. (\ref{eq:ChapB-RHD2})
into the following relaxation equations:
\begin{align}
  \label{eq:relax1}
  \Pi
  &= -\zeta  \theta
  \nonumber\\
  &- \tau_\Pi \frac{\partial}{\partial\tau}\Pi - \ell_{\Pi J} \nabla\cdot J
  \nonumber\\
  &+\kappa_{\Pi\Pi} \Pi\theta\nonumber\\
  &+ \kappa^{(1)}_{\Pi J}J_\rho \nabla^\rho T + \kappa^{(2)}_{\Pi J}J_\rho \nabla^\rho \frac{\mu}{T}
  \nonumber\\
  &+ \kappa_{\Pi\pi}\pi_{\rho\sigma}\sigma^{\rho\sigma}
  \nonumber\\
  &+ b_{\Pi\Pi\Pi}  \Pi^2 + b_{\Pi JJ} J^\rho J_\rho
  + b_{\Pi\pi\pi}\pi^{\rho\sigma}\pi_{\rho\sigma},
  \\
  \label{eq:relax2}
  J^\mu
  &= \lambda \frac{T^2}{h^2}  \nabla^\mu \frac{\mu}{T}
  \nonumber\\
  &- \tau_J \Delta^{\mu\rho}\frac{\partial}{\partial\tau}J_{\rho}
  - \ell_{J\Pi} \nabla^\mu \Pi
  - \ell_{J\pi}  \Delta^{\mu\rho} \nabla_\nu {\pi^\nu}_\rho
  \nonumber\\
  &+\kappa^{(1)}_{J\Pi}\Pi \nabla^\mu T + \kappa^{(2)}_{J\Pi}\Pi \nabla^\mu \frac{\mu}{T}
  \nonumber\\
  &+ \kappa^{(1)}_{JJ}J^\mu \theta 
  + \kappa^{(2)}_{JJ}J_\rho \sigma^{\mu\rho}
  + \kappa^{(3)}_{JJ}J_\rho \omega^{\mu\rho}
  \nonumber\\
  &+ \kappa^{(1)}_{J\pi}\pi^{\mu\rho}\nabla_\rho T
  + \kappa^{(2)}_{J\pi}\pi^{\mu\rho}\nabla_\rho \frac{\mu}{T}
  \nonumber\\
  &+ b_{J\Pi J} \Pi J^\mu + b_{JJ\pi} J_\rho \pi^{\rho\mu},
  \\
  \label{eq:relax3}
  \pi^{\mu\nu}
  &= 2\eta \sigma^{\mu\nu}
  \nonumber\\
  &- \tau_\pi \Delta^{\mu\nu\rho\sigma}\frac{\partial}{\partial\tau}\pi_{\rho\sigma}
  - \ell_{\pi J} \nabla^{\langle\mu} J^{\nu\rangle} 
  \nonumber\\
  &+\kappa_{\pi\Pi} \Pi \sigma^{\mu\nu}
  \nonumber\\
  &+ \kappa^{(1)}_{\pi J}J^{\langle\mu}\nabla^{\nu\rangle} T
  + \kappa^{(2)}_{\pi J}J^{\langle\mu}\nabla^{\nu\rangle} \frac{\mu}{T}
  \nonumber\\
  &+ \kappa^{(1)}_{\pi\pi}\pi^{\mu\nu} \theta
+ \kappa^{(2)}_{\pi\pi} {\pi_{\rho}}^{\langle\mu}\sigma^{\nu\rangle\rho}
  + \kappa^{(3)}_{\pi\pi} {\pi_{\rho}}^{\langle\mu}\omega^{\nu\rangle\rho}
  \nonumber\\
  &+ b_{\pi\Pi\pi} \Pi \pi^{\mu\nu}
  + b_{\pi JJ} J^{\langle\mu} J^{\nu\rangle}
  + b_{\pi\pi\pi}\pi^{\lambda\langle\mu} {\pi^{\nu\rangle}}_{\lambda},
\end{align}
where we have introduced the notation
 $A^{\langle\mu\nu\rangle}\equiv\Delta^{\mu\nu\rho\sigma}A_{\rho\sigma}$
for a traceless and symmetric tensor.
Here 
 $\theta\equiv\nabla\cdot u$,
$\sigma^{\mu\nu}\equiv\Delta^{\mu\nu\rho\sigma}\nabla_{\rho\sigma}$,
and  $\omega^{\mu\nu} \equiv \frac{1}{2}  (\nabla^\mu u^\nu - \nabla^\mu u^\nu)$
denote the scalar expansion,  shear tensor and  vorticity, respectively.
We refer to  Appendix \ref{sec:app2}
for the explicit definitions of 
many other hydrodynamic valuables introduced in (\ref{eq:relax1})-(\ref{eq:relax3}).

Now the physical meaning of each term  
in (\ref{eq:relax1})-(\ref{eq:relax3}) should be clear:
The first lines in Eqs. (\ref{eq:relax1})-(\ref{eq:relax3})
are identical with the so-called constitutive  equations,
which define the relations between 
the dissipative variables $\Pi$, $J^\mu$, and $\pi^{\mu\nu}$
and
the thermodynamic forces given by the gradients of $T$, $\mu$, and $u^\mu$.
Substituting the constitutive equations
into the conserved currents $J^{\mu\alpha}_{\mathrm{hydro}}$ in Eq. (\ref{eq:currentsbyRG}),
we have the first-order hydrodynamics in the Landau-Lifshitz frame.
The terms in the other lines
are the new terms appearing in the second-order hydrodynamics.
The second lines denote
the relaxation terms given by the temporal and spatial derivatives of the dissipative variables,
which describe the relaxation processes 
of the dissipative variables to the thermodynamic forces.
The third, fourth, and fifth lines are composed of
the products of the thermodynamic forces and dissipative variables,
among which we remark that the vorticity term appears.
The final lines give the non-linear terms of the dissipative variables.

Our approach is based on a kind of statistical physics, and thus give
microscopic expressions of the transport and relaxation coefficeints. Here we 
present the resultant microscopic representations 
of the transport coefficients
, i.e., the bulk viscosity $\zeta$, thermal conductivity $\lambda$, and shear viscosity $\eta$,
and some of the relaxation times
$\tau_\Pi$, $\tau_J$, and $\tau_\pi$;
\begin{align}
  \label{eq:TC1byRG}
\zeta &= - \frac{1}{T}
  \langle \hat{\Pi} ,\hat{L}^{-1} \hat{\Pi} \rangle\equiv   \zeta^{\mathrm{RG}},
  \\
  \label{eq:TC2byRG}
\lambda &= \frac{1}{3T^2} 
  \langle \hat{J}^\mu, \hat{L}^{-1} \hat{J}_\mu \rangle \equiv   \lambda^{\mathrm{RG}},
  \\
  \label{eq:TC3byRG}
  \eta &= -\frac{1}{10T}
  \langle \hat{\pi}^{\mu\nu}, \hat{L}^{-1} \hat{\pi}_{\mu\nu} \rangle\equiv \eta^{\mathrm{RG}},
  \\
  \label{eq:RT1byRG}
\tau_{\Pi}&= - \frac{\langle \hat{\Pi}, \hat{L}^{-2} \hat{\Pi} \rangle
  }{
  \langle \hat{\Pi}, \hat{L}^{-1} \hat{\Pi} \rangle}\equiv   \tau^{\mathrm{RG}}_\Pi,
  \\
  \label{eq:RT2byRG}
\tau_J&= - \frac{\langle \hat{J}^\mu , \hat{L}^{-2} \hat{J}_\mu \rangle
  }{
  \langle \hat{J}^\rho,\hat{L}^{-1} \hat{J}_\rho \rangle}\equiv   \tau^{\mathrm{RG}}_J ,
  \\
  \label{eq:RT3byRG}
\tau_{\pi}&= - \frac{\langle \hat{\pi}^{\mu\nu}, \hat{L}^{-2} \hat{\pi}_{\mu\nu} \rangle
  }{
  \langle \hat{\pi}^{\rho\sigma} ,\hat{L}^{-1} \hat{\pi}_{\rho\sigma} \rangle}\equiv   \tau^{\mathrm{RG}}_\pi.
\end{align}
We leave the microscopic expressions of other coefficients in Appendix \ref{sec:app2}.
We first note that
$\zeta^{\mathrm{RG}}$, $\lambda^{\mathrm{RG}}$, and  $\eta^{\mathrm{RG}}$
are perfectly in agreement with
those of the Chapman-Enskog (CE) expansion method \cite{de1980relatlvlatlc},
which we denote as $\zeta^{\mathrm{CE}}$, $\lambda^{\mathrm{CE}}$, and $\eta^{\mathrm{CE}}$.
Here it is noteworthy that our expressions of the transport coefficients can be nicely 
rewritten in the form of Green-Kubo formula \cite{Jeon:1994if,Jeon:1995zm,Hidaka:2010gh} 
in the linear response theory. To see this, we first introduce 
the ``time-evolved'' vectors defined by
\begin{align}
  \label{eq:ChapA-5-2-008}
  (\hat{\Pi}_p(s),\hat{J}^\mu_p(s),\hat{\pi}^{\mu\nu}_p(s))
  \equiv
  \int\mathrm{d}q \big[ \mathrm{e}^{s\hat{L}} \big]_{pq}
  (\hat{\Pi}_q,\hat{J}^\mu_q,\hat{\pi}^{\mu\nu}_q),
\end{align}
where the time-evolution operator is given by the linearized collision operator.
Then, we have 
\begin{align}
  \label{eq:ChapA-5-2-002}
  \zeta^{\mathrm{RG}} &= \frac{1}{T}\int_0^\infty \mathrm{d}s 
  \langle \hat{\Pi}(0),\hat{\Pi}(s) \rangle,
  \\
  \label{eq:ChapA-5-2-003}
  \lambda^{\mathrm{RG}} &= - \frac{1}{3T^2}\int_0^\infty \mathrm{d}s
  \langle \hat{J}^\mu(0),\hat{J}_\mu(s) \rangle,
  \\
  \label{eq:ChapA-5-2-004}
  \eta^{\mathrm{RG}} &= \frac{1}{10T}\int_0^\infty \mathrm{d}s
  \langle \hat{\pi}^{\mu\nu}(0),\hat{\pi}_{\mu\nu}(s) \rangle.
\end{align}
We note that the integrands in the formulae have the meanings of the relaxation functions or
time correlation functions;
\begin{align} 
R_{\Pi}(s)&\equiv \frac{1}{T} \langle  \hat{\Pi}(0),\hat{\Pi}(s)\rangle, \\
R_J(s) &\equiv - \frac{1}{3T^2} \langle \hat{J}^\mu(0),\hat{J}_\mu(s) \rangle, \\
R_{\pi}(s)&=\frac{1}{10T} \langle \hat{\pi}^{\mu\nu}(0),\hat{\pi}_{\mu\nu}(s) \rangle.
\end{align}
We stress that the results of the transport coefficients all show the reliability of our 
approach  based on the doublet scheme in the RG method.
We remark that the naive version of moment method by
 Israel and Stewart (IS) fails to give the Chapman-Enskog formulae \cite{Israel:1979wp}, as is well known.

Thus it may be a good news for us that the explicit formulae of
the relaxation times given above
also differ from those given by  
IS \cite{Israel:1979wp},
which read
\begin{align}
\tau_\Pi^{\mathrm{IS}}
 & \equiv -\frac{
    \langle \Pi,\Pi \rangle
  }{
    \langle \Pi , \hat{L}  \Pi \rangle
  },
  \\
\tau_J^{\mathrm{IS}}
 & \equiv -\frac{
    \langle\, J^\mu , J_\mu \rangle
  }{
    \langle J^\rho,\hat{L} J_\rho \rangle
  },
  \\
\tau_\pi^{\mathrm{IS}}
 & \equiv -\frac{
    \langle \pi^{\mu\nu}, \pi_{\mu\nu} \rangle
  }{
    \langle \pi^{\rho\sigma}, \hat{L} \pi_{\rho\sigma} \rangle
  }.
\end{align}
Indeed we shall now show that our formulae of the relaxation times
allow a natural interpretation of them.
To see this, we rewrite the expressions of the relaxation times given in
Eqs. (\ref{eq:RT1byRG})-(\ref{eq:RT3byRG})
in terms of the time-evolved vectors again:
\begin{align}
  \label{eq:relax-tau-pi1-0}
  \tau^{\mathrm{RG}}_\Pi &=
  \frac{
    \int_0^\infty \mathrm{d}s\,s R_\Pi(s)
  }{
    \int_0^\infty \mathrm{d}s R_\Pi(s)
  }
  ,\\
  \label{eq:relax-tau-J-0}
  \tau^{\mathrm{RG}}_J &=
  \frac{
    \int_0^\infty \mathrm{d}s\,s
    R_J(s)
  }{
    \int_0^\infty \mathrm{d}s
    R_J(s)
  }
  ,\\
  \label{eq:relax-tau-pi2-0}
  \tau^{\mathrm{RG}}_{\pi} &=
  \frac{
    \int_0^\infty \mathrm{d}s\,s
    R_\pi(s)
  }{
    \int_0^\infty \mathrm{d}s
    R_\pi(s)
  }.
\end{align}
It is noteworthy that
all the relaxation times 
are expressed in terms of
the relaxation functions 
$R_\Pi(s)$, 
$R_J(s)$, and
$R_\pi(s)$,
respectively.
Then the formulae (\ref{eq:relax-tau-pi1-0})-(\ref{eq:relax-tau-pi2-0}) 
allow the natural interpretation of the resepective relaxation times as
the  correlation times in the respective relaxation functions.
We  emphasize that
it is for the first time that
the relaxation times are expressed in terms of the relaxation functions in the context of the
derivation of the second-order relativistic hydrodynamic equation
from the relativistic Boltzmann equation.


\subsection{
  Discussions
}
We now examine the basic properties of the
resultant
hydrodynamic equations
(\ref{eq:balanceeqbyRG}) and (\ref{eq:relax1})-(\ref{eq:relax3}).
First we show that our equation is really causal in the sense that the velocities
of any fluctuation around the equilibrium is less than that of the light velocity
with a detailed proof is left to
Appendix \ref{sec:app4},
where the stability of 
the static solution is also prooved.
Next we compare our formulae of the relaxation equations
with those derived by the moment method. Then we give numerical results
of the transport coefficients and relaxation times given by
Eqs. (\ref{eq:TC1byRG})-(\ref{eq:TC3byRG}) and (\ref{eq:RT1byRG})-(\ref{eq:RT3byRG}), respectively,
and compare them with those by other methods.

\subsubsection{
  Causal property of hydrodynamic equations obtained by RG method
}

We give a brief account of the proof that
the velocities of hydrodynamic modes described
by the hydrodynamic equations (\ref{eq:balanceeqbyRG}) and (\ref{eq:relax1})-(\ref{eq:relax3})
do not exceed the speed of light, i.e., the unity.
We note that
the detailed proof is presented in Appendix \ref{sec:app4}.

First,
we linearize the hydrodynamic equations around
equilibrium state specified by constant temperature, constant chemical potential, and constant fluid flow,
as follows:
\begin{eqnarray}
  \label{eq:linearized-hydrodynamic-equation}
  ( \Lambda\,A^{\alpha\beta,\gamma\delta} - \tilde{B}^{\alpha\beta,\gamma\delta}(k) )
  \, \delta \tilde{X}_{\gamma\delta}(\Lambda\,;\,k)
  = 0,
\end{eqnarray}
where
the matrices $A^{\alpha\beta,\gamma\delta}$ and $\tilde{B}^{\alpha\beta,\gamma\delta}(k)$
are defined in Eqs. (\ref{eq:matrixA1})-(\ref{eq:matrixA4})
and
(\ref{eq:matrixB1})-(\ref{eq:matrixB4}), respectively,
and
the variables $\delta \tilde{X}_{\alpha\beta}(\Lambda\,;\,k)$
are Fourier-Laplace transformations of $\delta X_{\alpha\beta}(\tau\,;\,\sigma)$ given by
\begin{eqnarray}
  \delta X_{\mu\nu} &\equiv& \frac{\Delta_{\mu\nu}}{3\,T\,\zeta^{\mathrm{RG}}}\Bigg|_{\mathrm{eq}}\,\delta \Pi
  - \frac{1}{2\,T\,\eta^{\mathrm{RG}}}\Bigg|_{\mathrm{eq}}\,\delta \pi_{\mu\nu},\\
  \delta X_{\mu 4} &\equiv& \frac{h}{T^2\,\lambda^{\mathrm{RG}}}\Big|_{\mathrm{eq}}\,\delta J_{\mu},\\
  \delta X_{4\mu} &\equiv& - \frac{1}{T}\Big|_{\mathrm{eq}}\,\delta u_\mu
  + \frac{u_\mu}{T^2}\Big|_{\mathrm{eq}}\,\delta T,\\
  \delta X_{44} &\equiv& \frac{1}{T}\Big|_{\mathrm{eq}}\,\delta\mu - \frac{\mu}{T^2}\Big|_{\mathrm{eq}}\,\delta T,
\end{eqnarray}
with the arguments $(\tau\,;\,\sigma)$ being omitted.
Here,
$\delta T$, $\delta \mu$, $\delta u^\mu$, $\delta \Pi$, $\delta J^\mu$, and $\delta \pi^{\mu\nu}$
are fluctuations from the equilibrium state
and all coefficients take values at the equilibrium state.
Furthermore,
$i\Lambda$ and $k^\mu$ are frequency and wavelength
conjugate to $\tau$ and $\sigma^\mu$, respectively.
We note that
$k^\mu$ is space-like vector, $k^2 < 0$, for any $k^\mu \ne 0$,
which satisfies $k^\mu = \Delta^{\mu\nu} k_\nu$
because of $\sigma^\mu = \Delta^{\mu\nu} \sigma_\nu$.
We also note that
the condition $\delta \tilde{X} \ne 0$ into Eq. (\ref{eq:linearized-hydrodynamic-equation})
leads to the dispersion relation $\Lambda = \Lambda(k)$.

Then,
as a typical quantity used for the check of the causality,
we examine a character velocity $v_{\mathrm{ch}}$
that is defined as
\begin{eqnarray}
  v_{\mathrm{ch}} \equiv \lim_{-k^2 \rightarrow \infty} \, \sqrt{ \frac{\partial}{\partial k_\mu}\Lambda(k)
  \cdot \frac{\partial}{\partial k^\mu}\Lambda(k) }.
\end{eqnarray}
With the use of
the explicit definitions of  $A^{\alpha\beta,\gamma\delta}$ and $\tilde{B}^{\alpha\beta,\gamma\delta}(k)$,
we can show that
\begin{eqnarray}
  v_{\mathrm{ch}} \le 1,
\end{eqnarray}
is satisfied for any collision operator $\hat{L}_{pq}$, that is, any differential cross section.
We emphasize that
our hydrodynamic equations surely have the causal property,
and hence can be applied to various high-energy hydrodynamic systems.

\subsubsection{
  Relation between relaxation equations by RG method and those by the other formalisms
}
The relaxation equations \eqref{eq:relax1}-\eqref{eq:relax3} can be made into the different form by iteration.
Here,
let us focus on the relaxation equation for the stress tensor,
i.e., Eq. \eqref{eq:relax3},
by setting $\Pi=J^{\mu}=0$:
\begin{align}
  \label{const_pi}
  \pi^{\mu\nu}
  &= 2\eta^{\mathrm{RG}}\sigma^{\mu\nu}
- \tau^{\mathrm{RG}}_\pi \Delta^{\mu\nu\rho\sigma}\frac{\partial}{\partial\tau}\pi_{\rho\sigma}
+ b_{\pi\pi\pi}\pi^{\lambda\langle\mu} {\pi^{\nu\rangle}}_{\lambda}
  \nonumber\\
  &+ \kappa^{(1)}_{\pi\pi}\pi^{\mu\nu} \theta
  + \kappa^{(2)}_{\pi\pi} {\pi_{\rho}}^{\langle\mu}\sigma^{\nu\rangle\rho}
  + \kappa^{(3)}_{\pi\pi} {\pi_{\rho}}^{\langle\mu}\omega^{\nu\rangle\rho}.
\end{align}
By solving this equation with respect to $\pi^{\mu\nu}$
in an iterative manner
and using the equality
\begin{align}
 \Delta^{\mu\nu\rho\sigma}\frac{\partial}{\partial\tau}\sigma_{\rho\sigma}
 &=- \frac{\partial}{\partial\tau}u^{\langle\mu} \cdot \frac{\partial}{\partial\tau}u^{\nu\rangle}
 + \nabla^{\langle\mu}\frac{\partial}{\partial\tau}u^{\nu\rangle}
 -\frac{2}{3}\theta\sigma^{\mu\nu}\nonumber\\
 &-\sigma^{\lambda\langle\mu}{\sigma^{\nu\rangle}}_\lambda
 -\omega^{\lambda\langle\mu}{\omega^{\nu\rangle}}_\lambda
 -2\sigma^{\lambda\langle\mu}{\omega^{\nu\rangle}}_\lambda,
\end{align}
and the balance equation (\ref{eq:balanceeqbyRG}),
we find that the resultant equation includes the following terms
\begin{align}
 \label{iteration}
 &\sigma^{\lambda\langle\mu}{\sigma^{\nu\rangle}}_\lambda,\,\,\,
 \omega^{\lambda\langle\mu}{\omega^{\nu\rangle}}_\lambda,\,\,\,
 \sigma^{\lambda\langle\mu}{\omega^{\nu\rangle}}_\lambda,\,\,\,
 \theta\sigma^{\mu\nu},\nonumber\\
 &\nabla^{\langle\mu} T \cdot \nabla^{\nu\rangle} T,\,\,\, 
 \nabla^{\langle\mu} T \cdot \nabla^{\nu\rangle}\frac{\mu}{T},\,\,\,
 \nabla^{\langle\mu}\frac{\mu}{T} \cdot \nabla^{\nu\rangle}\frac{\mu}{T},
 \nonumber\\
 &\nabla^{\langle\mu}\nabla^{\nu\rangle} T,\,\,\, 
 \nabla^{\langle\mu}\nabla^{\nu\rangle}\frac{\mu}{T},
\end{align}
in addition to those given
in Eq. \eqref{const_pi}.
In this iterative manner,
our hydrodynamic equation apparently gets to have
 all the terms given by $\mathcal{K}^{\mu\nu}$ of Eq.~(73) in \cite{Denicol:2012cn}.
Notice, however, that the last two terms of Eq. \eqref{iteration} 
have a form of the second-order spatial derivatives of hydrodynamic 
variables,
which make the hydrodynamic equation 
parabolic and accordingly acausal.
Hence, we have an important observation that
the naive iteration may spoil 
the causal property of the original hydrodynamic equation,
and thus we must use the original form of the
relaxation
equations \eqref{const_pi} or \eqref{eq:relax1}-\eqref{eq:relax3}.
Furthermore, since the appearance of the nonlinear vortex term
$\omega^{\lambda\langle\mu}{\omega^{\nu\rangle}}_\lambda$ seems to be 
inevitably associated with that of the second-order spatial derivative terms, 
the explicit appearance of such a nonlinear vortex term  should
be avoided in the relaxation equation 
although its effect should be
included in Eq. \eqref{const_pi} implicitly.

\subsubsection{Numerical example:
  transport coefficients and relaxation times
}

In this subsection, we present numerical examples of the transport coefficients
and relaxation times using the microscopic expressions given in the present approach, 
and compare them with those in the previous works.
Note that the microscopic expressions are solely given in terms 
of the linearized collision operator $\hat{L}$, which is in turn uniquely
determined by the transition probability $\omega(p\,,\,p_1|p_2\,,\,p_3)$.
A general form of the transition probability
reads
\begin{eqnarray}
  \omega(p\,,\,p_1|p_2\,,\,p_3) = \delta^4(p+p_1-p_2-p_3) \, s \, \sigma(s,\,\theta),
\end{eqnarray}
where $\sigma(s,\,\theta)$ denotes a differential cross section,
$s \equiv (p+p_1)^2$ a total momentum squared,
and $\theta \equiv \cos^{-1} [(p-p_1)\cdot(p_2-p_3)/(p-p_1)^2]$ a scattering angle.
Here, we examine the  case of a constant cross section for simplicity;
\begin{eqnarray}
  \sigma(s,\,\theta) = \sigma_T / 4\pi
\end{eqnarray}
with $\sigma_T$ being a total cross section.

We focus on the shear viscosity $\eta$ and relaxation time $\tau_\pi$
for the stress tensor
in the classical and massless limits, i.e.,
$a = 0$ and $m/T = 0$.
The calculation of $\eta$ and $\tau_\pi$
can be reduced to that of $X_p \equiv \big[ \hat{L}^{-1}\hat{\pi}^{\mu\nu} \big]_p$,
which satisfies the linear equation $\big[ \hat{L} X \big]_p = \hat{\pi}^{\mu\nu}_p$.
The last equation can be solved numerically
in an exact manner without recourse to any ansatz for the functional form of $X_p$.

In Table \ref{tab:001},
we show the numerical results together with those of the previous works.
We confirm that
our formulae for $\eta$ and $\tau_\pi$ give results different from
those by the (naive) Israel-Stewart moment method \cite{Israel:1979wp}.
Furthermore,
our relaxation time differs from that of Denicol et al. \cite{Denicol:2012cn},
which is an improvement of the Israel-Stewart moment method adopting 41 moments,
although their result for the  shear viscosity 
tends to numerically agrees
with the Chapman-Enskog/RG value \cite{de1980relatlvlatlc}.

\begin{table}[h]
\caption{
  \label{tab:001}
  Values of the shear viscosity and relaxation time for the stress tensor
  for a classical gas with a constant cross section in the massless limit, in 
  the RG method,
  Israel-Stewart's 14-moment method \cite{Israel:1979wp},
  and Denicol et al.'s 41-moment method \cite{Denicol:2012cn}.
}
\begin{tabular}{ccccc}
\hline
\hline
 & RG & Israel-Stewart & Denicol et al. \\
\hline
$\eta$ [$T/\sigma_T$] & 1.27 & 1.2 & 1.267 \\
\hline
$\tau_\pi$ [$1/n\sigma_T$] & 1.66 & 1.8 & 2 \\
\hline
\hline
\end{tabular}
\end{table}



\section{
  Summary and concluding remarks
}
\label{sec:sec4}
In this paper,
we have derived
the second-order hydrodynamic equation
systematically from the relativistic Boltzmann equation
with the quantum statistical effect.
Our derivation is based on
a novel development of the renormalization-group (RG) method.
In this method,
we have solved the Boltzmann equation
faithfully in a way valid up to the mesoscopic scales of space and time,
and then have
reduced the solution to
a simpler equation describing the mesoscopic dynamics of the Boltzmann equation.
We have found that
our theory nicely gives a compact expression
of the deviation of the distribution function 
in terms of the linearized collision operator,
which is different from those used as
an ansatz in the conventional fourteen-moment method.
In fact,
in contrast to the ansatz in the fourteen-moment method,
our distribution function produces the transport coefficients
which have the same microscopic expressions
as those derived in the Chapman-Enskog expansion method.
Furthermore,
new microscopic expressions of the relaxation times
are obtained, which differ from
 those derived in any other formalisms such as the moment method.
We have shown that the present expressions of the relaxation times 
can be nicely rewritten in terms of the respective relaxation functions, which allow
a physically natural interpretation of the relaxation times, and thus assert the 
plausibility of our results.
 
The present asymptotic analysis utilizing a perturbation theory is 
based on the physical assumption that
only the spatial inhomogeneity
is the origin of the dissipation,
and the expansion parameter $\epsilon$ 
is introduced for characterizing the inhomogeneity, 
which may be identified with the Knudsen number:  
The inhomogeneity gives a deviation of the distribution function 
from the local equilibrium one $f^{\rm eq}$,
and accordingly the ratio of the deviation to $f^{\rm eq}$ is necessarily proportional to $\epsilon$.
We emphasize that 
the inhomogeneity and the ratio of the distribution functions are
necessarily of the same order in our asymptotic analysis.
It is worth emphasizing that
the present asymptotic analysis 
combined
with the 
perturbative
expansion successfully
solves the Boltzmann equation consistently
and leads to the mesoscopic dynamics including
the constitutive equations that relate the dissipative quantities
and the spatial gradients of the equilibrium quantities.

We have given a proof that
the propagating velocities of the fluctuations of the hydrodynamical variables
do not exceed the light velocity,
and hence our seconder-order equation ensures the desired causality.
We have also proved that
the equilibrium state
is stable for any perturbation described by our equation.
These results strongly suggest
the validity of our formulation based on the RG method. 

We have demonstrated  numerically that
the relaxation times differ from those given 
in the moment methods in the literature,
even in the sophisticated one  so as to numerically
reproduce the transport coefficients given in the Chapman-Enskog (and RG) method.
The calculation was done only in the case of a constant differential cross section.
It is interesting 
to extend the present calculation to more realistic cases 
with the differential cross section depending on the total momentum and  scattering angle,
which have immediate applications to  relativistic systems
consisting of quarks, gluons, and hadrons.
Then it is an imperative task 
to apply the present method to derive the multi-component relativistic 
hydrodynamic equation, which is now under way \cite{Kikuchi2015}.
It is of interest to use the resultant equations for phenomenological
analysis of relativistic heavy-ion collisions performed in  
 RHIC and LHC, although there exist multi-component hydrodynamic equations
deriven on the basis of the moment method \cite{Prakash:1993bt,Monnai:2010qp},
which was found to have unsatisfactory aspects for the single-component equation,
as was shown in the present article.



\begin{acknowledgements}
We would like to thank Y. Hatta, A. Monnai, G. S. Denicol, and R. Venugopalan for useful comments.
This work was supported in part 
by the Core Stage Back UP program in Kyoto University,
by the Grants-in-Aid for Scientific Research from JSPS
 (Nos.24340054, 
 24540271
)  
and by the Yukawa International Program for Quark-Hadron Sciences.
\end{acknowledgements}

\appendix

\setcounter{equation}{0}
\section{
  Detailed derivation of the excited modes and their explicit expressions
}
\label{sec:app3}
In this Appendix,
we derive the expression of Eq.\eqref{eq:varphi1},
whose calculation can be reduced to that of
\begin{align}
 \label{1st-PF}
 \big[Q_0F_0\big]_p
 = \big[F_0-P_0F_0\big]_p
 = F_{0p}- \varphi_{0p}^{\alpha}\eta^{-1}_{0\alpha\beta}
 \langle\varphi_{0}^{\beta},F_0\rangle,
\end{align}
with
\begin{align}
 F_{0p}
 = \frac{1}{p\cdot u}\Bigg[p^{\mu}p^{\nu}\nabla_{\mu}\frac{u_{\nu}}{T}
 -p^{\mu}\nabla_{\mu}\frac{\mu}{T}\Bigg].
\end{align}
Here, we have used Eq. (\ref{eq:F0})
and $f^{\mathrm{eq}}_p=1/[\mathrm{e}^{(p\cdot u-\mu)/T}-a]$.

We introduce the following quantities for later covenience:
\begin{align}
 a_\ell \equiv \int\mathrm{d}pf^{\mathrm{eq}}_p\bar{f}^{\mathrm{eq}}_p(p\cdot u)^\ell,\,\,\,\ell=0,\,1,\,\cdots.
\end{align}
Then the metric $\eta^{\alpha\beta}_0 = \langle \varphi_0^\alpha,\varphi_0^\beta\rangle$ 
are expressed as
\begin{align}
 \eta^{\mu\nu}_0
 &= a_3 u^{\mu}u^{\nu} + (m^2a_1-a_3) \frac{1}{3} \Delta^{\mu\nu},\\
 \eta^{\mu4}_0 = \eta^{4\mu}_0 &= a_2 u^{\mu},\\
 \eta^{44}_0 &= a_1,
\end{align}
while the inverse metric $\eta_{0\alpha\beta}^{-1}$ read
\begin{align}
 \label{1st-invmet8}
 \eta^{-1}_{0\mu\nu} &= \frac{a_1u^{\mu}u^{\nu}}{a_3 a_1 - a^2_2}
+ \frac{3\Delta^{\mu\nu}}{m^2 a_1-a_3},\\
 \label{1st-invmet9}
 \eta^{-1}_{0\mu4} = \eta^{-1}_{04\mu} &= \frac{-a_2 u^{\mu}}{a_3 a_1 - a^2_2},\\
 \label{1st-invmet10}
 \eta^{-1}_{044} &= \frac{a_3}{a_3 a_1 - a^2_2}.
\end{align}
The inner products $\langle\varphi_{0}^{\beta},F_0\rangle$
are evaluated as follows: 
\begin{align}
 \label{1st-inner2}
 \langle\varphi_{0}^{\mu},F_0\rangle
 &=\frac{m^2a_1-a_3}{3}\Bigg[ -\frac{1}{T^2}\nabla^{\mu}T
 +u^{\mu}\frac{1}{T}\nabla\cdot u \Bigg]\nonumber\\
 &-\frac{m^2a_0-a_2}{3}\nabla^{\mu}\frac{\mu}{T},\\
 \label{1st-inner3}
 \langle\varphi_{0}^{4},F_0\rangle
 &=\frac{m^2a_0-a_2}{3}\frac{1}{T}\nabla\cdot u.
\end{align}

Inserting the inverse metric \eqref{1st-invmet8}-\eqref{1st-invmet10} and
the inner products \eqref{1st-inner2} and \eqref{1st-inner3} into Eq. \eqref{1st-PF},
we have
\begin{align}
 \label{eq:Q0F0}
 \big[ Q_0 F_0 \big]_p
 &= \frac{1}{T} \frac{1}{p\cdot u} \Bigg[\Pi_p  (- \nabla\cdot u)
 - J^\mu_p T \frac{m^2 a_0 - a_2}{m^2 a_1 - a_3} \nabla_\mu \frac{\mu}{T}\nonumber\\
 &+ \pi^{\mu\nu}_p  \Delta_{\mu\nu\rho\sigma}\nabla^\rho u^\sigma \Bigg],
\end{align}
where $\Pi_p$, $J^\mu_p$, and $\pi^{\mu\nu}_p$ are given by
\begin{align}
 \Pi_p
 &\equiv -\frac{m^2(a_2a_0-a_1^2)}{3(a_3a_1-a_2^2)}(p\cdot u)^2\nonumber\\
 &+ \frac{m^2(a_3a_0-a_2a_1)}{3(a_3a_1-a_2^2)}(p\cdot u)
 -\frac{m^2}{3},\\
 J^\mu_p
 &\equiv
 -\Delta^{\mu\nu}p_\nu
 \Bigg[
 (p\cdot u) - \frac{m^2 a_1 - a_3}{m^2 a_0 - a_2}
 \Bigg],\\
 \pi^{\mu\nu}_p
 &\equiv
 \Delta^{\mu\nu\rho\sigma}p_\rho p_\sigma.
\end{align}
As is shown below, the following relatons hold;
\begin{align}
 -\frac{m^2(a_2a_0-a_1^2)}{3(a_3a_1-a_2^2)}
 &=
 \frac{1}{3}
 -
 \frac{
 \frac{\partial P}{\partial T}\frac{\partial n}{\partial \mu}
 - \frac{\partial P}{\partial \mu}\frac{\partial n}{\partial T}
 }{
 \frac{\partial e}{\partial T}\frac{\partial n}{\partial \mu}
 - \frac{\partial e}{\partial \mu}\frac{\partial n}{\partial T}
 }
 =
 \frac{1}{3}
 -
 \left.\frac{\partial P}{\partial e}\right|_{n},\\
 \frac{m^2(a_3a_0-a_2a_1)}{3(a_3a_1-a_2^2)}
 &= 
 \frac{
 \frac{\partial P}{\partial T}\frac{\partial e}{\partial \mu}
 - \frac{\partial P}{\partial \mu}\frac{\partial e}{\partial T}
 }{
 \frac{\partial n}{\partial T}\frac{\partial e}{\partial \mu}
 - \frac{\partial n}{\partial \mu}\frac{\partial e}{\partial T}
 }
 =
 \left.\frac{\partial P}{\partial n}\right|_{e},\\
 \frac{m^2 a_1 - a_3}{m^2 a_0 - a_2}
 &=
 T \, \frac{\frac{\partial P}{\partial T}}{\frac{\partial P}{\partial \mu}}
 + \mu
 = \frac{e+P}{n}.
\end{align}
Then we 
see that
$\big[ Q_0 F_0 \big]_p$ in Eq. (\ref{eq:Q0F0})
takes the form given in Eq. (\ref{eq:varphi1}).
In the derivation of the above relations, 
we have used the equations derived from the explicit forms of $n$, $e$, and $P$
given by Eqs. (\ref{eq:Chap0-004})-(\ref{eq:Chap0-006}), 
\begin{align}
\frac{\partial n}{\partial T}
&= -\frac{1}{T^2}a_2 + \frac{\mu}{T^2}a_1,\\
\frac{\partial n}{\partial \mu}
&= - \frac{1}{T}a_1,\\
\frac{\partial e}{\partial T}
&= -\frac{1}{T^2}a_3 + \frac{\mu}{T^2}a_2,\\
\frac{\partial e}{\partial \mu}
&= - \frac{1}{T}a_2,\\
\frac{\partial P}{\partial T}
&= \frac{1}{3T^2}\,(-a_3 + \mu a_2 + m^2 a_1 - m^2 \mu a_0),\\
\frac{\partial P}{\partial \mu}
&= -\frac{1}{3T}\,(a_2 - m^2 a_0),
\end{align}
and the relations derived from the Gibbs-Duhem equation $\mathrm{d}P = s\mathrm{d}T + n\mathrm{d}\mu$,
\begin{align}
 \label{eq:thermodynamic_relation1}
 \frac{\partial P}{\partial T} &= s = \frac{e + P - \mu n}{T},\\
 \label{eq:thermodynamic_relation2}
 \frac{\partial P}{\partial \mu} &= n,
\end{align}
with $s$ being the entropy density.
We note that
the relations (\ref{eq:thermodynamic_relation1})
and (\ref{eq:thermodynamic_relation2}) can be shown
not only by the Gibbs-Duhem equation
but also by a straightforward manipulation based on the explicit forms of $n$, $e$, and $P$.

%


\setcounter{equation}{0}
\section{
  Solution to the linear differential equation (\ref{eq:2nd-ordereq})
  with a time dependent inhomogeneous term
}
\label{sec:app1}
In this Appendix,
we present the detailed derivation of
the second-order solution (\ref{eq:2nd-ordersol}) and initial value (\ref{eq:2nd-orderinit}).
We rewrite the second-order equation (\ref{eq:2nd-ordereq}) into
\begin{align}
  \label{eq:ChapB-6-001}
  \frac{\partial}{\partial\tau}X(\tau) &= \hat{L} X(\tau) + K(\tau-\tau_0),
\end{align}
with $X(\tau) \equiv (f^{\mathrm{eq}} \bar{f}^{\mathrm{eq}})^{-1} \tilde{f}^{(2)}(\tau)$.

The solution reads
\begin{align}
  \label{eq:ChapB-6-002}
  X(\tau)
  &= \mathrm{e}^{\hat{L}(\tau-\tau_0)}X(\tau_0) +
  \int_{\tau_0}^\tau \mathrm{d}\tau^\prime \mathrm{e}^{\hat{L}(\tau-\tau^\prime)}K(\tau^\prime-\tau_0)
  \nonumber\\    
  &= \mathrm{e}^{\hat{L}(\tau-\tau_0)}X(\tau_0) +
  \int_{\tau_0}^\tau \mathrm{d}\tau^\prime P_0K(\tau^\prime-\tau_0)
  \nonumber\\
  &+
  \int_{\tau_0}^\tau \mathrm{d}\tau^\prime 
  \mathrm{e}^{\hat{L}(\tau-\tau^\prime)}Q_0K(\tau^\prime-\tau_0),
\end{align}
where we have inserted $1 = P_0 + Q_0$
in front of $K(\tau^\prime-\tau_0)$.
Substituting the Taylor expansion
\begin{align}
  K(\tau^\prime - \tau_0) =
  \mathrm{e}^{(\tau^\prime - \tau_0)\partial/\partial s} K(s)\Big|_{s=0},
\end{align}
into Eq. (\ref{eq:ChapB-6-002})
and carrying out integration with respect to $\tau^\prime$,
we have
\begin{align}
  \label{eq:ChapB-6-004}
  &X(\tau)\nonumber\\
  &= \mathrm{e}^{\hat{L}(\tau-\tau_0)}\,X(\tau_0)
  +
  \Bigg[ (1 - \mathrm{e}^{(\tau-\tau_0)\partial/\partial s})
  (- \partial/\partial s)^{-1} P_0
  \nonumber\\
  &+
  (\mathrm{e}^{\hat{L}(\tau-\tau_0)} -
  \mathrm{e}^{(\tau-\tau_0)\partial/\partial s})
  (\hat{L} - \partial/\partial s)^{-1} 
  Q_0\Bigg] K(s)\Big|_{s=0}
  \nonumber\\
  &= \mathrm{e}^{\hat{L}(\tau-\tau_0)}
  \Big[ 
  X(\tau_0)
  +
  Q_1 (\hat{L} - \partial/\partial s)^{-1}
  Q_0 K(s)\Big|_{s=0}
  \Big]
  \nonumber\\
  &+
  \Bigg[(1 - \mathrm{e}^{(\tau-\tau_0)\partial/\partial s})
  (- \partial/\partial s)^{-1} P_0
  \nonumber\\
  &+
  (\mathrm{e}^{\hat{L}(\tau-\tau_0)} -
  \mathrm{e}^{(\tau-\tau_0)\partial/\partial s})
  P_1  (\hat{L} - \partial/\partial s)^{-1}
  Q_0
  \nonumber\\
  &-
  \mathrm{e}^{(\tau-\tau_0)\partial/\partial s}
  Q_1 (\hat{L} - \partial/\partial s)^{-1}
  Q_0\Bigg] K(s)\Big|_{s=0},
\end{align}
where $1 = P_0 + P_1 + Q_1$ has been inserted in front of $(\hat{L} - \partial/\partial s)^{-1} Q_0 K(s)$
in the second line of Eq. (\ref{eq:ChapB-6-004}).
We note that
the contributions from the inhomogeneous term $K(\tau-\tau_0)$
are decomposed into two parts,
whose time dependencies are given by
$\mathrm{e}^{\hat{L}(\tau-\tau_0)}$
and
$\mathrm{e}^{(\tau-\tau_0)\partial/\partial s}$,
respectively.
The former gives a fast motion 
characterized by the eigenvalues  of $\hat{L}$ acting on the Q${}_0$ space,
while the time dependence of the latter is 
independent of the dynamics due to the absence of $\hat{L}$.
Since we are interested in the motion coming from the P${}_0$ and P${}_1$ spaces,
we can eliminate the former associated with the Q${}_1$ space
with a choice of the initial value $X(\tau_0)$ that has not yet
been specified as follows:
\begin{align}
  \label{eq:ChapB-6-005}
  X(\tau_0) =
  - Q_1 (\hat{L} - \partial/\partial s)^{-1} 
  Q_0  K(s)\Big|_{s=0},
\end{align}
which reduces Eq. (\ref{eq:ChapB-6-004}) to
\begin{align}
  \label{eq:ChapB-6-006}
  X(\tau) &=
  \Big[ (1 - \mathrm{e}^{(\tau-\tau_0)\partial/\partial s})
  (- \partial/\partial s)^{-1} P_0
  \nonumber\\
  &+(\mathrm{e}^{\hat{L}(\tau-\tau_0)} -
  \mathrm{e}^{(\tau-\tau_0)\partial/\partial s})
  P_1 (\hat{L} - \partial/\partial s)^{-1} Q_0
  \nonumber\\
  &-\mathrm{e}^{(\tau-\tau_0)\partial/\partial s}
  Q_1  (\hat{L} - \partial/\partial s)^{-1} Q_0\Big] K(s)\Big|_{s=0}.
\end{align}
Using
$X(\tau_0) = (f^{\mathrm{eq}} \bar{f}^{\mathrm{eq}})^{-1} f^{(2)}$,
we can convert Eqs. (\ref{eq:ChapB-6-005}) and (\ref{eq:ChapB-6-006})
into Eqs. (\ref{eq:2nd-orderinit}) and (\ref{eq:2nd-ordersol}),
respectively.


\setcounter{equation}{0}
\section{
  Detailed derivation of the relaxation equations
}
\label{sec:app2}

In this Appendix,
we present a detailed derivation of the relaxation equation
given by Eqs. (\ref{eq:relax1})-(\ref{eq:relax3}).

First, 
we introduce the differential operator given by
\begin{align}
  \label{eq:2nd-relax-detailed-004}
  \Bigg[ (p\cdot u) \frac{\partial}{\partial\tau}
  + \epsilon p\cdot\nabla \Bigg] \delta_{pq}
  =(p\cdot u) v^\alpha_{pq} D_\alpha,
\end{align}
where
\begin{align}
  \label{eq:2nd-relax-detailed-005}
  v^\alpha_{pq} &\equiv \left\{
  \begin{array}{ll}
    \displaystyle{
      v^\mu_{pq} \equiv \frac{1}{p\cdot u}\Delta^{\mu\nu}p_\nu\delta_{pq},
    }
    &
    \displaystyle{
      \alpha = \mu,
    } \\[2mm]
    \displaystyle{
      \delta_{pq},
    }
    &
    \displaystyle{
      \alpha = 4,
    }
  \end{array}
  \right.
  \\
  \label{eq:2nd-relax-detailed-006}
  D_\alpha &\equiv \left\{
  \begin{array}{ll}
    \displaystyle{
      \epsilon \, \nabla_\mu,
    }
    &
    \displaystyle{
      \alpha = \mu,
    } \\[2mm]
    \displaystyle{
      \frac{\partial}{\partial\tau},
    }
    &
    \displaystyle{
      \alpha = 4.
    }
  \end{array}
  \right.
\end{align}
Then Eq. (\ref{eq:ChapB-RHD2}) is converted into the following form:
\begin{align}
  \label{eq:relaxalleq}
  &\langle  \hat{L}^{-1}  \hat{\psi}^{i} ,
  (f^{\mathrm{eq}} \bar{f}^{\mathrm{eq}})^{-1}v^\alpha D_\alpha
  \Big[ f^{\mathrm{eq}}(1 + \epsilon \bar{f}^{\mathrm{eq}} \hat{L}^{-1}  \hat{\chi}^{j}  \psi_{j})\Big] \rangle
  \nonumber\\
  &= \epsilon \langle \hat{L}^{-1} \, \hat{\psi}^{i} ,
  \hat{\chi}^{j} \psi_{j} \rangle
  \nonumber\\
  &+ \epsilon^2 \frac{1}{2}\langle \hat{L}^{-1}  \hat{\psi}^{i} ,
  B \Big[ \hat{L}^{-1} \hat{\chi}^{j} \psi_{j} \Big]\Big[ \hat{L}^{-1} \hat{\chi}^{k} \psi_{k} \Big] \rangle
  + O(\epsilon^3),
\end{align}
where we have introduced the following ``vectors''
consisting of three components:
\begin{align}
  \hat{\psi}^{i}_p &\equiv (\hat{\Pi}_p,\hat{J}^\mu_p,\hat{\pi}^{\mu\nu}_p),
  \\
  \psi_{j} &\equiv (\Pi,J_\rho,\pi_{\rho\sigma}),
  \\
  \psi_{k} &\equiv (\Pi,J_\kappa,\pi_{\kappa\lambda}),
  \\
  \hat{\chi}^{j}_p &\equiv (\hat{\Pi}_p/(-T\zeta^{\mathrm{RG}}),
  h\hat{J}^\rho_p/(T^2\lambda^{\mathrm{RG}}),
  \hat{\pi}^{\rho\sigma}_p/(-2T\eta^{\mathrm{RG}})),
  \\
  \hat{\chi}^{k}_p &\equiv (\hat{\Pi}_p/(-T\zeta^{\mathrm{RG}}),
  h\hat{J}^\kappa_p/(T^2\lambda^{\mathrm{RG}}),
  \hat{\pi}^{\kappa\lambda}_p/(-2T\eta^{\mathrm{RG}})),
\end{align}
with $i$, $j$, and $k$ being indices
specifying the vector components.

We expand the left-hand sides of Eq. (\ref{eq:relaxalleq}) as
\begin{align}
  \label{eq:2nd-relax-detailed-007}
  &\langle \hat{L}^{-1} \hat{\psi}^{i} ,
  (f^{\mathrm{eq}}\bar{f}^{\mathrm{eq}})^{-1} 
v^\alpha D_\alpha\Big[ f^{\mathrm{eq}}(1 + \epsilon\bar{f}^{\mathrm{eq}}\hat{L}^{-1}
  \hat{\chi}^{j} \psi_{j})\Big] \rangle
  \nonumber\\
  &=
  \langle \hat{L}^{-1} \hat{\psi}^{i},
  (f^{\mathrm{eq}}\bar{f}^{\mathrm{eq}})^{-1}v^\alpha D_\alpha
  f^{\mathrm{eq}} \rangle
  \nonumber\\
  &+
  \epsilon\langle  \hat{L}^{-1} \hat{\psi}^{i},
  (f^{\mathrm{eq}} \bar{f}^{\mathrm{eq}})^{-1}
v^\alpha D_\alpha
  \Big[ f^{\mathrm{eq}}\bar{f}^{\mathrm{eq}}
  \hat{L}^{-1} \hat{\chi}^{j}\Big] \rangle \psi_{j}
  \nonumber\\
  &+
  \epsilon\langle \hat{L}^{-1} \hat{\psi}^{i},v^\alpha
  \hat{L}^{-1} \hat{\chi}^{j} \rangle D_\alpha \psi_{j}.
\end{align}
The first and third terms of Eq. (\ref{eq:2nd-relax-detailed-007}) are calculated to be
\begin{align}
  \label{eq:firstterm}
  &\langle \hat{L}^{-1} \hat{\psi}^{i},
  (f^{\mathrm{eq}} \bar{f}^{\mathrm{eq}})^{-1}v^\alpha D_\alpha
  f^{\mathrm{eq}} \rangle
  =\epsilon \langle \hat{L}^{-1} \hat{\psi}^{i} ,
  \hat{\chi}^{j} \rangle X^\prime_{j},\\
  \label{eq:thirdterm}
  & \epsilon\langle  \hat{L}^{-1} \hat{\psi}^{i},v^\alpha
  \hat{L}^{-1} \hat{\chi}^{j} \rangle D_\alpha \psi_{j}\nonumber\\
  &=\epsilon\langle \hat{L}^{-1} \hat{\psi}^{i},
  \hat{L}^{-1} \hat{\chi}^{j} \rangle \frac{\partial}{\partial\tau} \psi_{j}
  +\epsilon^2 \langle \hat{L}^{-1} \hat{\psi}^{i},v^\mu
  \hat{L}^{-1} \hat{\chi}^{j} \rangle \nabla_\mu \psi_{j},
\end{align}
respectively.
In Eq. (\ref{eq:firstterm}), we have introduced
\begin{align}
  X^\prime_{i} \equiv (-\zeta^{\mathrm{RG}}\nabla\cdot u,
  \lambda^{\mathrm{RG}}T^2h^{-2}\nabla_\mu (\mu/T),
  2\eta^{\mathrm{RG}}\nabla_\mu u_\nu).
\end{align}
Substituting Eq. (\ref{eq:2nd-relax-detailed-007}) combined with
Eqs. (\ref{eq:firstterm}) and (\ref{eq:thirdterm}) into Eq. (\ref{eq:relaxalleq}),
we have
\begin{align}
  \label{eq:relaxalleq2}
  &\epsilon \langle \hat{L}^{-1} \hat{\psi}^{i},
  \hat{\chi}^{j} \rangle \psi_{j}
  \nonumber\\
  &=\epsilon\langle \hat{L}^{-1} \hat{\psi}^{i},\hat{\chi}^{j} \rangle X^\prime_{j}
  +\epsilon \langle \hat{L}^{-1} \hat{\psi}^{i},
  \hat{L}^{-1} \hat{\chi}^{j} \rangle \frac{\partial}{\partial\tau} \psi_{j}
  \nonumber\\
  &+\epsilon^2\langle \hat{L}^{-1} \hat{\psi}^{i},v^\mu
  \hat{L}^{-1} \hat{\chi}^{j} \rangle \nabla_\mu \psi_{j}
  \nonumber\\
  &+ \epsilon^2 \frac{1}{2}M^{i,j,k}\psi_{j}\psi_{k}
  + \epsilon N^{i,j} \psi_{j}
  + O(\epsilon^3),
\end{align}
with
\begin{align}
  M^{i,j,k} &\equiv
  - \langle \hat{L}^{-1} \hat{\psi}^{i},
  B \Big[ \hat{L}^{-1} \hat{\chi}^{j} \Big]
  \Big[ \hat{L}^{-1} \hat{\chi}^{k} \Big]
  \rangle,
  \\
  N^{i,j} &\equiv
  \langle \hat{L}^{-1} \hat{\psi}^{i},
  (f^{\mathrm{eq}} \bar{f}^{\mathrm{eq}})^{-1} 
v^\alpha D_\alpha
  \Big[ f^{\mathrm{eq}}\bar{f}^{\mathrm{eq}} \hat{L}^{-1} \hat{\chi}^{j}\Big] \rangle.
\end{align}
Some coefficients in Eq. (\ref{eq:relaxalleq2}) can be easily calculated as
\begin{widetext}
\begin{align}
  \langle \hat{L}^{-1} \hat{\psi}^{i},\hat{\chi}^{j} \rangle
  &= \left(
  \begin{array}{ccc}
    \displaystyle{
    \frac{\langle \hat{L}^{-1} \hat{\Pi},\hat{\Pi} \rangle}{-T\zeta^{\mathrm{RG}}}
    }
    &
    \displaystyle{
    \frac{h\langle \hat{L}^{-1} \hat{\Pi},\hat{J}^\rho \rangle}{T^2\lambda^{\mathrm{RG}}}
    }
    &
    \displaystyle{
    \frac{\langle \hat{L}^{-1} \hat{\Pi},\hat{\pi}^{\rho\sigma} \rangle}{-2T\eta^{\mathrm{RG}}}
    }
    \\
    \displaystyle{
    \frac{\langle \hat{L}^{-1} \hat{J}^\mu,\hat{\Pi} \rangle}{-T\zeta^{\mathrm{RG}}}
    }
    &
    \displaystyle{
    \frac{h\langle \hat{L}^{-1} \hat{J}^\mu,\hat{J}^\rho \rangle}{T^2\lambda^{\mathrm{RG}}}
    }
    &
    \displaystyle{
    \frac{\langle \hat{L}^{-1} \hat{J}^\mu,\hat{\pi}^{\rho\sigma} \rangle}{-2T\eta^{\mathrm{RG}}}
    }
    \\
    \displaystyle{
    \frac{\langle \hat{L}^{-1} \hat{\pi}^{\mu\nu},\hat{\Pi} \rangle}{-T\zeta^{\mathrm{RG}}}
    }
    &
    \displaystyle{
    \frac{h\langle \hat{L}^{-1} \hat{\pi}^{\mu\nu},\hat{J}^\rho \rangle}{T^2\lambda^{\mathrm{RG}}}
    }
    &
    \displaystyle{
    \frac{\langle \hat{L}^{-1} \hat{\pi}^{\mu\nu},\hat{\pi}^{\rho\sigma} \rangle}{-2T\eta^{\mathrm{RG}}}
    }
  \end{array}
  \right)
  = \left(
  \begin{array}{ccc}
    1 & 0 & 0\\
    0 & h\Delta^{\mu\rho} & 0\\
    0 & 0 & \Delta^{\mu\nu\rho\sigma}
  \end{array}
  \right),\\
  \langle \hat{L}^{-1} \hat{\psi}^{i},\hat{L}^{-1}\,\hat{\chi}^{j} \rangle
  &= \left(
  \begin{array}{ccc}
    \displaystyle{
    \frac{\langle \hat{L}^{-1} \hat{\Pi},\hat{L}^{-1} \hat{\Pi} \rangle}{-T\zeta^{\mathrm{RG}}}
    }
    &
    \displaystyle{
    \frac{h\langle \hat{L}^{-1} \hat{\Pi},\hat{L}^{-1} \hat{J}^\rho \rangle}{T^2\lambda^{\mathrm{RG}}}
    }
    &
    \displaystyle{
    \frac{\langle \hat{L}^{-1} \hat{\Pi},\hat{L}^{-1} \hat{\pi}^{\rho\sigma} \rangle}{-2T\eta^{\mathrm{RG}}}
    }
    \\
    \displaystyle{
    \frac{\langle \hat{L}^{-1} \hat{J}^\mu,\hat{L}^{-1} \hat{\Pi} \rangle}{-T\zeta^{\mathrm{RG}}}
    }
    &
    \displaystyle{
    \frac{h\langle \hat{L}^{-1} \hat{J}^\mu,\hat{L}^{-1} \hat{J}^\rho \rangle}{T^2\lambda^{\mathrm{RG}}}
    }
    &
    \displaystyle{
    \frac{\langle \hat{L}^{-1} \hat{J}^\mu,\hat{L}^{-1} \hat{\pi}^{\rho\sigma} \rangle}{-2T\eta^{\mathrm{RG}}}
    }
    \\
    \displaystyle{
    \frac{\langle \hat{L}^{-1} \hat{\pi}^{\mu\nu},\hat{L}^{-1} \hat{\Pi} \rangle}{-T\zeta^{\mathrm{RG}}}
    }
    &
    \displaystyle{
    \frac{h\langle \hat{L}^{-1} \hat{\pi}^{\mu\nu},\hat{L}^{-1} \hat{J}^\rho \rangle}{T^2\lambda^{\mathrm{RG}}}
    }
    &
    \displaystyle{
    \frac{\langle \hat{L}^{-1} \hat{\pi}^{\mu\nu},\hat{L}^{-1} \hat{\pi}^{\rho\sigma} \rangle}{-2T\eta^{\mathrm{RG}}}
    }
  \end{array}
  \right)\nonumber\\
  &= \left(
  \begin{array}{ccc}
    - \tau_\Pi & 0 & 0\\
    0 & -h\tau_J \Delta^{\mu\rho} & 0\\
    0 & 0 & -\tau_\pi \Delta^{\mu\nu\rho\sigma}
  \end{array}
  \right),\\
  \langle \hat{L}^{-1} \hat{\psi}^{i},v^a \hat{L}^{-1}\,\hat{\chi}^{j} \rangle
  &= \left(
  \begin{array}{ccc}
    \displaystyle{
    \frac{\langle \hat{L}^{-1} \hat{\Pi},v^a \hat{L}^{-1} \hat{\Pi} \rangle}{-T\zeta^{\mathrm{RG}}}
    }
    &
    \displaystyle{
    \frac{h\langle \hat{L}^{-1} \hat{\Pi},v^a \hat{L}^{-1} \hat{J}^\rho \rangle}{T^2\lambda^{\mathrm{RG}}}
    }
    &
    \displaystyle{
    \frac{\langle \hat{L}^{-1} \hat{\Pi},v^a \hat{L}^{-1} \hat{\pi}^{\rho\sigma} \rangle}{-2T\eta^{\mathrm{RG}}}
    }
    \\
    \displaystyle{
    \frac{\langle \hat{L}^{-1} \hat{J}^\mu,v^a \hat{L}^{-1} \hat{\Pi} \rangle}{-T\zeta^{\mathrm{RG}}}
    }
    &
    \displaystyle{
    \frac{h\langle \hat{L}^{-1} \hat{J}^\mu,v^a \hat{L}^{-1} \hat{J}^\rho \rangle}{T^2\lambda^{\mathrm{RG}}}
    }
    &
    \displaystyle{
    \frac{\langle \hat{L}^{-1} \hat{J}^\mu,v^a \hat{L}^{-1} \hat{\pi}^{\rho\sigma} \rangle}{-2T\eta^{\mathrm{RG}}}
    }
    \\
    \displaystyle{
    \frac{\langle \hat{L}^{-1} \hat{\pi}^{\mu\nu},v^a \hat{L}^{-1} \hat{\Pi} \rangle}{-T\zeta^{\mathrm{RG}}}
    }
    &
    \displaystyle{
    \frac{h\langle \hat{L}^{-1} \hat{\pi}^{\mu\nu},v^a \hat{L}^{-1} \hat{J}^\rho \rangle}{T^2\lambda^{\mathrm{RG}}}
    }
    &
    \displaystyle{
    \frac{\langle \hat{L}^{-1} \hat{\pi}^{\mu\nu},v^a \hat{L}^{-1} \hat{\pi}^{\rho\sigma} \rangle}{-2T\eta^{\mathrm{RG}}}
    }
  \end{array}
  \right)\nonumber\\
  &= \left(
  \begin{array}{ccc}
    0 & -h\ell_{\Pi J}\Delta^{a\rho} & 0\\
    -\ell_{J \Pi}\Delta^{\mu a} & 0 & -\ell_{J \pi}\Delta^{\mu a\rho\sigma}\\
    0 & -h\ell_{\pi J}\Delta^{\mu\nu a\rho} & 0
  \end{array}
  \right),
\end{align}
\end{widetext}
Here,
we have introduced the relaxation times $\tau_\Pi$, $\tau_J$, and $\tau_\pi$
and the relaxation lengths $\ell_{\Pi J}$, $\ell_{J\Pi}$, $\ell_{J\pi}$, and $\ell_{\pi J}$.
They are defined as follows,
\begin{align}
  \label{eq:2nd-relax-detailed-011}
  \tau_\Pi &\equiv \frac{\langle \hat{L}^{-1}\hat{\Pi},\hat{L}^{-1}\hat{\Pi} \rangle}{T\zeta^{\mathrm{RG}}},
  \\
  \label{eq:2nd-relax-detailed-012}
  \tau_J &\equiv - \frac{\langle \hat{L}^{-1}\hat{J}^\mu,\hat{L}^{-1}\hat{J}_\mu \rangle}{3T^2\lambda^{\mathrm{RG}}},
  \\
  \label{eq:2nd-relax-detailed-013}
  \tau_\pi &\equiv \frac{\langle \hat{L}^{-1}\hat{\pi}^{\mu\nu},\hat{L}^{-1}\,\hat{\pi}_{\mu\nu} \rangle}{10T\eta^{\mathrm{RG}}},
  \\
  \label{eq:2nd-relax-detailed-014}
  \ell_{\Pi J} &\equiv - \frac{\langle \hat{L}^{-1}\hat{\Pi},v^\mu \hat{L}^{-1}\hat{J}_\mu \rangle}{3T^2\lambda^{\mathrm{RG}}},
  \\
  \label{eq:2nd-relax-detailed-015}
  \ell_{J\Pi} &\equiv \frac{\langle \hat{L}^{-1}\hat{J}^\mu,v_\mu \hat{L}^{-1}\hat{\Pi} \rangle}{3T\zeta^{\mathrm{RG}}},
  \\
  \label{eq:2nd-relax-detailed-016}
  \ell_{J\pi} &\equiv \frac{\langle \hat{L}^{-1}\hat{J}^\mu,v^\nu \hat{L}^{-1}\hat{\pi}_{\mu\nu} \rangle}{10T \eta^{\mathrm{RG}}},
  \\
  \label{eq:2nd-relax-detailed-017}
  \ell_{\pi J} &\equiv - \frac{\langle \hat{L}^{-1}\hat{\pi}^{\mu\nu},v_\mu \hat{L}^{-1}\hat{J}_{\nu} \rangle}{5T^2\lambda^{\mathrm{RG}}}.
\end{align}
We note that
$\tau_\Pi$, $\tau_J$, and $\tau_\pi$ 
are denoted as
$\tau^{\mathrm{RG}}_\Pi$, $\tau^{\mathrm{RG}}_J$, and $\tau^{\mathrm{RG}}_\pi$ 
in the text and given in Eqs. (\ref{eq:RT1byRG})-(\ref{eq:RT3byRG}).

From now on,
we examine the terms associated with $M^{i,j,k}$ and $N^{i,j}$ in Eq. (\ref{eq:relaxalleq2}).
We write down the useful formulae for space-like tensors $A$ for later convenience:
\begin{widetext}
\begin{align}
 \label{eq:f1}
 \langle A^{\mu\nu} \rangle
 &=\frac{1}{3}\Delta^{\mu\nu}\langle {A^{\rho}}_\rho \rangle,
 \\
 \label{eq:f2}
 \langle A^{\langle\mu\nu\rangle\rho\sigma} \rangle
 &=\frac{1}{5}\Delta^{\mu\nu\rho\sigma}\langle {A^{\langle\alpha\beta\rangle}}_{\langle\alpha\beta\rangle} \rangle,
 \\
 \label{eq:f3}
 \langle A^{\mu\nu\rho\sigma} \rangle
 &=\frac{1}{3}\Delta^{\mu\nu}\langle {{A^{\alpha}}_\alpha}^{\rho\sigma} \rangle
 + \langle A^{\langle\mu\nu\rangle\rho\sigma} \rangle + \langle A^{(\mu\nu)\rho\sigma} \rangle
 \nonumber\\
 &=\frac{1}{9}\Delta^{\mu\nu}\Delta^{\rho\sigma}\langle {{{A^{\alpha}}_\alpha}^{\beta}}_\beta \rangle
 + \frac{1}{5}\Delta^{\mu\nu\rho\sigma}\langle {A^{\langle\alpha\beta\rangle}}_{\langle\alpha\beta\rangle} \rangle 
 + \frac{1}{3}\Omega^{\mu\nu\rho\sigma}\langle {A^{(\alpha\beta)}}_{(\alpha\beta)} \rangle,
 \\
 \label{eq:f4}
 \langle A^{\langle\mu\nu\rangle\langle\rho\sigma\rangle\langle\alpha\beta\rangle} \rangle
 &=\frac{12}{35}\Delta^{\mu\nu\gamma\delta}{\Delta^{\rho\sigma\lambda}}_\gamma{\Delta^{\alpha\beta}}_{\lambda\delta}
 \langle {{{A^{\langle\tau\eta\rangle}}_{\langle\tau}}^{\kappa\rangle}}_{\langle\kappa\eta\rangle} \rangle,
 \\
 \label{eq:f5}
 \langle A^{\langle\mu\nu\rangle\langle\rho\sigma\rangle\alpha\beta} \rangle
 &=\frac{1}{3}\Delta^{\alpha\beta}\langle {A^{\langle\mu\nu\rangle\langle\rho\sigma\rangle\lambda}}_\lambda \rangle
 + \langle A^{\langle\mu\nu\rangle\langle\rho\sigma\rangle\langle\alpha\beta\rangle} \rangle 
 + \langle A^{\langle\mu\nu\rangle\langle\rho\sigma\rangle(\alpha\beta)} \rangle
 \nonumber\\
 &=\frac{1}{15}\Delta^{\mu\nu\rho\sigma}\Delta^{\alpha\beta}\langle {{{A^{\langle\gamma\delta\rangle}}_{\langle\gamma\delta\rangle}}^{\lambda}}_\lambda \rangle
 + \frac{12}{35}\Delta^{\mu\nu\gamma\delta}{\Delta^{\rho\sigma\lambda}}_\gamma{\Delta^{\alpha\beta}}_{\lambda\delta}
 \langle {{{A^{\langle\tau\eta\rangle}}_{\langle\tau}}^{\kappa\rangle}}_{\langle\kappa\eta\rangle} \rangle
 \nonumber\\
 &+ \frac{4}{15}\Delta^{\mu\nu\gamma\delta}{\Delta^{\rho\sigma\lambda}}_\gamma{\Omega^{\alpha\beta}}_{\lambda\delta}
 \langle {{{A^{\langle\tau\eta\rangle}}_{\langle\tau}}^{\kappa\rangle}}_{(\kappa\eta)} \rangle.
\end{align}
\end{widetext}
where we have defined $\Omega^{\mu\nu\rho\sigma}\equiv \frac{1}{2}(\Delta^{\mu\rho}\Delta^{\nu\sigma} - \Delta^{\mu\sigma}\Delta^{\nu\rho})$ and $A^{(\mu\nu)} \equiv \Omega^{\mu\nu\rho\sigma}A_{\rho\sigma}$ for an arbitrary tensor $A^{\mu\nu}$. In the first equality of Eq.~\eqref{eq:f3} and the first equality of Eq.~\eqref{eq:f5}, we have used the fact that a space-like rank-two tensor $B^{\mu\nu}$ \cite{footnote:spacelike} 
is decomposed to be
\begin{align}
 \label{eq:decomposition}
 B^{\mu\nu}=\Delta^{\mu\nu}{B^{\rho}}_\rho/3+B^{\langle\mu\nu\rangle}+B^{(\mu\nu)}.
\end{align} 
The numerical factors may be verified by contracting both sides of equations.
To see how to use these formulae,
let us consider 
$\langle A^{\langle\rho\sigma\rangle\alpha\beta} \rangle\psi_{\langle\rho\sigma\rangle} \chi_{\alpha\beta}$, 
which is found in the fifth line after the first equality of Eq.~\eqref{eq:Pichi}, for instance.
\begin{align}
 \langle A^{\langle\rho\sigma\rangle\alpha\beta} \rangle\psi_{\langle\rho\sigma\rangle} \chi_{\alpha\beta}
 &=\frac{1}{5}\Delta^{\rho\sigma\alpha\beta}\langle {A^{\langle\gamma\delta\rangle}}_{\langle\gamma\delta\rangle} \rangle
 \psi_{\langle\rho\sigma\rangle} \chi_{\alpha\beta}
 \nonumber\\
 &=\frac{1}{5}\langle {A^{\langle\gamma\delta\rangle}}_{\langle\gamma\delta\rangle} \rangle
 \psi^{\langle\rho\sigma\rangle} \chi_{\langle\rho\sigma\rangle},
\end{align}
where we have used Eq.~\eqref{eq:f2} in the second equality.

Using the formulae \eqref{eq:f1}, \eqref{eq:f2}, and \eqref{eq:f4}, the non-linear terms of $M^{i,j,k}$ for $\hat{\psi}^i =\hat{\Pi},\ \hat{J}^\mu,\ \hat{\pi}^{\mu\nu}$ can be reduced to
\begin{align}
 &\underline{\hat{\psi}^i=\hat{\Pi}}
 \nonumber\\[5pt]
 &-\frac{\epsilon^2}{2}\big<L^{-1}\hat{\Pi},B[L^{-1}\hat{\chi}^j,L^{-1}\hat{\chi}^k]\big>\psi_j\psi_k
 \nonumber\\
 &=-\epsilon^2\frac{\big<L^{-1}\hat{\Pi},B[L^{-1}\hat{\Pi},L^{-1}\hat{\Pi}]\big>}{2(T\zeta)^2}\Pi^2
 \nonumber\\
 &-\epsilon^2\frac{\big<L^{-1}\hat{\Pi},B[L^{-1}\hat{J}^\rho,L^{-1}\hat{J}^\sigma]\big>}{2(T^2\lambda/h)^2}J_\rho J_\sigma
 \nonumber\\
 &-\epsilon^2\frac{\big<L^{-1}\hat{\Pi},B[L^{-1}\hat{\pi}^{\rho\sigma},L^{-1}\hat{\pi}^{\alpha\beta}]\big>}{2(2T\eta)^2}\pi_{\rho\sigma}\pi_{\alpha\beta}
 \nonumber\\
 &=\epsilon^2\Bigg(
 -\frac{\big<L^{-1}\hat{\Pi},B[L^{-1}\hat{\Pi},L^{-1}\hat{\Pi}]\big>}{2(T\zeta)^2}\Pi^2
 \nonumber\\
 &-\frac{\big<L^{-1}\hat{\Pi},B[L^{-1}\hat{J}^{\mu}, L^{-1}\hat{J}_\mu]\big>}{6(T^2\lambda/h)^2}J^{\rho}J_\rho
 \nonumber\\
 &-\frac{\big<L^{-1}\hat{\Pi},B[L^{-1}\hat{\pi}^{\mu\nu}, L^{-1}\hat{\pi}_{\mu\nu}]\big>}{10(2T\eta)^2}\pi^{\rho\sigma}\pi_{\rho\sigma}
 \Bigg)
 \nonumber\\
 &=\epsilon^2(b_{\Pi\Pi\Pi}\Pi^2+b_{\Pi JJ}J^{\rho}J_\rho+b_{\Pi\pi\pi}\pi^{\rho\sigma}\pi_{\rho\sigma}),
 \\[10pt]
 &\underline{\hat{\psi}^i=\hat{J}^\mu}
 \nonumber\\[5pt]
 &-\frac{\epsilon^2}{2}\big<L^{-1}\hat{J}^{\mu},B[L^{-1}\hat{\psi}^j, L^{-1}\hat{\psi}^k]\big>\chi_j\chi_k
 \nonumber\\
 &=\epsilon^2\frac{\big<L^{-1}\hat{J}^{\mu},B[L^{-1}\hat{\Pi}, L^{-1}\hat{J}^\rho]\big>}{(T\zeta)(T^2\lambda/h)}\Pi J_\rho
 \nonumber\\
 &+\epsilon^2\frac{\big<L^{-1}\hat{J}^{\mu},B[L^{-1}\hat{J}^\rho, L^{-1}\hat{\pi}^{\alpha\beta}]\big>}{(T^2\lambda/h)(2T\eta)}J_\rho\pi_{\alpha\beta}
 \nonumber\\
 &=\epsilon^2\Bigg(
 \frac{\big<L^{-1}\hat{J}^{\mu},B[L^{-1}\hat{\Pi}, L^{-1}\hat{J}_\mu]\big>}{3(T\zeta)(T^2\lambda/h)}\Pi J^{\mu}
 \nonumber\\
 &+\frac{\big<L^{-1}\hat{J}^{\mu},B[L^{-1}\hat{J}^{\nu}, L^{-1}\hat{\pi}_{\mu\nu}]\big>}{5(T^2\lambda/h)(2T\eta)}J^\rho{\pi_\rho}^\mu
 \Bigg)
 \nonumber\\
 &=\epsilon^2(b_{J\Pi J}\Pi J^{\mu}+b_{JJ\pi}J^\rho{\pi_\rho}^\mu),
 \\[10pt]
 \label{eq:piBpsipsi}
 &\underline{\hat{\psi}^i=\hat{\pi}^{\mu\nu}}
 \nonumber\\[5pt]
 &-\frac{\epsilon^2}{2}\big<L^{-1}\hat{\pi}^{\mu\nu},B[L^{-1}\hat{\psi}^j, L^{-1}\hat{\psi}^k]\big>\chi_j\chi_k
 \nonumber\\
 &=-\epsilon^2\frac{\big<L^{-1}\hat{\pi}^{\mu\nu},B[L^{-1}\hat{\Pi}, L^{-1}\hat{\pi}^{\rho\sigma}]\big>}{(T\zeta)(2T\eta)}\Pi\pi_{\rho\sigma}
 \nonumber\\
 &-\epsilon^2\frac{\big<L^{-1}\hat{\pi}^{\mu\nu},B[L^{-1}\hat{J}^\rho, L^{-1}\hat{J}^\sigma]\big>}{2(T^2\lambda/h)^2}J_\rho J_\sigma
 \nonumber\\
 &-\epsilon^2\frac{\big<L^{-1}\hat{\pi}^{\mu\nu},B[L^{-1}\hat{\pi}^{\rho\sigma}, L^{-1}\hat{\pi}^{\alpha\beta}]\big>}{2(2T\eta)^2}\pi_{\rho\sigma}\pi_{\alpha\beta}
 \nonumber\\
 &=\epsilon^2\Bigg(
 -\frac{\big<L^{-1}\hat{\pi}^{\mu\nu},B[L^{-1}\hat{\Pi}, L^{-1}\hat{\pi}_{\mu\nu}]\big>}{5(T\zeta)(2T\eta)}\Pi\pi^{\mu\nu}
 \nonumber\\
 &-\frac{\big<L^{-1}\hat{\pi}^{\mu\nu},B[L^{-1}\hat{J}_\mu, L^{-1}\hat{J}_\nu]\big>}{10(T^2\lambda/h)^2}J^{\langle\mu} J^{\nu\rangle}
 \nonumber\\
 &-\frac{\big<L^{-1}\hat{\pi}^{\mu\nu},
 B[L^{-1}{\hat{\pi}_{\mu}}^{\lambda},L^{-1}\hat{\pi}_{\lambda\nu}]\big>}{(35/6)(2T\eta)^2}\pi^{\rho\langle\mu}{\pi^{\nu\rangle}}_\rho
 \Bigg)
 \nonumber\\
 &=\epsilon^2(b_{\pi\Pi\pi}\Pi\pi^{\mu\nu}+b_{\pi JJ}J^{\langle\mu} J^{\nu\rangle}
 +b_{\pi\pi\pi}\pi^{\rho\langle\mu}{\pi^{\nu\rangle}}_\rho),
\end{align}
where the coefficients
$b_{\Pi\Pi\Pi}$, $b_{\Pi JJ}$, $b_{\Pi \pi\pi}$,
$b_{J \Pi J}$, $b_{J J \pi}$, $b_{\pi \Pi \pi}$, $b_{\pi JJ}$, and $b_{\pi\pi\pi}$
are given by
\begin{align}
 &b_{\Pi\Pi\Pi} \equiv -\frac{\big<L^{-1}\hat{\Pi},B[L^{-1}\hat{\Pi}, L^{-1}\hat{\Pi}]\big>}{2(T\zeta)^2},
 \\
 &b_{\Pi JJ} \equiv -\frac{\big<L^{-1}\hat{\Pi},B[L^{-1}\hat{J}^{\mu}, L^{-1}\hat{J}_\mu]\big>}{6(T^2\lambda/h)^2},
 \\
 &b_{\Pi\pi\pi} \equiv -\frac{\big<L^{-1}\hat{\Pi},B[L^{-1}\hat{\pi}^{\mu\nu}, L^{-1}\hat{\pi}_{\mu\nu}]\big>}{10(2T\eta)^2},
 \\
 &b_{J\Pi J} \equiv \frac{\big<L^{-1}\hat{J}^{\mu},B[L^{-1}\hat{\Pi}, L^{-1}\hat{J}_\mu]\big>}{3(T\zeta)(T^2\lambda/h)},
 \\
 &b_{JJ\pi} \equiv \frac{\big<L^{-1}\hat{J}^{\mu},B[L^{-1}\hat{J}^{\nu}, L^{-1}\hat{\pi}_{\mu\nu}]\big>}{5(T^2\lambda/h)(2T\eta)},
 \\
 &b_{\pi\Pi\pi} \equiv -\frac{\big<L^{-1}\hat{\pi}^{\mu\nu},B[L^{-1}\hat{\Pi}, L^{-1}\hat{\pi}_{\mu\nu}]\big>}{5(T\zeta)(2T\eta)},
 \\
 &b_{\pi JJ} \equiv -\frac{\big<L^{-1}\hat{\pi}^{\mu\nu},B[L^{-1}\hat{J}_\mu, L^{-1}\hat{J}_\nu]\big>}{10(T^2\lambda/h)^2},
 \\
 &b_{\pi\pi\pi} \equiv -\frac{\big<L^{-1}\hat{\pi}^{\mu\nu},
 B[L^{-1}{\hat{\pi}_{\mu}}^{\lambda},L^{-1}\hat{\pi}_{\lambda\nu}]\big>}{(35/6)(2T\eta)^2}.
\end{align}

Next, we rewrite $N^{i.j}$ as follows:
\begin{align}
 &\epsilon\Big<L^{-1}\hat{\psi}^i,(f^{\mathrm{eq}}\bar{f}^{\mathrm{eq}})^{-1}
 \left[\frac{\partial}{\partial\tau}+\epsilon v\cdot\nabla\right]
 f^{\mathrm{eq}}\bar{f}^{\mathrm{eq}}L^{-1}\hat{\chi}^j\Big>\psi_j
 \nonumber\\
 &=\epsilon\Big<L^{-1}\hat{\psi}^i,(f^{\mathrm{eq}}\bar{f}^{\mathrm{eq}})^{-1}
 \frac{\partial}{\partial T}[f^{\mathrm{eq}}\bar{f}^{\mathrm{eq}}L^{-1}\hat{\chi}^j]\Big>\psi_j
 \frac{\partial}{\partial\tau}T
 \nonumber\\
 &+\epsilon^2\Big<L^{-1}\hat{\psi}^i,(f^{\mathrm{eq}}\bar{f}^{\mathrm{eq}})^{-1}
 v^{\beta}\frac{\partial}{\partial T}[f^{\mathrm{eq}}\bar{f}^{\mathrm{eq}}L^{-1}\hat{\chi}^j]\Big>\psi_j
 \nabla_{\beta}T
 \nonumber\\
 &+\epsilon\Big<L^{-1}\hat{\psi}^i,(f^{\mathrm{eq}}\bar{f}^{\mathrm{eq}})^{-1}
 \frac{\partial}{\partial\frac{\mu}{T}}[f^{\mathrm{eq}}\bar{f}^{\mathrm{eq}}L^{-1}\hat{\chi}^j]\Big>\psi_j
 \frac{\partial}{\partial\tau}\frac{\mu}{T}
 \nonumber\\
 &+\epsilon^2\Big<L^{-1}\hat{\psi}^i,(f^{\mathrm{eq}}\bar{f}^{\mathrm{eq}})^{-1}
 v^{\beta}\frac{\partial}{\partial\frac{\mu}{T}}[f^{\mathrm{eq}}\bar{f}^{\mathrm{eq}}
 L^{-1}\hat{\chi}^j]\Big>\psi_j\nabla_{\beta}\frac{\mu}{T}
 \nonumber\\
 &+\epsilon\Big<L^{-1}\hat{\psi}^i,(f^{\mathrm{eq}}\bar{f}^{\mathrm{eq}})^{-1}
 \frac{\partial}{\partial u^{\beta}}[f^{\mathrm{eq}}\bar{f}^{\mathrm{eq}}L^{-1}\hat{\chi}^j]\Big>\psi_j
 \frac{\partial}{\partial\tau}u^{\beta}
 \nonumber\\
 &+\epsilon^2\Big<L^{-1}\hat{\psi}^i,(f^{\mathrm{eq}}\bar{f}^{\mathrm{eq}})^{-1}
 v^{\beta}\frac{\partial}{\partial u^{\alpha}}[f^{\mathrm{eq}}\bar{f}^{\mathrm{eq}}
 L^{-1}\hat{\chi}^j]\Big>\psi_j\nabla_{\beta}u^{\alpha}.
 \label{M-relax5}
\end{align}
The temporal derivative of $T$, $\mu/T$, and $u^{\mu}$ are rewritten by using the 
balance equations up to the first order with respect to $\epsilon$, which correspond to 
the Euler equation:
\begin{align}
 \label{eq:euler1}
 \frac{\partial}{\partial\tau}T
 &=-T\left.\frac{\partial P}{\partial e}\right|_{n}\epsilon\theta
 +O(\epsilon^2),
 \\
 \label{eq:euler2}
 \frac{\partial}{\partial\tau}\frac{\mu}{T}
 &=-\left.\frac{\partial P}{\partial n}\right|_{e}\epsilon\theta
 +O(\epsilon^2),
 \\
 \label{eq:euler3}
 \frac{\partial}{\partial\tau}u^{\mu}
 &=\frac{1}{T}\epsilon\nabla^{\mu}T+\frac{T}{h}\epsilon\nabla^{\mu}\frac{\mu}{T}
 +O(\epsilon^2).
\end{align}
Using the formulae \eqref{eq:f1}-\eqref{eq:f5} and Euler equation \eqref{eq:euler1}-\eqref{eq:euler3},  we convert Eq.~\eqref{M-relax5} into the following forms:
\begin{widetext}
\begin{align}
 \label{eq:Pichi}
 &\underline{\hat{\psi}^i=\hat{\Pi}}
 \nonumber\\[5pt]
 &\epsilon\Big<L^{-1}\hat{\Pi},(f^{\mathrm{eq}}\bar{f}^{\mathrm{eq}})^{-1}
 \left[\frac{\partial}{\partial\tau}+\epsilon v\cdot\nabla\right]
 f^{\mathrm{eq}}\bar{f}^{\mathrm{eq}}L^{-1}\hat{\chi}^j\Big>\psi_j
 \nonumber\\
 &=\epsilon^2\Big<L^{-1}\hat{\Pi},(f^{\mathrm{eq}}\bar{f}^{\mathrm{eq}})^{-1}
 \left[-T\left.\frac{\partial P}{\partial e}\right|_{n}\frac{\partial}{\partial T}
 -\left.\frac{\partial P}{\partial n}\right|_{e}\frac{\partial}{\partial\frac{\mu}{T}}\right]
 \frac{f^{\mathrm{eq}}\bar{f}^{\mathrm{eq}}L^{-1}\hat{\Pi}}{-T\zeta}\Big>
 \Pi \theta
 \nonumber\\
 &+\epsilon^2\Big<L^{-1}\hat{\Pi},(f^{\mathrm{eq}}\bar{f}^{\mathrm{eq}})^{-1}
 \left[v^{\beta}\frac{\partial}{\partial T}+\frac{1}{T}\frac{\partial}{\partial u_{\beta}}\right]
 \frac{f^{\mathrm{eq}}\bar{f}^{\mathrm{eq}}L^{-1}\hat{J}^\rho}{T^2\lambda/h}\Big>
 J_\rho \nabla_{\beta}T
 \nonumber\\
 &+\epsilon^2\Big<L^{-1}\hat{\Pi},(f^{\mathrm{eq}}\bar{f}^{\mathrm{eq}})^{-1}
 \left[v^{\beta}\frac{\partial}{\partial\frac{\mu}{T}}+\frac{T}{h}\frac{\partial}{\partial u_{\beta}}\right]
 \frac{f^{\mathrm{eq}}\bar{f}^{\mathrm{eq}}L^{-1}\hat{J}^\rho}{T^2\lambda/h}\Big>
 J_\rho \nabla_{\beta}\frac{\mu}{T}
 \nonumber\\
 &+\epsilon^2\Big<L^{-1}\hat{\Pi},(f^{\mathrm{eq}}\bar{f}^{\mathrm{eq}})^{-1}
 v^{\beta}\frac{\partial}{\partial u^{\alpha}}\frac{f^{\mathrm{eq}}\bar{f}^{\mathrm{eq}}
 L^{-1}\hat{\Pi}}{-T\zeta}\Big>
 \Pi \nabla_{\beta}u^{\alpha}
 \nonumber\\
 &+\epsilon^2\Big<L^{-1}\hat{\Pi},(f^{\mathrm{eq}}\bar{f}^{\mathrm{eq}})^{-1}
 v^{\beta}\frac{\partial}{\partial u^{\alpha}}\frac{f^{\mathrm{eq}}\bar{f}^{\mathrm{eq}}
 L^{-1}\hat{\pi}^{\rho\sigma}}{-2T\eta}\Big>
 \pi_{\rho\sigma} \nabla_{\beta}u^{\alpha}
 \nonumber\\
 &=\epsilon^2\Bigg[
 \Big<L^{-1}\hat{\Pi},(f^{\mathrm{eq}}\bar{f}^{\mathrm{eq}})^{-1}
 \left[-T\left.\frac{\partial P}{\partial e}\right|_{n}\frac{\partial}{\partial T}
 -\left.\frac{\partial P}{\partial n}\right|_{e}\frac{\partial}{\partial\frac{\mu}{T}}\right]
 \frac{f^{\mathrm{eq}}\bar{f}^{\mathrm{eq}}L^{-1}\hat{\Pi}}{-T\zeta}\Big>\Pi \theta
 \nonumber\\
 &+\frac{\Delta^{\alpha\beta}}{3}
 \Big<L^{-1}\hat{\Pi},(f^{\mathrm{eq}}\bar{f}^{\mathrm{eq}})^{-1}
 \left[v_\alpha \frac{\partial}{\partial T}+\frac{1}{T}\frac{\partial}{\partial u^\alpha}\right]
 \frac{f^{\mathrm{eq}}\bar{f}^{\mathrm{eq}}L^{-1}\hat{J}_\beta}{T^2\lambda/h}\Big>J^\rho\nabla_{\rho}T
 \nonumber\\
 &+\frac{\Delta^{\alpha\beta}}{3}
 \Big<L^{-1}\hat{\Pi},(f^{\mathrm{eq}}\bar{f}^{\mathrm{eq}})^{-1}
 \left[v_\alpha \frac{\partial}{\partial\frac{\mu}{T}}+\frac{T}{h}\frac{\partial}{\partial u^\alpha}\right]
 \frac{f^{\mathrm{eq}}\bar{f}^{\mathrm{eq}}L^{-1}\hat{J}_\beta}{T^2\lambda/h}\Big>J^\rho\nabla_{\rho}\frac{\mu}{T}
 \nonumber\\
 &+\frac{\Delta^{\alpha\beta}}{3}\Big<L^{-1}\hat{\Pi},(f^{\mathrm{eq}}\bar{f}^{\mathrm{eq}})^{-1}
 v_\alpha \frac{\partial}{\partial u^\beta}
 \frac{f^{\mathrm{eq}}\bar{f}^{\mathrm{eq}}L^{-1}\hat{\Pi}}{-T\zeta}\Big>\Pi \theta
 \nonumber\\
 &+\frac{\Delta^{\alpha\beta\gamma\delta}}{5}\Big<L^{-1}\hat{\Pi},(f^{\mathrm{eq}}\bar{f}^{\mathrm{eq}})^{-1}
 v_\alpha \frac{\partial}{\partial u^\beta}\frac{f^{\mathrm{eq}}\bar{f}^{\mathrm{eq}}
 L^{-1}\hat{\pi}_{\gamma\delta}}{-2T\eta}\Big>\pi^{\rho\sigma}\sigma_{\rho\sigma}
 \Bigg]
 \nonumber\\
 &=\epsilon^2\Big[
 \kappa_{\Pi\Pi}\Pi \theta
 +\kappa^{(1)}_{\Pi J}J^\rho\nabla_{\rho}T
 +\kappa^{(2)}_{\Pi J}J^\rho\nabla_{\rho}\frac{\mu}{T}
 +\kappa_{\Pi\pi}\pi^{\rho\sigma}\sigma_{\rho\sigma}
 \Big],
 \\[10pt]
 &\underline{\hat{\psi}^i=\hat{J}^\mu}
 \nonumber\\[5pt]
 &\epsilon\Big<L^{-1}\hat{J}^{\mu},(f^{\mathrm{eq}}\bar{f}^{\mathrm{eq}})^{-1}
 \left[\frac{\partial}{\partial\tau}+\epsilon v\cdot\nabla\right]
 f^{\mathrm{eq}}\bar{f}^{\mathrm{eq}}L^{-1}\hat{\psi}^j\Big>\chi_j
 \nonumber\\
 &=\epsilon^2\Big<L^{-1}\hat{J}^\mu,(f^{\mathrm{eq}}\bar{f}^{\mathrm{eq}})^{-1}
 \left[-T\left.\frac{\partial P}{\partial e}\right|_{n}\frac{\partial}{\partial T}
 -\left.\frac{\partial P}{\partial n}\right|_{e}\frac{\partial}{\partial\frac{\mu}{T}}\right]
 \frac{f^{\mathrm{eq}}\bar{f}^{\mathrm{eq}}L^{-1}\hat{J}^\nu}{T^2\lambda/h}\Big>
 J_\nu \theta
 \nonumber\\
 &+\epsilon^2\Big<L^{-1}\hat{J}^\mu,(f^{\mathrm{eq}}\bar{f}^{\mathrm{eq}})^{-1}
 \left[v^{\beta}\frac{\partial}{\partial T}+\frac{1}{T}\frac{\partial}{\partial u_{\beta}}\right]
 \frac{f^{\mathrm{eq}}\bar{f}^{\mathrm{eq}}L^{-1}\hat{\Pi}}{-T\zeta}\Big>
 \Pi \nabla_{\beta}T
 \nonumber\\
 &+\epsilon^2\Big<L^{-1}\hat{J}^\mu,(f^{\mathrm{eq}}\bar{f}^{\mathrm{eq}})^{-1}
 \left[v^{\beta}\frac{\partial}{\partial T}+\frac{1}{T}\frac{\partial}{\partial u_{\beta}}\right]
 \frac{f^{\mathrm{eq}}\bar{f}^{\mathrm{eq}}L^{-1}\hat{\pi}^{\rho\sigma}}{-2T\eta}\Big>
 \pi_{\rho\sigma} \nabla_{\beta}T
 \nonumber\\
 &+\epsilon^2\Big<L^{-1}\hat{J}^\mu,(f^{\mathrm{eq}}\bar{f}^{\mathrm{eq}})^{-1}
 \left[v^{\beta}\frac{\partial}{\partial\frac{\mu}{T}}+\frac{T}{h}\frac{\partial}{\partial u_{\beta}}\right]
 \frac{f^{\mathrm{eq}}\bar{f}^{\mathrm{eq}}L^{-1}\hat{\Pi}}{-T\zeta}\Big>
 \Pi \nabla_{\beta}\frac{\mu}{T}
 \nonumber\\
 &+\epsilon^2\Big<L^{-1}\hat{J}^\mu,(f^{\mathrm{eq}}\bar{f}^{\mathrm{eq}})^{-1}
 \left[v^{\beta}\frac{\partial}{\partial\frac{\mu}{T}}+\frac{T}{h}\frac{\partial}{\partial u_{\beta}}\right]
 \frac{f^{\mathrm{eq}}\bar{f}^{\mathrm{eq}}L^{-1}\hat{\pi}^{\rho\sigma}}{-2T\eta}\Big>
 \pi_{\rho\sigma} \nabla_{\beta}\frac{\mu}{T}
 \nonumber\\
 &+\epsilon^2\Big<L^{-1}\hat{J}^\mu,(f^{\mathrm{eq}}\bar{f}^{\mathrm{eq}})^{-1}
 v^{\beta}\frac{\partial}{\partial u^{\alpha}}\frac{f^{\mathrm{eq}}\bar{f}^{\mathrm{eq}}
 L^{-1}\hat{J}^\rho}{T^2\lambda/h}\Big>
 J_\rho \nabla_{\beta}u^{\alpha}
 \nonumber\\
 &=\epsilon^2\Bigg[
 \frac{\Delta^{\rho\sigma}}{3}
 \Big<L^{-1}\hat{J}_\rho,(f^{\mathrm{eq}}\bar{f}^{\mathrm{eq}})^{-1}
 \left[-T\left.\frac{\partial P}{\partial e}\right|_{n}\frac{\partial}{\partial T}
 -\left.\frac{\partial P}{\partial n}\right|_{e}\frac{\partial}{\partial\frac{\mu}{T}}\right]
 \frac{f^{\mathrm{eq}}\bar{f}^{\mathrm{eq}}L^{-1}\hat{J}_\sigma}{T^2\lambda/h}\Big>J^\mu \theta
 \nonumber\\
 &+\frac{\Delta^{\rho\sigma}}{3}
 \Big<L^{-1}\hat{J}_\rho,(f^{\mathrm{eq}}\bar{f}^{\mathrm{eq}})^{-1}
 \left[v_\sigma\frac{\partial}{\partial T}+\frac{1}{T}\frac{\partial}{\partial u^\sigma}\right]
 \frac{f^{\mathrm{eq}}\bar{f}^{\mathrm{eq}}L^{-1}\hat{\Pi}}{-T\zeta}\Big>\Pi\nabla^{\mu}T
 \nonumber\\
 &+\frac{\Delta^{\alpha\beta\gamma\delta}}{5}
 \Big<L^{-1}\hat{J}_\alpha,(f^{\mathrm{eq}}\bar{f}^{\mathrm{eq}})^{-1}
 \left[v_\beta \frac{\partial}{\partial T}+\frac{1}{T}\frac{\partial}{\partial u^\beta}\right]
 \frac{f^{\mathrm{eq}}\bar{f}^{\mathrm{eq}}L^{-1}\hat{\pi}_{\gamma\delta}}{-2T\eta}\Big>\pi^{\mu\rho}\nabla_{\rho}T
 \nonumber\\
 &+\frac{\Delta^{\rho\sigma}}{3}
 \Big<L^{-1}\hat{J}_\rho,(f^{\mathrm{eq}}\bar{f}^{\mathrm{eq}})^{-1}
 \left[v_\sigma \frac{\partial}{\partial\frac{\mu}{T}}+\frac{T}{h}\frac{\partial}{\partial u^\sigma}\right]
 \frac{f^{\mathrm{eq}}\bar{f}^{\mathrm{eq}}L^{-1}\hat{\Pi}}{-T\zeta}\Big>\Pi\nabla^{\mu}\frac{\mu}{T}
 \nonumber\\
 &+\frac{\Delta^{\alpha\beta\gamma\delta}}{5}
 \Big<L^{-1}\hat{J}_\alpha,(f^{\mathrm{eq}}\bar{f}^{\mathrm{eq}})^{-1}
 \left[v_\beta\frac{\partial}{\partial\frac{\mu}{T}}+\frac{T}{h}\frac{\partial}{\partial u^\beta}\right]
 \frac{f^{\mathrm{eq}}\bar{f}^{\mathrm{eq}}L^{-1}\hat{\pi}_{\gamma\delta}}{-2T\eta}\Big>\pi^{\mu\rho}\nabla_{\rho}\frac{\mu}{T}
 \nonumber\\
 &+\frac{\Delta^{\rho\sigma}\Delta^{\alpha\beta}}{9}
 \Big<L^{-1}\hat{J}_\rho,(f^{\mathrm{eq}}\bar{f}^{\mathrm{eq}})^{-1}
 v_\alpha \frac{\partial}{\partial u^\beta}
 \frac{f^{\mathrm{eq}}\bar{f}^{\mathrm{eq}}L^{-1}\hat{J}_\sigma}{T^2\lambda/h}\Big>J^\mu \theta
 \nonumber\\
 &+\frac{\Delta^{\alpha\beta\gamma\delta}}{5}
 \Big<L^{-1}\hat{J}_\alpha,(f^{\mathrm{eq}}\bar{f}^{\mathrm{eq}})^{-1}
 v_\gamma\frac{\partial}{\partial u^{\delta}}\frac{f^{\mathrm{eq}}\bar{f}^{\mathrm{eq}}
 L^{-1}\hat{J}_\beta}{T^2\lambda/h}\Big>J^\rho{\sigma^\mu}_\rho
 \nonumber\\
 &+\frac{\Omega^{\alpha\beta\gamma\delta}}{3}
 \Big<L^{-1}\hat{J}_\alpha,(f^{\mathrm{eq}}\bar{f}^{\mathrm{eq}})^{-1}
 v_\gamma \frac{\partial}{\partial u^{\delta}}\frac{f^{\mathrm{eq}}\bar{f}^{\mathrm{eq}}
 L^{-1}\hat{J}_\beta}{T^2\lambda/h}\Big>J^\rho{\omega^\mu}_\rho
 \Bigg]
 \nonumber\\
 &=\epsilon^2\Big[
 \kappa^{(1)}_{J\Pi}\Pi\nabla^{\mu}T
 +\kappa^{(2)}_{J\Pi}\Pi\nabla^{\mu}\frac{\mu}{T}
 +\kappa^{(1)}_{JJ}J^\mu \theta
 +\kappa^{(2)}_{JJ}J^\rho{\sigma^\mu}_\rho
 +\kappa^{(3)}_{JJ}J^\rho{\omega^\mu}_\rho
 +\kappa^{(1)}_{J\pi}\pi^{\mu\rho}\nabla_{\rho}T
 +\kappa^{(2)}_{J\pi}\pi^{\mu\rho}\nabla_{\rho}\frac{\mu}{T}
 \Big],
 \\[10pt]
 &\underline{\hat{\psi}^i=\hat{\pi}^{\mu\nu}}
 \nonumber\\[5pt]
 &\epsilon\Big<L^{-1}\hat{\pi}^{\mu\nu},(f^{\mathrm{eq}}\bar{f}^{\mathrm{eq}})^{-1}
 \left[\frac{\partial}{\partial\tau}+\epsilon v\cdot\nabla\right]
 f^{\mathrm{eq}}\bar{f}^{\mathrm{eq}}L^{-1}\hat{\psi}^j\Big>\chi_j
 \nonumber\\
 &=\epsilon^2\Big<L^{-1}\hat{\pi}^{\mu\nu},(f^{\mathrm{eq}}\bar{f}^{\mathrm{eq}})^{-1}
 \left[-T\left.\frac{\partial P}{\partial e}\right|_{n}\frac{\partial}{\partial T}
 -\left.\frac{\partial P}{\partial n}\right|_{e}\frac{\partial}{\partial\frac{\mu}{T}}\right]
 \frac{f^{\mathrm{eq}}\bar{f}^{\mathrm{eq}}L^{-1}\hat{\pi}^{\rho\sigma}}{-2T\eta}]\Big>
 \pi_{\rho\sigma} \theta 
 \nonumber\\
 &+\epsilon^2\Big<L^{-1}\hat{\pi}^{\mu\nu},(f^{\mathrm{eq}}\bar{f}^{\mathrm{eq}})^{-1}
 \left[v^{\beta}\frac{\partial}{\partial T}+\frac{1}{T}\frac{\partial}{\partial u_{\beta}}\right]
 \frac{f^{\mathrm{eq}}\bar{f}^{\mathrm{eq}}L^{-1}\hat{J}^\rho}{T^2\lambda/h}\Big>
 J_\rho \nabla_{\beta}T
 \nonumber\\
 &+\epsilon^2\Big<L^{-1}\hat{\pi}^{\mu\nu},(f^{\mathrm{eq}}\bar{f}^{\mathrm{eq}})^{-1}
 \left[v^{\beta}\frac{\partial}{\partial\frac{\mu}{T}}+\frac{T}{h}\frac{\partial}{\partial u_{\beta}}\right]
 \frac{f^{\mathrm{eq}}\bar{f}^{\mathrm{eq}}L^{-1}\hat{J}^\rho}{T^2\lambda/h}\Big>
 J_\rho \nabla_{\beta}\frac{\mu}{T}
 \nonumber\\
 &+\epsilon^2\Big<L^{-1}\hat{\pi}^{\mu\nu},(f^{\mathrm{eq}}\bar{f}^{\mathrm{eq}})^{-1}
 v^{\beta}\frac{\partial}{\partial u^{\alpha}}\frac{f^{\mathrm{eq}}\bar{f}^{\mathrm{eq}}
 L^{-1}\hat{\Pi}}{-T\zeta}\Big>
 \Pi \nabla_{\beta}u^{\alpha}
 \nonumber\\
 &+\epsilon^2\Big<L^{-1}\hat{\pi}^{\mu\nu},(f^{\mathrm{eq}}\bar{f}^{\mathrm{eq}})^{-1}
 v^{\beta}\frac{\partial}{\partial u^{\alpha}}\frac{f^{\mathrm{eq}}\bar{f}^{\mathrm{eq}}
 L^{-1}\hat{\pi}^{\rho\sigma}}{-2T\eta}\Big>
 \pi_{\rho\sigma} \nabla_{\beta}u^{\alpha}
 \nonumber\\
 &=\epsilon^2\Bigg[
 \frac{\Delta^{\rho\sigma\alpha\beta}}{5}
 \Big<L^{-1}\hat{\pi}_{\rho\sigma},(f^{\mathrm{eq}}\bar{f}^{\mathrm{eq}})^{-1}
 \left[-T\left.\frac{\partial P}{\partial e}\right|_{n}\frac{\partial}{\partial T}
 -\left.\frac{\partial P}{\partial n}\right|_{e}\frac{\partial}{\partial\frac{\mu}{T}}\right]
 \frac{f^{\mathrm{eq}}\bar{f}^{\mathrm{eq}}L^{-1}\hat{\pi}_{\alpha\beta}}{-2T\eta}\Big>\pi^{\mu\nu} \theta
 \nonumber\\
 &+\frac{\Delta^{\rho\sigma\alpha\beta}}{5}
 \Big<L^{-1}\hat{\pi}_{\rho\sigma},(f^{\mathrm{eq}}\bar{f}^{\mathrm{eq}})^{-1}
 v_\alpha\frac{\partial}{\partial u^{\beta}}\frac{f^{\mathrm{eq}}\bar{f}^{\mathrm{eq}}
 L^{-1}\hat{\Pi}}{-T\zeta}\Big>\Pi \sigma^{\mu\nu}
 \nonumber\\
 &+\frac{\Delta^{\rho\sigma\alpha\beta}}{5}
 \Big<L^{-1}\hat{\pi}_{\rho\sigma},(f^{\mathrm{eq}}\bar{f}^{\mathrm{eq}})^{-1}
 \left[v_\alpha \frac{\partial}{\partial T}+\frac{1}{T}\frac{\partial}{\partial u^\alpha}\right]
 \frac{f^{\mathrm{eq}}\bar{f}^{\mathrm{eq}}L^{-1}\hat{J}_\beta}{T^2\lambda/h}\Big>J^{\langle\mu}\nabla^{\nu\rangle} T
 \nonumber\\
 &+\frac{\Delta^{\rho\sigma\alpha\beta}}{5}
 \Big<L^{-1}\hat{\pi}_{\rho\sigma},(f^{\mathrm{eq}}\bar{f}^{\mathrm{eq}})^{-1}
 \left[v_\alpha \frac{\partial}{\partial\frac{\mu}{T}}
 +\frac{T}{h}\frac{\partial}{\partial u^\alpha}\right]
 \frac{f^{\mathrm{eq}}\bar{f}^{\mathrm{eq}}L^{-1}\hat{J}_\beta}{T^2\lambda/h}\Big>J^{\langle\mu}\nabla^{\nu\rangle} \frac{\mu}{T}
 \nonumber\\
 &+\frac{\Delta^{\rho\sigma\alpha\beta}}{5}
 \Big<L^{-1}\hat{\pi}_{\rho\sigma},(f^{\mathrm{eq}}\bar{f}^{\mathrm{eq}})^{-1}
 v_\alpha\frac{\partial}{\partial u^{\beta}}\frac{f^{\mathrm{eq}}\bar{f}^{\mathrm{eq}}
 L^{-1}\hat{\Pi}}{-T\zeta}\Big>\Pi \sigma^{\mu\nu}
 \nonumber\\
 &+\frac{\Delta^{\rho\sigma\alpha\beta}\Delta^{\gamma\delta}}{15}
 \Big<L^{-1}\hat{\pi}_{\rho\sigma},(f^{\mathrm{eq}}\bar{f}^{\mathrm{eq}})^{-1}
 v_\gamma \frac{\partial}{\partial u^\delta}
 \frac{f^{\mathrm{eq}}\bar{f}^{\mathrm{eq}}L^{-1}\hat{\pi}_{\alpha\beta}}{-2T\eta}\Big>\pi^{\mu\nu} \theta
 \nonumber\\
 &+\frac{12}{35}\Delta^{\tau\eta\gamma\delta}{\Delta^{\kappa\sigma\lambda}}_\gamma{\Delta^{\alpha\beta}}_{\lambda\delta}
 \Big<L^{-1}\hat{\pi}_{\tau\eta},(f^{\mathrm{eq}}\bar{f}^{\mathrm{eq}})^{-1}
 v_\alpha \frac{\partial}{\partial u^\beta}\frac{f^{\mathrm{eq}}\bar{f}^{\mathrm{eq}}
 L^{-1}\hat{\pi}_{\kappa\sigma}}{-2T\eta}\Big> \pi^{\rho\langle\mu}{\sigma^{\nu\rangle}}_\rho
 \nonumber\\
 &+\frac{4}{15}\Delta^{\tau\eta\gamma\delta}{\Delta^{\kappa\sigma\lambda}}_\gamma{\Omega^{\alpha\beta}}_{\lambda\delta}
 \Big<L^{-1}\hat{\pi}_{\tau\eta},(f^{\mathrm{eq}}\bar{f}^{\mathrm{eq}})^{-1}
 v_\alpha\frac{\partial}{\partial u^\beta}\frac{f^{\mathrm{eq}}\bar{f}^{\mathrm{eq}}
 L^{-1}\hat{\pi}_{\kappa\sigma}}{-2T\eta}\Big> \pi^{\rho\langle\mu}{\omega^{\nu\rangle}}_\rho
 \Bigg]
 \nonumber\\
 &=\epsilon^2\Big[
 \kappa_{\pi\Pi} \Pi \sigma^{\mu\nu}
 + \kappa^{(1)}_{\pi J}J^{\langle\mu}\nabla^{\nu\rangle} T
 + \kappa^{(2)}_{\pi J}J^{\langle\mu}\nabla^{\nu\rangle} \frac{\mu}{T}
 + \kappa^{(1)}_{\pi\pi}\pi^{\mu\nu} \theta
 + \kappa^{(2)}_{\pi\pi} \pi^{\rho\langle\mu}{\sigma^{\nu\rangle}}_\rho
 + \kappa^{(3)}_{\pi\pi} \pi^{\rho\langle\mu}{\omega^{\nu\rangle}}_\rho
 \Big].
\end{align}
The coefficients 
$\kappa_{\Pi\Pi}$, $\kappa^{(1)}_{\Pi J}$, $\kappa^{(2)}_{\Pi J}$, $\kappa_{\Pi\pi}$, $\kappa^{(1)}_{J\Pi}$, $\kappa^{(2)}_{J\Pi}$, $\kappa^{(1)}_{JJ}$,
$\kappa^{(2)}_{JJ}$, $\kappa^{(3)}_{JJ}$, $\kappa^{(1)}_{J\pi}$, $\kappa^{(2)}_{J\pi}$, $\kappa_{\pi\Pi}$, $\kappa^{(1)}_{\pi J}$,
$\kappa^{(2)}_{\pi J}$, $\kappa^{(1)}_{\pi\pi}$, $\kappa^{(2)}_{\pi\pi}$, and $\kappa^{(3)}_{\pi\pi}$
are defined by
\begin{align}
 \kappa_{\Pi\Pi}
 &\equiv 
 \Big<L^{-1}\hat{\Pi},(f^{\mathrm{eq}}\bar{f}^{\mathrm{eq}})^{-1}
 \left[-T\left.\frac{\partial P}{\partial e}\right|_{n}\frac{\partial}{\partial T}
 -\left.\frac{\partial P}{\partial n}\right|_{e}\frac{\partial}{\partial\frac{\mu}{T}}
 +\frac{1}{3}v^\mu \frac{\partial}{\partial u^\mu}\right]
 \frac{f^{\mathrm{eq}}\bar{f}^{\mathrm{eq}}L^{-1}\hat{\Pi}}{-T\zeta}\Big>,
 \\
 \kappa_{\Pi J}^{(1)}
 &\equiv 
 \frac{\Delta^{\mu\nu}}{3}
 \Big<L^{-1}\hat{\Pi},(f^{\mathrm{eq}}\bar{f}^{\mathrm{eq}})^{-1}
 \left[v_\mu \frac{\partial}{\partial T}+\frac{1}{T}\frac{\partial}{\partial u^\mu}\right]
 \frac{f^{\mathrm{eq}}\bar{f}^{\mathrm{eq}}L^{-1}\hat{J}_\nu}{T^2\lambda/h}\Big>,
 \\
 \kappa_{\Pi J}^{(2)}
 &\equiv 
 \frac{\Delta^{\mu\nu}}{3}
 \Big<L^{-1}\hat{\Pi},(f^{\mathrm{eq}}\bar{f}^{\mathrm{eq}})^{-1}
 \left[v_\mu \frac{\partial}{\partial\frac{\mu}{T}}+\frac{T}{h}\frac{\partial}{\partial u^\mu}\right]
 \frac{f^{\mathrm{eq}}\bar{f}^{\mathrm{eq}}L^{-1}\hat{J}_\nu}{T^2\lambda/h}\Big>,
 \\
 \kappa_{\Pi\pi}
 &\equiv 
 \frac{\Delta^{\mu\nu\rho\sigma}}{5}\Big<L^{-1}\hat{\Pi},(f^{\mathrm{eq}}\bar{f}^{\mathrm{eq}})^{-1}
 v_\mu \frac{\partial}{\partial u^\nu}\frac{f^{\mathrm{eq}}\bar{f}^{\mathrm{eq}}
 L^{-1}\hat{\pi}_{\rho\sigma}}{-2T\eta}\Big>,
 \\
 \kappa_{J\Pi}^{(1)}
 &\equiv
 \frac{1}{3}
 \Big<L^{-1}\hat{J}^\mu,(f^{\mathrm{eq}}\bar{f}^{\mathrm{eq}})^{-1}
 \left[v_\mu\frac{\partial}{\partial T}+\frac{1}{T}\frac{\partial}{\partial u^\mu}\right]
 \frac{f^{\mathrm{eq}}\bar{f}^{\mathrm{eq}}L^{-1}\hat{\Pi}}{-T\zeta}\Big>,
 \\
 \kappa_{J\Pi}^{(2)}
 &\equiv 
 \frac{1}{3}
 \Big<L^{-1}\hat{J}^\mu,(f^{\mathrm{eq}}\bar{f}^{\mathrm{eq}})^{-1}
 \left[v_\mu \frac{\partial}{\partial\frac{\mu}{T}}+\frac{T}{h}\frac{\partial}{\partial u^\mu}\right]
 \frac{f^{\mathrm{eq}}\bar{f}^{\mathrm{eq}}L^{-1}\hat{\Pi}}{-T\zeta}\Big>,
 \\
 \kappa_{JJ}^{(1)}
 &\equiv 
 \frac{\Delta^{\mu\nu}}{3}
 \Big<L^{-1}\hat{J}_\mu,(f^{\mathrm{eq}}\bar{f}^{\mathrm{eq}})^{-1}
 \left[-T\left.\frac{\partial P}{\partial e}\right|_{n}\frac{\partial}{\partial T}
 -\left.\frac{\partial P}{\partial n}\right|_{e}\frac{\partial}{\partial\frac{\mu}{T}}
 +\frac{1}{3}v^\rho \frac{\partial}{\partial u^\rho}\right]
 \frac{f^{\mathrm{eq}}\bar{f}^{\mathrm{eq}}L^{-1}\hat{J}_\nu}{T^2\lambda/h}\Big>,
 \\
 \kappa_{JJ}^{(2)}
 &\equiv 
 \frac{\Delta^{\mu\nu\rho\sigma}}{5}
 \Big<L^{-1}\hat{J}_\mu,(f^{\mathrm{eq}}\bar{f}^{\mathrm{eq}})^{-1}
 v_\rho\frac{\partial}{\partial u^{\sigma}}\frac{f^{\mathrm{eq}}\bar{f}^{\mathrm{eq}}
 L^{-1}\hat{J}_\nu}{T^2\lambda/h}\Big>,
 \\
 \kappa_{JJ}^{(3)}
 &\equiv 
 \frac{\Omega^{\mu\nu\rho\sigma}}{3}
 \Big<L^{-1}\hat{J}_\mu,(f^{\mathrm{eq}}\bar{f}^{\mathrm{eq}})^{-1}
 v_\rho\frac{\partial}{\partial u^{\sigma}}\frac{f^{\mathrm{eq}}\bar{f}^{\mathrm{eq}}
 L^{-1}\hat{J}_\nu}{T^2\lambda/h}\Big>,
 \\
 \kappa_{J\pi}^{(1)}
 &\equiv 
 \frac{\Delta^{\mu\nu\rho\sigma}}{5}
 \Big<L^{-1}\hat{J}_\mu,(f^{\mathrm{eq}}\bar{f}^{\mathrm{eq}})^{-1}
 \left[v_\nu \frac{\partial}{\partial T}+\frac{1}{T}\frac{\partial}{\partial u^\nu}\right]
 \frac{f^{\mathrm{eq}}\bar{f}^{\mathrm{eq}}L^{-1}\hat{\pi}_{\rho\sigma}}{-2T\eta}\Big>,
 \\
 \kappa_{J\pi}^{(2)}
 &\equiv 
 \frac{\Delta^{\mu\nu\rho\sigma}}{5}
 \Big<L^{-1}\hat{J}_\mu,(f^{\mathrm{eq}}\bar{f}^{\mathrm{eq}})^{-1}
 \left[v_\nu \frac{\partial}{\partial\frac{\mu}{T}}+\frac{T}{h}\frac{\partial}{\partial u^\nu}\right]
 \frac{f^{\mathrm{eq}}\bar{f}^{\mathrm{eq}}L^{-1}\hat{\pi}_{\rho\sigma}}{-2T\eta}\Big>,
 \\
 \kappa_{\pi\Pi}
 &\equiv 
 \frac{\Delta^{\mu\nu\rho\sigma}}{5}
 \Big<L^{-1}\hat{\pi}_{\mu\nu},(f^{\mathrm{eq}}\bar{f}^{\mathrm{eq}})^{-1}
 v_\rho\frac{\partial}{\partial u^{\sigma}}\frac{f^{\mathrm{eq}}\bar{f}^{\mathrm{eq}}
 L^{-1}\hat{\Pi}}{-T\zeta}\Big>,
 \\
 \kappa_{\pi J}^{(1)}
 &\equiv
 \frac{\Delta^{\mu\nu\rho\sigma}}{5}
 \Big<L^{-1}\hat{\pi}_{\mu\nu},(f^{\mathrm{eq}}\bar{f}^{\mathrm{eq}})^{-1}
 \left[v_\rho \frac{\partial}{\partial T}+\frac{1}{T}\frac{\partial}{\partial u^\rho}\right]
 \frac{f^{\mathrm{eq}}\bar{f}^{\mathrm{eq}}L^{-1}\hat{J}_\sigma}{T^2\lambda/h}\Big>,
 \\
 \kappa_{\pi J}^{(2)}
 &\equiv 
 \frac{\Delta^{\mu\nu\rho\sigma}}{5}
 \Big<L^{-1}\hat{\pi}_{\mu\nu},(f^{\mathrm{eq}}\bar{f}^{\mathrm{eq}})^{-1}
 \left[v_\rho \frac{\partial}{\partial\frac{\mu}{T}}+\frac{T}{h}\frac{\partial}{\partial u^\rho}\right]
 \frac{f^{\mathrm{eq}}\bar{f}^{\mathrm{eq}}L^{-1}\hat{J}_\sigma}{T^2\lambda/h}\Big>,
 \\
 \kappa_{\pi\pi}^{(1)}
 &\equiv 
 \frac{\Delta^{\mu\nu\rho\sigma}}{5}
 \Big<L^{-1}\hat{\pi}_{\mu\nu},(f^{\mathrm{eq}}\bar{f}^{\mathrm{eq}})^{-1}
 \left[-T\left.\frac{\partial P}{\partial e}\right|_{n}\frac{\partial}{\partial T}
 -\left.\frac{\partial P}{\partial n}\right|_{e}\frac{\partial}{\partial\frac{\mu}{T}}
 +\frac{1}{3}v^\mu \frac{\partial}{\partial u^\mu}\right]
 \frac{f^{\mathrm{eq}}\bar{f}^{\mathrm{eq}}L^{-1}\hat{\pi}_{\rho\sigma}}{-2T\eta}\Big>,
 \\
 \kappa_{\pi\pi}^{(2)}
 &\equiv 
 \frac{12}{35}\Delta^{\mu\nu\gamma\delta}{\Delta^{\rho\sigma\lambda}}_\gamma{\Delta^{\alpha\beta}}_{\lambda\delta}
 \Big<L^{-1}\hat{\pi}_{\mu\nu},(f^{\mathrm{eq}}\bar{f}^{\mathrm{eq}})^{-1}
 v_\alpha \frac{\partial}{\partial u^\beta}\frac{f^{\mathrm{eq}}\bar{f}^{\mathrm{eq}}
 L^{-1}\hat{\pi}_{\rho\sigma}}{-2T\eta}\Big>,
 \\
 \kappa_{\pi\pi}^{(3)}
 &\equiv 
 \frac{4}{15}\Delta^{\mu\nu\gamma\delta}{\Delta^{\rho\sigma\lambda}}_\gamma{\Omega^{\alpha\beta}}_{\lambda\delta}
 \Big<L^{-1}\hat{\pi}_{\mu\nu},(f^{\mathrm{eq}}\bar{f}^{\mathrm{eq}})^{-1}
 v_\alpha\frac{\partial}{\partial u^\beta}\frac{f^{\mathrm{eq}}\bar{f}^{\mathrm{eq}}
 L^{-1}\hat{\pi}_{\rho\sigma}}{-2T\eta}\Big>.
\end{align}
\end{widetext}

Substituting the above equations into Eqs. (\ref{eq:relaxalleq2})
and setting $\epsilon$ equal to $1$,
we arrive at
the explicit form of the relaxation equations (\ref{eq:relax1})-(\ref{eq:relax3}).


\setcounter{equation}{0}
\section{
  Proofs of stability and causality
}
\label{sec:app4}
\subsubsection{
  Proof of stability around the static solution
}

In this Appendix, we show that the static solution, especially the equilibrium solution,
of the second-order relativistic hydrodynamic equation given by the pair of Eqs. (\ref{eq:ChapB-RHD1}) and (\ref{eq:ChapB-RHD2})
is stable against a small perturbation.

A generic constant solution reads
\begin{eqnarray}
  \label{eq:2nd-causal-001}
  T(\sigma\,;\,\tau) &=& T_0,\\
  \label{eq:2nd-causal-002}
  \mu(\sigma\,;\,\tau) &=& \mu_0,\\
  \label{eq:2nd-causal-003}
  u^\mu(\sigma\,;\,\tau) &=& u^\mu_0,\\
  \label{eq:2nd-causal-004}
  \Pi(\sigma\,;\,\tau) &=& 0,\\
  \label{eq:2nd-causal-0041}
  J^\mu(\sigma\,;\,\tau) &=& 0,\\
  \label{eq:2nd-causal-0042}
  \pi^{\mu\nu}(\sigma\,;\,\tau) &=& 0,
\end{eqnarray}
where $T_0$, $\mu_0$, and $u^\mu_0$ are constant.
We remark that
the equilibrium state
is correspondent to the special case of $u^\mu_0 = (1,\,0,\,0,\,0)$.

To show the stability of the constant solution,
we apply the linear stability analysis to
the second-order relativistic hydrodynamic equation (\ref{eq:ChapB-RHD1}) and (\ref{eq:ChapB-RHD2}).
We expand $T$, $\mu$, $u^\mu$, $\Pi$, $J^\mu$, and $\pi^{\mu\nu}$
around the constant solution as follows:
\begin{eqnarray}
  \label{eq:2nd-causal-005}
  T(\sigma\,;\,\tau) &=& T_0 + \delta T(\sigma\,;\,\tau),\\
  \label{eq:2nd-causal-006}
  \mu(\sigma\,;\,\tau) &=& \mu_0 + \delta\mu(\sigma\,;\,\tau),\\
  \label{eq:2nd-causal-007}
  u^\mu(\sigma\,;\,\tau) &=& u^\mu_0 + \delta u^\mu(\sigma\,;\,\tau),\\
  \label{eq:2nd-causal-008}
  \Pi(\sigma\,;\,\tau) &=& \delta \Pi(\sigma\,;\,\tau),\\
  J^\mu(\sigma\,;\,\tau) &=& \delta J^\mu(\sigma\,;\,\tau),\\
  \pi^{\mu\nu}(\sigma\,;\,\tau) &=& \delta \pi^{\mu\nu}(\sigma\,;\,\tau).
\end{eqnarray}
We assume that
the higher term than second order in terms of
$\delta T$, $\delta \mu$, $\delta u^\mu$, $\delta \Pi$, $\delta J^\mu$, and $\delta \pi^{\mu\nu}$ 
can be neglected since these quantities are small.

Instead of $\delta T$, $\delta \mu$, and $\delta u^\mu$ which are not 
independent of each other
because $\delta u_\mu \, u^\mu_0 = 0$,
we use
the following variables as the five independent variables composed of
$\delta T$, $\delta \mu$, and $\delta u^\mu$;
\begin{eqnarray}
  \label{eq:2nd-causal-009}
  \delta X_{4\mu} &\equiv& -\delta (u_\mu/T) = - \delta u_\mu/T_0 + \delta T \, u_{0\mu}/T^2_0,\\
  \delta X_{44} &\equiv& \delta (\mu/T) = \delta\mu/T_0 - \delta T \, \mu_0/T^2_0.
\end{eqnarray}
In the following,
we suppress the subscript ``0" in $T_0$, $\mu_0$, and $u_0^\mu$.
Furthermore,
we 
introduce the following variables
\begin{eqnarray}
  \delta X_{\mu\nu} &\equiv& \frac{- \Delta_{\mu\nu}\,\delta \Pi/3}{\langle\, \hat{\Pi}\,,\,\hat{L}^{-1}\,\hat{\Pi} \rangle}
  + \frac{\delta \pi_{\mu\nu}}{\frac{1}{5}\,\langle\, \hat{\pi}^{\rho\sigma}\,,\,\hat{L}^{-1}\,\hat{\pi}_{\rho\sigma} \rangle},\\
  \delta X_{\mu 4} &\equiv& \frac{h\,\delta J_{\mu}}{\frac{1}{3}\,\langle\, \hat{J}^{\rho}\,,\,\hat{L}^{-1}\,\hat{J}_{\rho} \rangle},
\end{eqnarray}
which 
are expressed in terms of $\delta\Pi$, $\delta J^\mu$, and $\delta\pi^{\mu\nu}$.
We treat $\delta X_{\alpha\beta} = (\delta X_{\mu\nu},\,\delta X_{\mu 4},\,\delta X_{4\nu},\,\delta X_{44})$
as the fundamental variables.

Substituting Eqs. (\ref{eq:2nd-causal-005})-(\ref{eq:2nd-causal-008}) into
the second-order relativistic hydrodynamic equation (\ref{eq:ChapB-RHD1}) and (\ref{eq:ChapB-RHD2}),
we obtain the linearized equation governing $\delta X_{\alpha\beta}$ as
\begin{eqnarray}
  \label{eq:2nd-causal-010}
  &&\langle\,\varphi_{0}^\alpha \,,\,
  \varphi^\beta_0 \rangle
  \frac{\partial}{\partial\tau}\delta X_{4\beta}
  +
  \langle\,\varphi_{0}^\alpha\,,\,
  \hat{L}^{-1}\,\varphi^{\nu\beta}_1 \rangle
  \frac{\partial}{\partial \tau}
  \delta X_{\nu\beta}\nonumber\\
  &&{}+
  \langle\,\varphi_{0}^{\alpha} \,,\,
  v^\rho\,
  \varphi^\beta_0 \rangle
  \nabla_\rho \delta X_{4\beta}\nonumber\\
  &&{}+
  \langle\,\varphi_{0}^{\alpha}\,,\,
  v^\rho\,
  \hat{L}^{-1}\,\varphi^{\nu\beta}_1\rangle
  \nabla_\rho \delta X_{\nu\beta}\nonumber\\
  &=& 0,\\
  \label{eq:2nd-causal-011}
  &&\langle\,\hat{L}^{-1}\,\varphi^{\mu\alpha}_1\,,\,\varphi^{\beta}_0 \rangle
  \frac{\partial}{\partial \tau}
  \delta X_{4\beta}\nonumber\\
  &&{}+
  \langle\,\hat{L}^{-1}\,\varphi^{\mu\alpha}_1\,,\,\hat{L}^{-1}\,\varphi^{\nu\beta}_1 \rangle
  \frac{\partial}{\partial \tau}
  \delta X_{\nu\beta}
  \nonumber\\
  &&{}+\langle\,\hat{L}^{-1}\,\varphi_{1}^{\mu\alpha}\,,\,
  v^\rho \, \varphi^{\beta}_0
  \rangle\nabla_\rho \delta X_{4\beta}\nonumber\\
  &&{}+
  \langle\,\hat{L}^{-1}\,\varphi^{\mu\alpha}_1\,,\,v^\rho \, \hat{L}^{-1}\,\varphi^{\nu\beta}_1 \rangle
  \nabla_\rho \delta X_{\nu\beta}\nonumber\\
  &=& \langle\,\hat{L}^{-1}\,\varphi_{1}^{\mu\alpha}\,,\,
  \hat{L}\,\hat{L}^{-1}\,\varphi^{\nu\beta}_1 \rangle \delta X_{\nu\beta}.
\end{eqnarray}
In the derivation of Eqs. (\ref{eq:2nd-causal-010}) and (\ref{eq:2nd-causal-011}),
we have used the fact that
%
\begin{eqnarray}
  \label{eq:2nd-causal-013}
  \delta (f^{\mathrm{eq}}_p) &=& f^{\mathrm{eq}}_p \, \bar{f}^{\mathrm{eq}}_p \, \varphi^\alpha_{0p} \, \delta X_{4\alpha},\\
  \label{eq:2nd-causal-014}
  \delta (\Psi_p) &=& \big[ \hat{L}^{-1} \, \varphi^{\mu\alpha}_1 \big]_p \, \delta X_{\mu\alpha},
\end{eqnarray}
with
\begin{eqnarray}
  \varphi^{\mu\alpha}_{1p}
  \equiv \left\{
  \begin{array}{ll}
    \displaystyle{
       - \Delta^{\mu\nu}\,\hat{\Pi}_p + \hat{\pi}^{\mu\nu}_p,
    }
    & \displaystyle{
      \alpha = \nu,
    } \\[2mm]
    \displaystyle{
       \hat{J}^\mu_p,
    }
    & \displaystyle{
      \alpha = 4.
    }
  \end{array}
  \right.
\end{eqnarray}

We can reduce Eqs. (\ref{eq:2nd-causal-010}) and (\ref{eq:2nd-causal-011}) to
\begin{eqnarray}
  \label{eq:2nd-causal-015}
  A^{\alpha\beta,\gamma\delta}\,\frac{\partial}{\partial\tau}\delta X_{\gamma\delta} +
  B^{\alpha\beta,\gamma\delta}\,\delta X_{\gamma\delta} = 0,
\end{eqnarray}
where $A^{\alpha\beta,\gamma\delta}$ and $B^{\alpha\beta,\gamma\delta}$
are defined as
\begin{eqnarray}
  \label{eq:matrixA1}
  A^{\mu\beta,\nu\delta} &\equiv& \langle\, \hat{L}^{-1}\,\varphi^{\mu\beta}_1 \,,\, \hat{L}^{-1}\,\varphi^{\nu\delta}_1 \rangle,\\
  \label{eq:matrixA2}
  A^{\mu\beta,4\delta} &\equiv& \langle\, \hat{L}^{-1}\,\varphi^{\mu\beta}_1 \,,\, \varphi^{\delta}_0 \rangle,\\
  \label{eq:matrixA3}
  A^{4\beta,\nu\delta} &\equiv& \langle\, \varphi^{\beta}_0 \,,\, \hat{L}^{-1} \, \varphi^{\nu\delta}_1 \rangle,\\
  \label{eq:matrixA4}
  A^{4\beta,4\delta} &\equiv& \langle\, \varphi^{\beta}_0 \,,\, \varphi^{\delta}_0 \rangle,\\
  B^{\mu\beta,\nu\delta} &\equiv&- \langle\, \varphi^{\mu\beta}_1 \,,\, \hat{L}^{-1}\,\varphi^{\nu\delta}_1\rangle\nonumber\\
  &&{}+ \langle\, \hat{L}^{-1}\,\varphi^{\mu\beta}_1 \,,\, v^\rho \, \hat{L}^{-1}\,\varphi^{\nu\delta}_1 \rangle \nabla_\rho,\\
  B^{\mu\beta,4\delta} &\equiv&\langle\, \hat{L}^{-1}\,\varphi^{\mu\beta}_1 \,,\, v^\rho \, \varphi^{\delta}_0 \rangle \nabla_\rho,\\
  B^{4\beta,\nu\delta} &\equiv&\langle\, \varphi^{\beta}_0 \,,\, v^\rho \, \hat{L}^{-1}\,\varphi^{\nu\delta}_1\rangle \nabla_\rho,\\
  B^{4\beta,4\delta} &\equiv&\langle\, \varphi^\beta_0 \,,\, v^\rho \, \varphi^\delta_0 \rangle \nabla_\rho.
\end{eqnarray}

We convert Eq. (\ref{eq:2nd-causal-015}) into the algebraic equation,
using the Fourier and Laplace transformations with respect to the spatial variable $\sigma^\mu$
and the temporal variable $\tau$, respectively.
By substituting
\begin{eqnarray}
  \label{eq:2nd-causal-018}
  \delta X_{\alpha\beta}(\sigma\,;\,\tau)
  = \delta \tilde{X}_{\alpha\beta}(k\,;\,\Lambda) \,
  \mathrm{e}^{ik\cdot\sigma - \Lambda\tau},
\end{eqnarray}
into Eq. (\ref{eq:2nd-causal-015}),
we have
\begin{eqnarray}
  \label{eq:2nd-causal-019}
  ( \Lambda\,A^{\alpha\beta,\gamma\delta} - \tilde{B}^{\alpha\beta,\gamma\delta} )
  \, \delta \tilde{X}_{\gamma\delta}
  = 0,
\end{eqnarray}
where $\tilde{B}^{\alpha\beta,\gamma\delta}$ is defined as
\begin{eqnarray}
  \label{eq:matrixB1}
  \tilde{B}^{\mu\beta,\nu\delta} &\equiv&- \langle\, \varphi^{\mu\beta}_1 \,,\, \hat{L}^{-1}\,\varphi^{\nu\delta}_1\rangle\nonumber\\
  &&{}+ \langle\, \hat{L}^{-1}\,\varphi^{\mu\beta}_1 \,,\, v^\rho \, \hat{L}^{-1}\,\varphi^{\nu\delta}_1 \rangle i \, k_\rho,\\
  \label{eq:matrixB2}
  \tilde{B}^{\mu\beta,4\delta} &\equiv&\langle\, \hat{L}^{-1}\,\varphi^{\mu\beta}_1 \,,\, v^\rho \, \varphi^{\delta}_0 \rangle i \, k_\rho,\\
  \label{eq:matrixB3}
  \tilde{B}^{4\beta,\nu\delta} &\equiv&\langle\, \varphi^{\beta}_0 \,,\, v^\rho \, \hat{L}^{-1}\,\varphi^{\nu\delta}_1 \rangle i \, k_\rho,\\
  \label{eq:matrixB4}
  \tilde{B}^{4\beta,4\delta} &\equiv&\langle\, \varphi^\beta_0 \,,\, v^\rho \, \varphi^\delta_0 \rangle i \, k_\rho,
\end{eqnarray}
We note that
$k^\mu$ is a space-like vector satisfying $k^\mu = \Delta^{\mu\nu}\,k_\nu$.
In the rest of this section,
we use the matrix representation
when no misunderstanding is expected.

Since we are interested in a solution other than $\delta \tilde{X} = 0$,
we can impose
\begin{eqnarray}
  \label{eq:2nd-causal-021}
  \det ( \Lambda\,A - \tilde{B} ) = 0.
\end{eqnarray}
It is noted that
Eq. (\ref{eq:2nd-causal-021}) leads to the dispersion relation
\begin{eqnarray}
  \label{eq:2nd-causal-022}
  \Lambda = \Lambda(k).
\end{eqnarray}
The stability of the constant solution
given by Eqs. (\ref{eq:2nd-causal-001})-(\ref{eq:2nd-causal-0042}) against a small perturbation
is equivalent to that $\delta X$ becomes close to the zero with time evolution.
Therefore,
our task is to show that
the real part of $\Lambda(k)$ is positive for any $k^\mu$.

We show that
$A$ is a real symmetric positive-definite matrix as follows:
\begin{eqnarray}
  \label{eq:2nd-causal-023}
  &&w_{\alpha\beta}\,A^{\alpha\beta,\gamma\delta}\,w_{\gamma\delta}\nonumber\\
  &=& \langle\, w_{\mu\beta}\, \hat{L}^{-1}\,\varphi^{\mu\beta}_1 + w_{4\beta}\, \varphi^{\beta}_0 \,,\,
  w_{\nu\delta} \, \hat{L}^{-1}\,\varphi^{\nu\delta}_1 + w_{4\delta} \, \varphi^{\delta}_0 \rangle\nonumber\\
  &=& \langle\, \chi \,,\, \chi \rangle
  > 0, \,\,\,\,\,\,w_{\alpha\beta} \neq 0,
\end{eqnarray}
with
$\chi_p \equiv w_{\mu\alpha}\, \big[\hat{L}^{-1}\,\varphi^{\mu\alpha}_1\big]_p + w_{4\alpha}\, \varphi^{\alpha}_{0p}$.
In Eq. (\ref{eq:2nd-causal-023}),
we have used
the positive-definite property of the inner product (\ref{eq:inner_positive}).

Equation (\ref{eq:2nd-causal-023}) means that
the inverse matrix $A^{-1}$ exists,
and $A^{-1}$ is also a real symmetric positive-definite matrix.
Thus, with the use of the Cholesky decomposition,
we can represent $A^{-1}$ as
\begin{eqnarray}
  \label{eq:2nd-causal-024}
  A^{-1} = {}^tU\,U,
\end{eqnarray}
where $U$ denotes a real upper triangular matrix and ${}^tU$ which is a transposed matrix of $U$.
Substituting Eq. (\ref{eq:2nd-causal-024}) into Eq. (\ref{eq:2nd-causal-021}),
we have
\begin{eqnarray}
  \label{eq:2nd-causal-025}
  \det ( \Lambda\,I - U \, \tilde{B}  \, {}^tU ) = 0,
\end{eqnarray}
where $I$ denotes the unit matrix.
It is noted that
$\Lambda(k)$ is an eigenvalue of $U \, \tilde{B} \, {}^tU$.

We find that
the real part of $\Lambda(k)$ is positive for any $k^\mu$
when
$\mathrm{Re}(U \, \tilde{B} \, {}^tU)$
is a positive definite matrix
where
$\mathrm{Re}(M) \equiv (M + M^\dagger)/2$.
In fact, we can show that
$\mathrm{Re}(U \, \tilde{B} \, {}^tU)$ is positive definite as follows:
\begin{eqnarray}
  \label{eq:2nd-causal-026}
  &&w_{\alpha\beta} \,
  [ \mathrm{Re}(U \, \tilde{B} \, {}^tU) ]^{\alpha\beta,\gamma\delta}
  \, w_{\gamma\delta}\nonumber\\
  &=&
  w_{\alpha\beta} \, [ U \, \mathrm{Re}(\tilde{B}) \, {}^tU ]^{\alpha\beta,\gamma\delta}
  \, w_{\gamma\delta}\nonumber\\
  &=& [w \, U]_{\alpha\beta} \,
  [\mathrm{Re}(\tilde{B})]^{\alpha\beta,\gamma\delta}
  \, [w \, U]_{\gamma\delta}\nonumber\\
  &=& - [w \, U]_{\mu\beta} \, \langle\, \varphi^{\mu\beta}_1\,,\,
  \hat{L}^{-1} \, \varphi^{\nu\delta}_1
  \rangle [w \, U]_{\nu\delta}\nonumber\\
  &=& - \langle\, \psi \,,\,
  \hat{L}^{-1} \, \psi \rangle > 0,
  \,\,\,\,\,\,w_{\alpha\beta} \neq 0,
\end{eqnarray}
with $\psi_p \equiv [w \, U]_{\mu\alpha} \, \varphi^{\mu\alpha}_{1p}$.
The inequality in the final line is satisfied
because the vector $\psi_p$ belongs to the Q${}_0$ space
spanned by the eigenvectors correspondent to the negative eigenvalues of $\hat{L}_{pq}$.
Therefore, we conclude that
the constant solution
given by Eqs. (\ref{eq:2nd-causal-001})-(\ref{eq:2nd-causal-0042})
is stable against a small perturbation
around the general constant solution.

\subsubsection{
  Proof of causality
}

Here, we show that
the propagation speed of the fluctuation $\delta X_{\alpha\beta}$
is not beyond the unity, i.e., the speed of light.
Here,
we suppose that
the propagation speed of $\delta X_{\alpha\beta}$
is given by a character speed,
whose Lorentz-invariant form may be given by 
\begin{eqnarray}
  \label{eq:2nd-causal-027}
  v_{\mathrm{ch}} \equiv \sqrt{- \Delta_{\mu\nu} \, v_{\mathrm{ch}}^\mu \, v_{\mathrm{ch}}^\nu}.
\end{eqnarray}
Here,
we have introduced
the space-like vector $v_{\mathrm{ch}}^\mu$ defined in terms of $\Lambda(k)$ 
given in \eqref{eq:2nd-causal-022} as
\begin{eqnarray}
  \label{eq:2nd-causal-028}
  v_{\mathrm{ch}}^\mu \equiv \lim_{-k^2 \rightarrow \infty} \, \Bigg[ - i\,\frac{\partial}{\partial k_\mu}\Lambda(k)\Bigg].
\end{eqnarray}

By differentiating Eq. (\ref{eq:2nd-causal-025}) with respect to $i\,k_\mu$,
we find that $v_\mathrm{ch}^\mu$ is an eigenvalue of $U\,C^\mu\,{}^tU$, i.e.,
\begin{eqnarray}
  \label{eq:2nd-causal-029}
  \det \Big[ v_\mathrm{ch}^\mu\,I - U\,C^\mu\,{}^tU \Big] = 0,
\end{eqnarray}
with
\begin{eqnarray}
  \label{eq:2nd-causal-030}
  \big[ C^{\rho} \big]^{\alpha\beta,\gamma\delta} \equiv
  \lim_{-k^2 \rightarrow \infty} \, \Bigg[ - i\,\frac{\partial}{\partial k_\rho}\tilde{B}^{\alpha\beta,\gamma\delta}\Bigg],
\end{eqnarray}
whose components are given by
\begin{eqnarray}
  \big[ C^{\rho} \big]^{\mu\beta,\nu\delta} &=&  \langle\, \hat{L}^{-1}\,\varphi^{\mu\beta}_1 \,,\, v^\rho \, \hat{L}^{-1}\,\varphi^{\nu\delta}_1
  \rangle,\\
  \big[ C^{\rho} \big]^{\mu\beta,4\delta} &=&  
  \langle\, \hat{L}^{-1}\,\varphi^{\mu\beta}_1 \,,\, v^\rho \, \varphi^{\delta}_0 \rangle,\\
  \big[ C^{\rho} \big]^{4\beta,\nu\delta} &=&  
  \langle\, \varphi^{\beta}_0 \,,\, v^\rho \, \hat{L}^{-1}\,\varphi^{\nu\delta}_1 \rangle,\\
  \big[ C^{\rho} \big]^{4\beta,4\delta} &=&
  \langle\, \varphi^\beta_0 \,,\, v^\rho \, \varphi^\delta_0\rangle.
\end{eqnarray}
An expectation value of $U\,C^\mu\,{}^tU$
with respect to an arbitrary vector $w^\prime \equiv {}^t(U^{-1})\,w$ 
can be written as
\begin{align}
  \label{eq:2nd-causal-031}
  &\frac{
    \big[w\,U^{-1}\big]_{\alpha\beta} \,
    \big[ U \, C^\mu \, {}^tU \big]^{\alpha\beta,\gamma\delta}
    \, \big[{}^t(U^{-1})\,w\big]_{\gamma\delta}
  }{
    w_{\alpha^\prime\beta^\prime} \,
    \big[U^{-1}\,{}^t(U^{-1})\big]^{\alpha^\prime\beta^\prime,\gamma^\prime\delta^\prime}
    \, w_{\gamma^\prime\delta^\prime}
  }\nonumber\\
  &=
  \frac{
    w_{\alpha\beta} \,
    \big[ C^{\mu} \big]^{\alpha\beta,\gamma\delta}
    \, w_{\gamma\delta}
  }{
    w_{\alpha^\prime\beta^\prime} \,
    A^{\alpha^\prime\beta^\prime,\gamma^\prime\delta^\prime}
    \, w_{\gamma^\prime\delta^\prime}
  }
  =
  \frac{
    \langle\, \chi \,,\, v^\mu \, \chi \rangle
  }{
    \langle\, \chi \,,\, \chi \rangle
  }
  =
  \langle\,v^\mu\,{\rangle}_{\chi},
\end{align}
with
$\chi_p = w_{\mu\alpha}\,\big[ \hat{L}^{-1} \, \varphi^{\mu\alpha}_1 \big]_p + w_{4\alpha}\,\varphi^{\alpha}_{0p}$.
Here,
we have introduced
\begin{eqnarray}
  \label{eq:2nd-causal-032}
  \langle\,O\,{\rangle}_{\chi} \equiv
  \frac{
    \langle\, \chi \,,\, O \, \chi \rangle
  }{
    \langle\, \chi \,,\, \chi  \rangle
  },
\end{eqnarray}
with $O$ being an arbitrary operator.

It is important to note that
if the inequality
\begin{eqnarray}
  \label{eq:2nd-causal-033}
  \sqrt{- \Delta_{\mu\nu} \, \langle \, v^\mu \, {\rangle}_{\chi} \,
  \langle \, v^\nu \, {\rangle}_{\chi}}
  \le 1,
\end{eqnarray}
are satisfied for any $\chi_p$,
we can conclude
\begin{eqnarray}
  \label{eq:2nd-causal-034}
  v_{\mathrm{ch}} = \sqrt{- \Delta_{\mu\nu} \, v_{\mathrm{ch}}^\mu \, v_{\mathrm{ch}}^\nu} \le 1.
\end{eqnarray}
Indeed, we can show 
that the inequality (\ref{eq:2nd-causal-033}) is satisfied in this case.
The proof is given as follows:
First, with the use of the identities
\begin{eqnarray}
  \label{eq:2nd-causal-034-2}
  - \Delta_{\mu\nu} \, v^\mu_p \, v^\nu_p &=& \frac{(p\cdot u)^2 - m^2}{(p\cdot u)^2} \le 1,\\
  \label{eq:2nd-causal-035}
  \langle\,1\,{\rangle}_{\chi} &=& 1,
\end{eqnarray}
we obtain
\begin{eqnarray}
  \label{eq:2nd-causal-036}
  \langle \, - \Delta_{\mu\nu} \, v^\mu \, v^\nu \, {\rangle}_{\chi} \le 1.
\end{eqnarray}
Then,
we notice
\begin{align}
  \label{eq:2nd-causal-037}
  &\langle \, - \Delta_{\mu\nu} \, v^\mu \, v^\nu \, {\rangle}_{\chi}\nonumber\\
  &=
  - \Delta_{\mu\nu} \, \langle \,v^\mu\,{\rangle}_{\chi} \,
  \langle \,v^\nu\,{\rangle}_{\chi}
  +
  \langle \, - \Delta_{\mu\nu} \, \delta v^\mu \,
  \delta v^\nu \, {\rangle}_{\chi}\nonumber\\
  &\ge
  - \Delta_{\mu\nu} \, \langle \,v^\mu\,{\rangle}_{\chi} \,
  \langle \,v^\nu\,{\rangle}_{\chi},
\end{align}
where
$\delta v^\mu_{pq} \equiv \delta v^\mu_{p} \, \delta_{pq}$
with
$\delta v^\mu_p \equiv v^\mu_p - \langle\,v^\mu\,{\rangle}_{\chi}$,
because
\begin{eqnarray}
  \label{eq:2nd-causal-038}
 - \Delta_{\mu\nu} \, \delta v^\mu_p \, \delta v^\nu_p \ge 0,
\end{eqnarray}
due to the fact that $\delta v^\mu_p$ is also a space-like vector.
By combing Eq. (\ref{eq:2nd-causal-037}) with Eq. (\ref{eq:2nd-causal-036}),
we complete the proof.

Thus, our fourteen-moment equation given by Eqs. (\ref{eq:ChapB-RHD1}) and (\ref{eq:ChapB-RHD2})
respects the causality
in the linear analysis around the homogeneous steady state (\ref{eq:2nd-causal-001})-(\ref{eq:2nd-causal-0042}),
in addition to the stability around the static solution.





\begin{thebibliography}{83}
\bibitem[{foo()}]{footnote:vector}
\bibinfo{note}{We call functions of $p$ ``vector" in this article}.
\bibitem [{foo()}]{footnote:spacelike}
  \BibitemOpen
  \bibinfo {note} {Generally, the decomposition of a rank-two
  tensor includes terms proportional to $u^\mu$ or/and $u^\nu$, i.e., time-like
  components. Such terms, however, identically vanish since a space-like tensor is
  defined as a tensor which satisfies both $u_\mu B^{\mu\nu}=0$ and $u_\nu
  B^{\mu\nu}=0$.
  For example, $\nabla^\mu u^\nu$ is decomposed as
  $\nabla^{\mu}u^{\nu}=\Delta^{\mu\nu}\theta/3+\sigma^{\mu\nu}+\omega^{\mu\nu}$.}
\end{thebibliography}
\end{document}